\documentclass[final,letter]{article}

\usepackage{cmap}
\usepackage[T1]{fontenc}
\usepackage{lmodern}
\usepackage[absolute]{textpos}

\usepackage{caption}

\usepackage[margin=1.2in]{geometry}
\usepackage{amssymb,amscd}
\usepackage{mathdots}
\usepackage{mathtools}
\usepackage{suffix}
\usepackage{physics}
\usepackage{subfig}
\usepackage{graphicx,calc}

\usepackage{array,multirow}

\usepackage{enumitem}
\usepackage{dcolumn}
\usepackage{booktabs}

\usepackage{xcolor} 
\usepackage{xstring}
\usepackage{etoolbox}
\usepackage{ifthen}

\usepackage[absolute]{textpos}

\usepackage{pgfplots,pgfplotstable}

\usepackage[bottom]{footmisc}
\usepackage{dsfont}

\usepackage{cleveref}

\setlist[description]{leftmargin=\parindent,labelindent=\parindent}

\crefformat{footnote}{#2\footnotemark[#1]#3}

\newcolumntype{w}[1]{>{\arraybackslash}p{#1}}

\newcommand{\secorsub}[1]{\subsection{#1}}
\newcommand{\suborsubsub}[1]{\subsubsection{#1}}

\newcommand{\fcaption}[1]{\caption{\small #1}}
\newcommand{\tcaption}[1]{\caption{\small #1}}
\newcommand{\smalllineskip}{\baselineskip=10pt}

\newenvironment{richtabular}[1]%
  {
  \centering
  \small\smalllineskip
  \begin{tabular}{#1}}%
  {\end{tabular}\par}

\newlength{\heightperqubit}
\newcommand{\basecircuit}[2][4]{%
  \includegraphics[%
      width=\textwidth,%
      height=#1\heightperqubit,%
      keepaspectratio]%
    {#2}%
  }

\newcommand{\includecircuit}[2][4]{%
  \centerline{%
  \basecircuit[#1]{#2}%
  }}

\newcommand{\incsubcirc}[2][4]{%
  \subfloat[]{\label{fig:#2}\basecircuit[#1]{#2}}%
  }

\def\clap#1{\hbox to 0pt{\hss#1\hss}}

\newcommand{\mathdisp}[2]{\mathchoice{#1}{#2}{#2}{#2}}
\newcommand{\makebig} [1]{\vphantom{\big(\big)}\smash{#1}}

\newcommand{\stilt}[1][\big]{\vphantom{#1(#1)}}
\newcommand{\stiltdisp}[1][\big]{\mathdisp{\stilt[#1]}{}}

\let\norm\undefined
\let\abs\undefined

\DeclareMathOperator*{\argmin}{arg\,min}

\DeclarePairedDelimiter\norm{\lVert}{\rVert}
\DeclarePairedDelimiter\tnorm{\lVert}{\rVert_{tr}}
\DeclarePairedDelimiter\abs{\lvert}{\rvert}

\newcommand{\qreg}[1]{\ifmmode{\text{\qreg{#1}}}\else{\textsc{#1}}\fi}
\newcommand{\qcirc}[1]{\ifmmode{\text{\qcirc{#1}}}\else{\textsc{#1}}\fi}
\def\@OPER#1{\ensuremath{\hat{#1}}}

\newcommand{\NEWOP}[2]{%
  \NewDocumentCommand#1{s}{%
    \IfBooleanTF##1
      {#2^\dagger}
      {#2}
  }
}

\newcommand{\GATE}[1]{\ifmmode{\text{\GATE{#1}}}\else{\textsc{#1}}\fi}
\def\PX/{\ensuremath{\hat{\sigma}_x}}
\def\PY/{\ensuremath{\hat{\sigma}_y}}
\def\PZ/{\ensuremath{\hat{\sigma}_z}}

\def\I/{\@OPER{\mathds{1}}}
\def\U/{\@OPER{U}}
\def\X/{\@OPER{X}}
\def\Y/{\@OPER{Y}}
\def\Z/{\@OPER{Z}}
\def\H/{\@OPER{H}}
\def\T/{\@OPER{T}}
\def\S/{\@OPER{S}}
\def\MSx/{\GATE{MS}_x}
\def\MSy/{\GATE{MS}_y}
\def\LAM/{\@OPER{\Lambda}}

\newcommand{\Haml} {\hat{\Lambda}}

\NEWOP{\Lam}{\hat{\Lambda}}

\def\NOT/{\GATE{not}}
\def\QFT/{\GATE{QFT}}
\def\QFTd/{\ensuremath{\GATE{QFT}^\dagger}}
\def\CNOT/{\GATE{cnot}}
\def\SWAP/{\GATE{swap}}
\def\Toffoli/{\GATE{Toffoli}}
\def\Fredkin/{\GATE{Fredkin}}

\def\CX/{\CNOT/}
\def\CCX/{\Toffoli/}

\newcommand{\R}[1]{\@OPER{R}_{#1}}

\WithSuffix\newcommand\U*{\U/^{\dagger}}

\NewDocumentCommand\Rx{s}{%
  \IfBooleanTF#1
    {\trigbraces{\@OPER{R}_x^\dagger}}
    {\trigbraces{\@OPER{R}_x}}
}
\NewDocumentCommand\Ry{s}{%
  \IfBooleanTF#1
    {\trigbraces{\@OPER{R}_y^\dagger}}
    {\trigbraces{\@OPER{R}_y}}
}
\NewDocumentCommand\Rz{s}{%
  \IfBooleanTF#1
    {\trigbraces{\@OPER{R}_z^\dagger}}
    {\trigbraces{\@OPER{R}_z}}
}

\def\RX/{\ensuremath{\Rx}}
\def\RY/{\ensuremath{\Ry}}
\def\RZ/{\ensuremath{\Rz}}

\newcommand{\SU}[1]{\ensuremath{\text{SU}(#1)}}
\newcommand{\reals}{\mathbb{R}}
\newcommand{\cmplx}{\mathbb{C}}
\newcommand{\ints}{\mathbb{Z}}

\newcommand{\paulis}{\mathcal{P}}

\newcommand{\e}[1]{10^{-#1}}
\WithSuffix\newcommand\e-{\e}

\newcommand{\python}{\texttt{python}}
\newcommand{\cpp}{C\texttt{++}}

\newcommand{\sm}[1]{\!#1\!}

\newcommand{\fft}{\mathfrak{F}}
\WithSuffix\newcommand\fft*{\fft^{-1}}
 
\newcommand{\defeq}   {{{}\triangleq{}}}

\newcommand{\ceil}[1]	{\ensuremath{\lceil	{#1}\rceil}}

\newcommand{\clog}[2] {\ensuremath{\ceil{\log_{#1}#2}}}

\newcommand{\lra}[1][\hspace{1em}]{\ensuremath{{}\xrightarrow{#1}{}}{}}

\newcommand{\cc}[1]{} 

\newcommand{\todo}[1]{\ifthenelse{\isundefined{\draft}}{}%
                     {\textbf{\textcolor{red}{!!!}}}}
\newcommand{\TODO}[1]{\ifthenelse{\isundefined{\draft}}{}%
                     {\textcolor{red}{\hfill(\textit{todo: #1})}}}
\newcommand{\ifthesis}[2]{\ifthenelse{\isundefined{\isthesis}}{#2}{#1}}

\DeclareRobustCommand{\plot}[1]{\cref{plot:#1}}
\DeclareRobustCommand{\fig} [1]{\cref{fig:#1}}

\DeclareRobustCommand{\tab} [1]{\cref{tab:#1}}
\DeclareRobustCommand{\sect}[1]{\cref{sec:#1}}
\DeclareRobustCommand{\eq}  [1]{\cref{eq:#1}}
\DeclareRobustCommand{\apx} [1]{\cref{apx:#1}}
\DeclareRobustCommand{\step}[1] {\labelcref{step:#1}}

\DeclareRobustCommand{\Plot}[1]{\Cref{plot:#1}}

\DeclareRobustCommand{\Tab} [1]{\Cref{tab:#1}}

\DeclareRobustCommand{\Eq}  [1]{\Cref{eq:#1}}
\DeclareRobustCommand{\Apx} [1]{\Cref{apx:#1}}

\newcommand{\sfig}[1]{(\labelcref{fig:#1})}

\newcommand{\nops}{L}
\newcommand{\nbits}{d}
\newcommand{\nsteps}{m}
\newcommand{\nslots}{R}
\newcommand{\nerr}{N_\epsilon}
\newcommand{\npauli}{\omega_\Lambda}

\newcommand{\noerr}{\eprob\!=\!0}

\newcommand{\polyeps}{\varepsilon}
\newcommand{\maxeps}{\varepsilon_{ja}}
\newcommand{\numeps}{\varepsilon_{num}}
\newcommand{\asyeps}{\varepsilon_{asy}}

\newcommand{\eprob}{p_{\epsilon}}

\newcommand{\fprob}{p_f}
\newcommand{\tdist}{\delta_{tr}}
\newcommand{\infid}{\delta_{F}}

\newcommand{\optnsteps}{\nsteps^*}
\newcommand{\optfprob}{p_f^*}

\newcommand{\optinfid}{\infid^*}

\newcommand{\avgfprob}{\overline{p}_{f}}
\newcommand{\avginfid}{\overline{\delta}_{F}}
\newcommand{\estfprob}{\avgfprob}
\newcommand{\estinfid}{\avginfid}

\newcommand{\linfprob}{\chi}
\newcommand{\linfidel}{\zeta}

\newcommand{\ULam} {\hat{\Lambda}}
\newcommand{\ulam} {\lambda}
\newcommand{\thlam}{\theta_\lambda}
\newcommand{\alam} {\stiltdisp{\alpha_\lambda^+}}
\newcommand{\alampm} {\stiltdisp{\alpha_\lambda^\pm}}
\newcommand{\alammp} {\stiltdisp{\alpha_\lambda^\mp}}
\newcommand{\alamd}{\alpha_\lambda^-}
\newcommand{\blam}[1]{\stiltdisp{\beta_\lambda^{#1}}}

\newcommand{\phis} {\phi_0,...,\phi_{\nsteps}}
\newcommand{\philist}{\{\phis\}}
\newcommand{\phivec} {\vec{\phi}}

\WithSuffix\newcommand\phis*{\phi_1,...,\phi_{\nsteps}}
\WithSuffix\newcommand\philist*{\{\phis*\}}

\newcommand{\E}[1]{\@OPER{E}_{#1}}
\WithSuffix\newcommand\E/{\@OPER{E}}

\newcommand{\besselconst}{{\Gamma_0}}

\newcommand{\toolstep}[3]{\textbf{#1:} #2 $\;\;\longrightarrow\;\;$ #3}
\newcommand{\textopt}[1]{\emph{#1}}

\newcommand{\qopt}{\textsc{q}}
\newcommand{\copt}{\textsc{t}}

\newcommand{\initstate} {\psi_0}
\newcommand{\finalstate}{\psi'}

\newcommand{\idealstate}{\psi^*}

\newcommand{\olap}{q}

\newcommand{\ang}{\varsigma}

\newcommand{\uit}  {\hat{V}}
\newcommand{\uemb} {\hat{U}_{\Lambda}}
\newcommand{\uselv}{\hat{W}_\Lambda}
\newcommand{\ucz}  {\hat{W}_{0}}
\newcommand{\urefl}{\ensuremath{\hat{W}_\alpha}}
\newcommand{\uprep}{\ensuremath{\hat{\Pi}_\alpha}}
\newcommand{\uqsp}{\hat{f}_\nsteps\big[\Lam\big]}
\newcommand{\uq}{\trigbraces{\hat{Q}}}

\newcommand{\eLam}{\widetilde{\Lambda}}
\newcommand{\eprep}{\ensuremath{\widetilde{\Pi}_\alpha}}

\newcommand{\uappf}{\trigbraces{\hat{F}_{\nsteps}}}
\newcommand{\uappg}{\trigbraces{\hat{G}_{\nsteps}}}
\newcommand{\uf}{\trigbraces{\hat{F}}}
\newcommand{\ug}{\trigbraces{\hat{G}}}

\WithSuffix\newcommand\uprep*{\hat{\Pi}^{\smash{\dagger}}_\alpha}
\WithSuffix\newcommand\eprep*{\widetilde{\Pi}^{\smash{\dagger}}_\alpha}
\WithSuffix\newcommand\uit*  {\hat{V}^{\dagger}}
\WithSuffix\newcommand\uemb* {\hat{U}^{\dagger}_{\Lambda}}
\WithSuffix\newcommand\uselv2{\hat{W}^{2}_\Lambda}
\WithSuffix\newcommand\uselv*{\hat{W}^{\dagger}_\Lambda}

\newcommand{\dprep}{\dyad{\alpha}{0}}
\WithSuffix\newcommand\dprep*{\dyad{0}{\alpha}}

\newcommand{\corrprep}{\ensuremath{\hat{C}_\Pi}}
\newcommand{\corrselv}{\ensuremath{\hat{C}_\Lambda}}

\newcommand{\upr}{\ensuremath{\hat{T}}}

\newcommand{\fth}{\trigbraces{\tilde{f}}}

\newcommand{\appf}{\trigbraces{\tilde{f}_{\nsteps}}}
\newcommand{\appg}{\trigbraces{\tilde{g}_{\nsteps}}}
\newcommand{\appa}{\trigbraces{\tilde{a}_{\nsteps}}}

\newcommand{\appc}{\trigbraces{\tilde{c}_{\nsteps}}}

\newcommand{\cqsp}{\ensuremath{\mathcal{C}_{f_\nsteps[\Lambda]}}}

\newcommand{\errchan}{\mathcal{E}}

\newcommand{\ecqsp}{\widetilde{\mathcal{C}}_{f_\nsteps[\Lambda]}}

\def\PREP/{\ensuremath{\uprep}}
\def\UPREP/{\smash{\ensuremath{\uprep*}}}
\def\REFL/{\ensuremath{\urefl}}
\def\SELV/{\ensuremath{\uselv}}

\def\CTL/{\qreg{ctl}}
\def\PHS/{\qreg{phs}}
\def\TGT/{\qreg{tgt}}

\newcommand{\tref}[1]{\includegraphics{#1}} 

\newcommand{\boundsplot}[1]{
\begin{figure}[!htb]
  \centering 
  \includegraphics{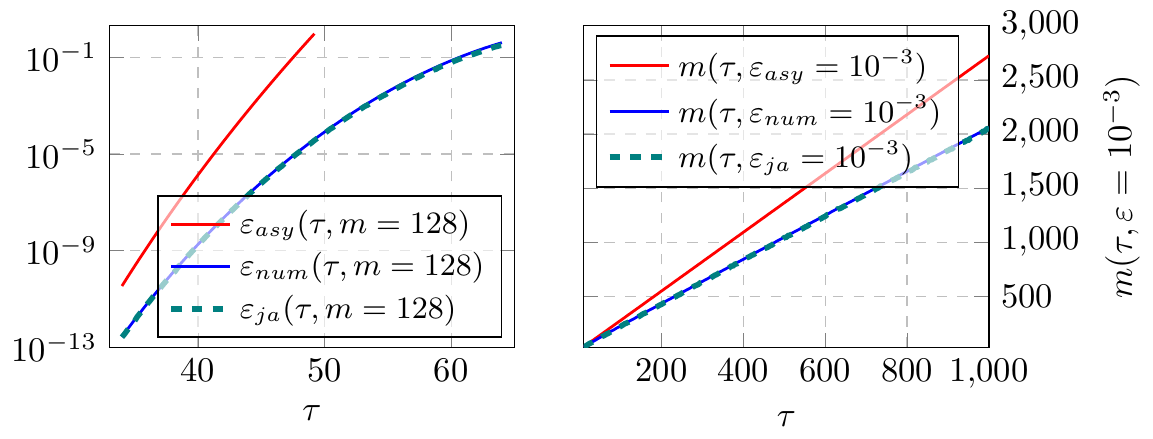}
    \fcaption{Comparison of the closed-form asymptotic error bound ($\asyeps$) and the tighter numerical bound ($\numeps$) computed from the finite sums in \eq{alg:haml:numeps}.
     Left-hand plot shows the magnitude of each bound as a function of $\tau$ at query depth $\nsteps=128$.  Right-hand plot shows the minimum query depth $\nsteps$ such that $\polyeps(\tau,\nsteps)\le10^{-3}$.  In both cases we also show an estimate of the true error $\maxeps$ (dashed lines) computed via numerical maximization of \eq{alg:haml:maxeps},
     which is difficult to distinguish from $\numeps$ (the lines corresponding to $\maxeps$ and $\numeps$ are mostly overlayed)
     }
    \label{plot:#1}
\end{figure}
}

\newcommand{\timingplot}[1]{
\begin{figure}[!htb]
  \begin{minipage}[c]{0.6\textwidth}
  \centering
  \includegraphics{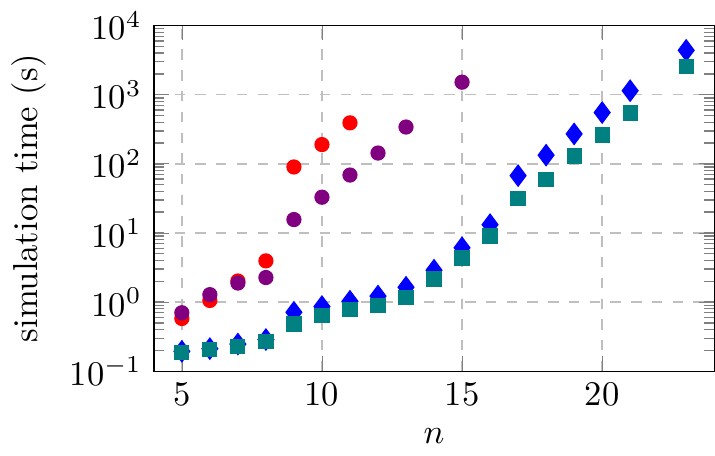}
  \end{minipage}
  \begin{minipage}[c]{0.6\textwidth}
  \end{minipage}\hfill
  \begin{minipage}[r]{.36\textwidth}
    \tref{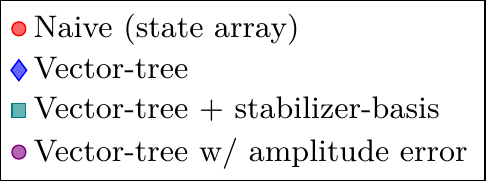}
  \end{minipage}
    \fcaption{Simulation runtimes for error-free $\nsteps=64$ QSP circuits with sizes $5\le n\le23$, run with each simulation enginer.  For the vector-tree simulator we also plot runtimes when systematic amplitude errors are applied (in which case the hybrid vector-tree+stabilizer-basis simulator offers no advantage.
    The jumps in runtime after $n=8$ and $n=16$ correspond to the jump in the size of the \CTL/ register.
    At smaller $n$, the tree and hybrid simulators perform comparably, while the latter reduces runtime by about a factor of two for $n\ge9$.  The simplistic array-style simulator becomes prohibitively slow after $n=11$.  Measured on a
    Dell Precision Tower 5810 with Intel Xeon E5-1660 v3 at 3.00GHz and 64GB RAM at 2133 MHz}
  \label{plot:#1}
\end{figure}
}

\newcommand{\gatesplot}[2]{
\begin{figure}[!htb]
  \centering 
  \includegraphics{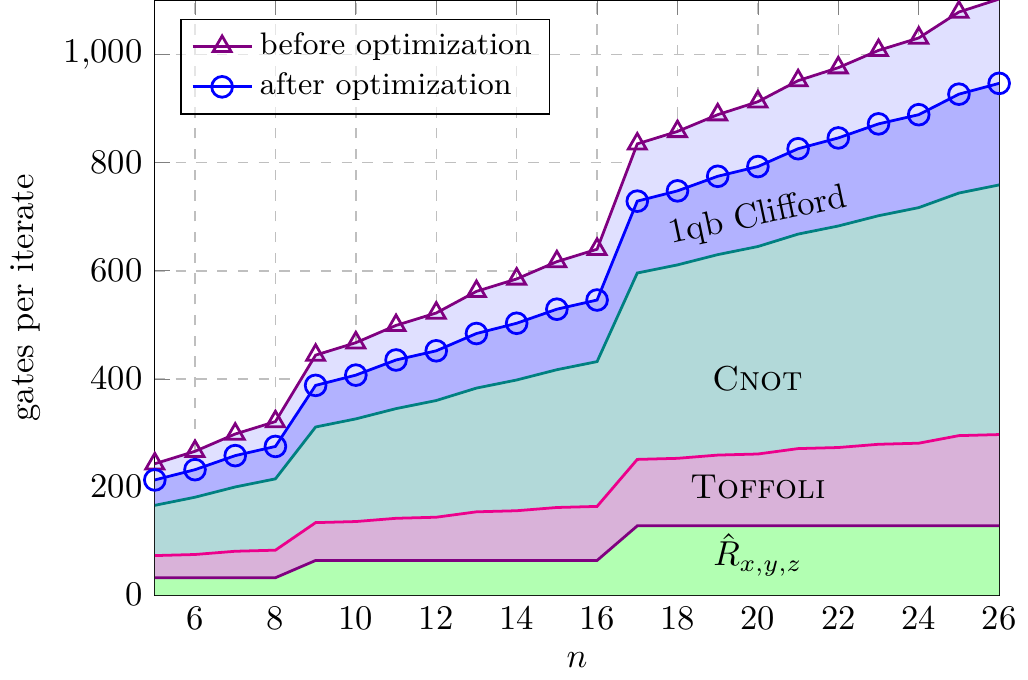}
  \fcaption{#2}
  \label{plot:#1}
\end{figure}
}

\newcommand{\rsltsnplot}[1]{
\begin{figure}[!htb]
  \centering 
  \includegraphics{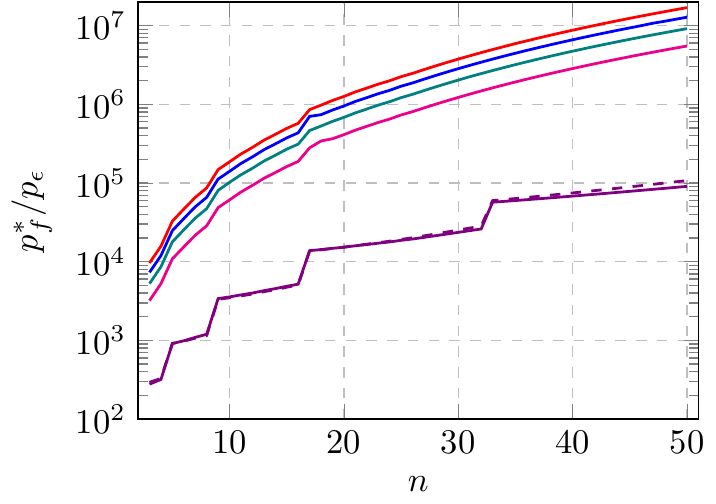}
  \begin{minipage}[c]{0.60\textwidth}
  \end{minipage}
  \hfill
  \begin{minipage}[c]{0.35\textwidth}
  \tref{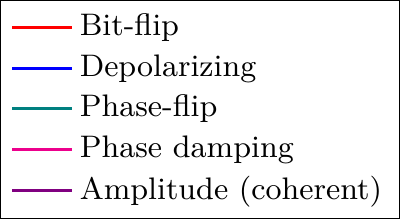}
  \end{minipage}
  \fcaption{Expected at-capacity failure probability $\optfprob(\tau,\errchan)$ (normalized by error strength) of `meaningful' Hamiltonian simulation experiment (i.e. with $\tau=n^2$).  Multiplying the $y$-axis by the error rate $\eprob$ of a given device (where $\epsilon^2=\eprob$ in the case of coherent error) gives the expected failure rate of a meaningful experiment as a function of system size $n$}
  \label{plot:#1}
\end{figure}
}

\newcommand{\rsltsplot}[1]{
\begin{figure}[!htb]
  \centering 
  \includegraphics{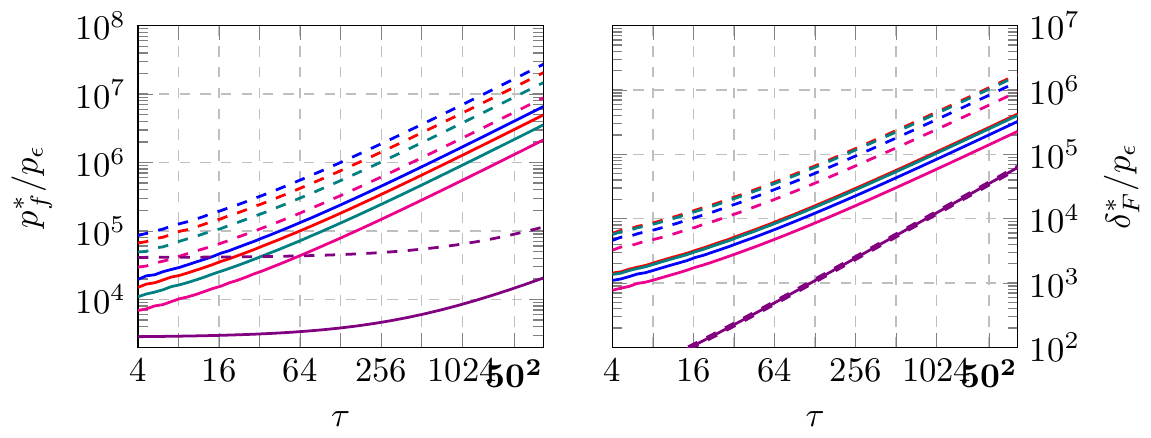}
  \tref{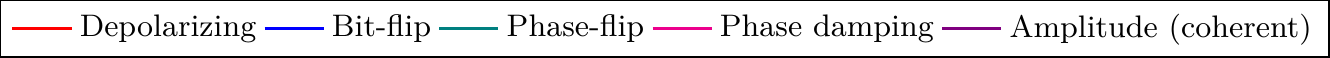}
  \fcaption{Simulation capacity models for $n=11$ (solid lines) and $n=50$ (dashed lines) QSP circuits subject to each error model (see \sect{mthd:error}).  The results shown for $n=11$ are derived directly from empirical simulation results, while the $n=50$ results are estimated by extrapolating from empirical models of the size dependence of the simulation capacity of circuits subject to stochastic and coherent errors}
  \label{plot:#1}
\end{figure}
}

\newcommand{\explot}[1]{
\begin{figure}[!htb]
  \centering
  \includegraphics{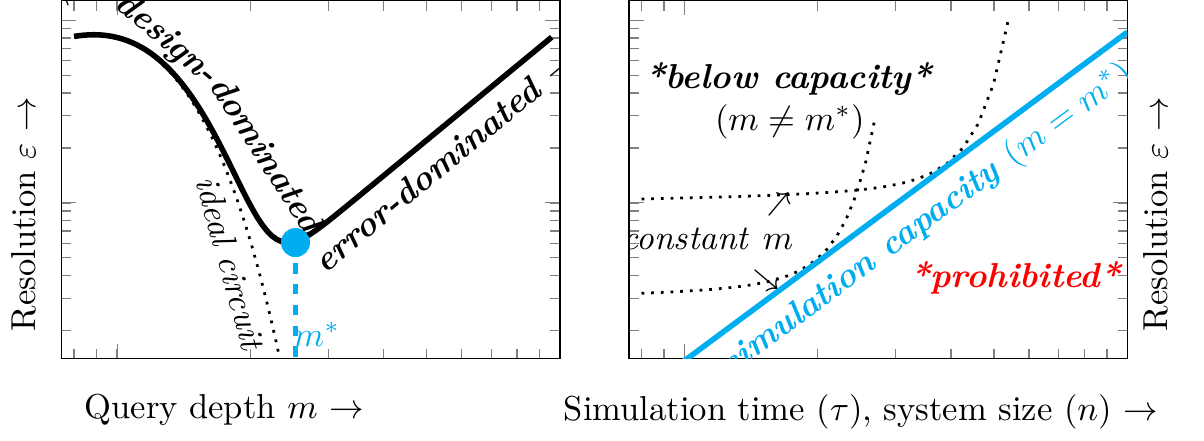}
    \vfill
    \fcaption{Left: a typical configuration plot of a faulty QSP experiment with fixed $(n,\tau,\errchan)$, demonstrating the optimal query depth $\optnsteps(n,\tau,\errchan)$ which balances inherent design error (at $\nsteps<\optnsteps)$ and the depth-dependent accumulation of applied errors.  Right: a typical simulation capacity model, showing the best-possible (i.e. configured with $\nsteps=\optnsteps$) QSP circuit performance as a function of either $\tau$ or $n$}
  \label{plot:#1}
\end{figure}
}

\newcommand{\theorymplot}[1]{
\begin{figure}[!htb]
  \centering
  \includegraphics{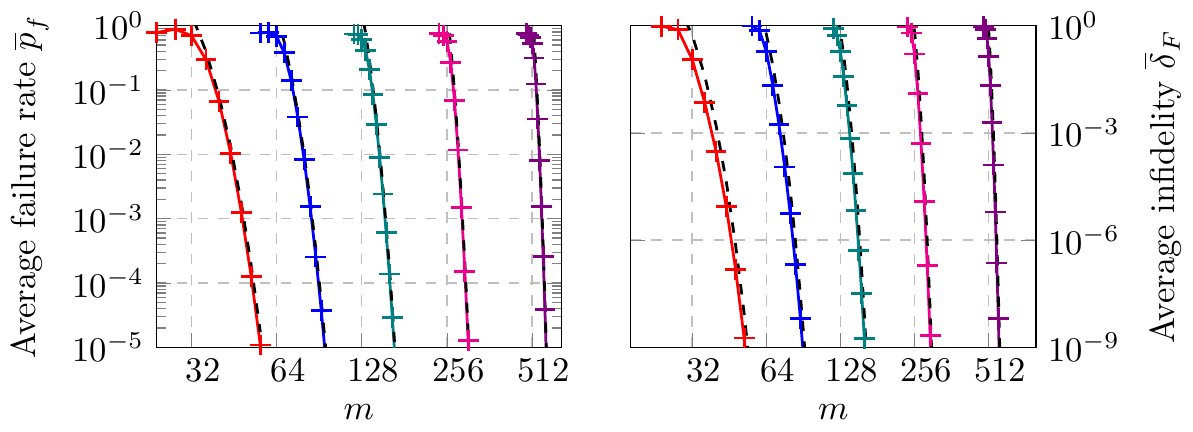}
  \centering 
  \tref{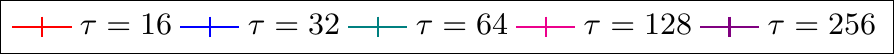}
  \fcaption{Configuration plots of error-free $n=11$ QSP circuits, averaged over 64 randomly generated Hamiltonians and input states.  Dashed lines indicate theoretical upper bounds $\fprob\le4\polyeps$ and $\infid\le(2\polyeps)^2$ described in \sect{alg:err}, where $\polyeps=\numeps(\tau,\nsteps)$ is the numerically computed numerical bound (\eq{alg:haml:numeps}) which was `baked in' to the circuit implementation through the calculation of \PHS/-qubit rotation phases $\phis$}
  \label{plot:#1}
\end{figure}
}

\newcommand{\optplot}[1]{
\begin{figure}[!htb]
  \centering 
  \includegraphics{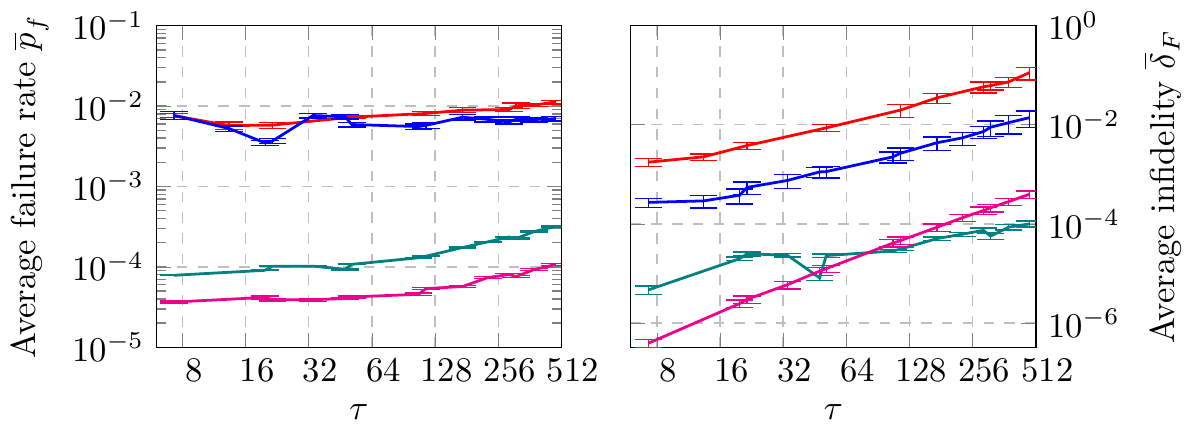}
  \tref{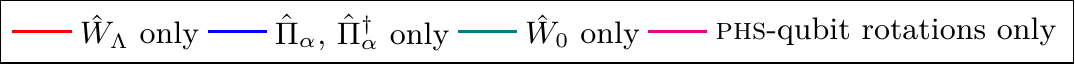}
  \fcaption{Simulation capacity contributions of each subcircuit of an $n=11$ QSP experiment subject to $\epsilon^2=\e6$ systematic amplitude errors, prior to the coherent error optimizations described in \apx{opt}}
  \label{plot:#1}
\end{figure}
}

\newcommand{\subcircplot}[1]{
\begin{figure}[!htb]
  \centering 
  \includegraphics{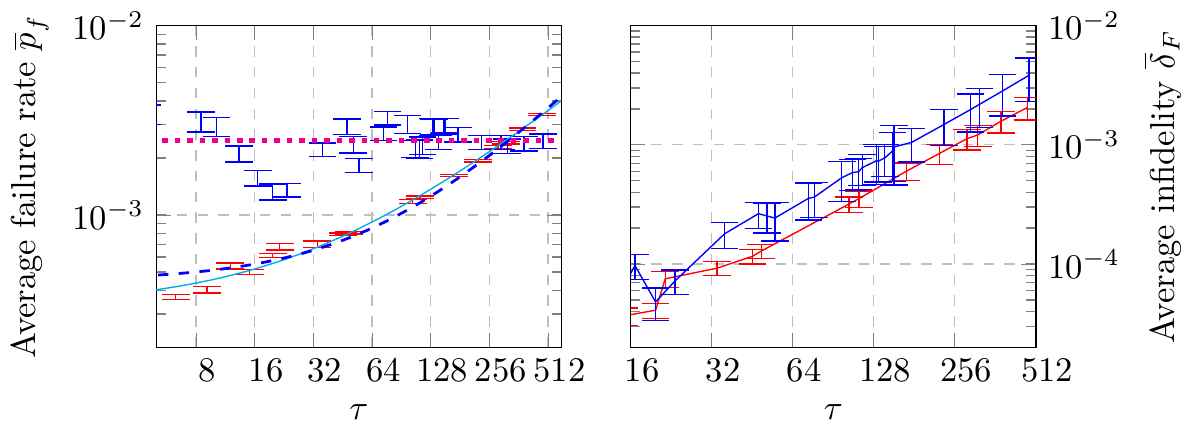}
  \tref{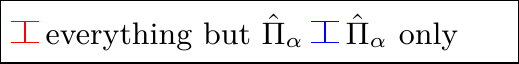}
  \fcaption{Simulation capacity plots of an $n=11$ QSP experiment subject to $\epsilon^2=\e6$ systematic amplitude errors, with errors restricted to either just the $\uprep$ subcircuit, or to every circuit element except the $\uprep$ subcircuit.  Model fits for each contribution (\cref{eq:analysis:prep,eq:analysis:noprep:pow,eq:analysis:noprep:lin}) are also shown, where for the non-$\uprep$ contribution we include both the power-law model (\eq{analysis:noprep:pow}, solid line) and more conservative linear model (\eq{analysis:noprep:lin}, dashed line)}
  \label{plot:#1}
\end{figure}
}

\newcommand{\selvplot}[1]{
\begin{figure}[!htb]
  \centering 
  \includegraphics{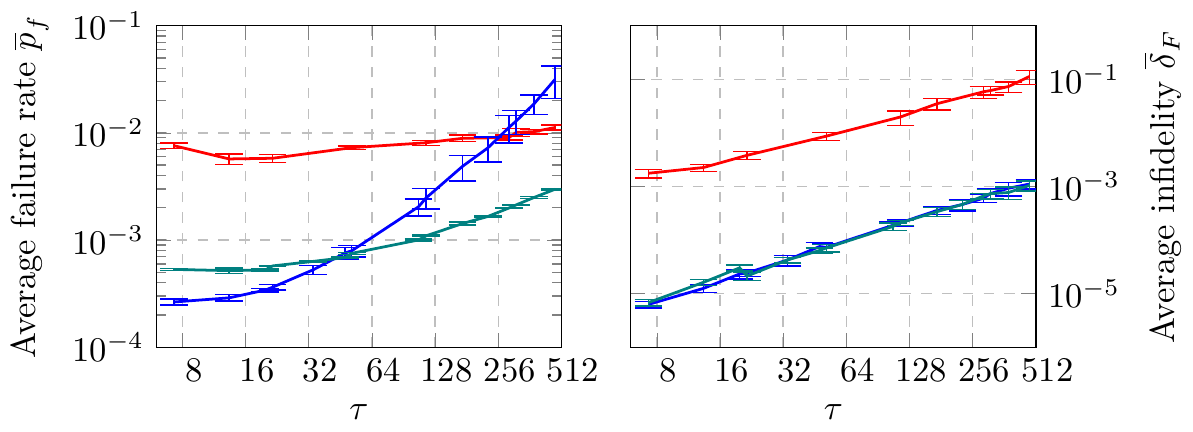}
  \tref{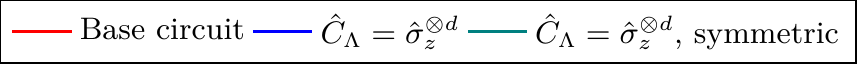}
  \fcaption{Simulation capacity plots of $n=11$ QSP circuits with $\epsilon^2=\e6$ systematic amplitude errors restricted to just the $\uselv$ subcircuit, before and after the coherent error optimizations described in \apx{opt}.  Results are shown for circuits with no optimization, with $\corrselv=\PZ/^{\otimes\nbits}$ echo operators, and with symmetrized $\corrselv=\PZ/^{\otimes\nbits}$ echo operators (see text).  Note that prior to symmetrization, the echo operators significantly reduce the at-capacity failure rate at small $\tau$ but also introduce a strong $\tau$ dependence which makes the circuit worse for sufficiently long simulations}
  \label{plot:#1}
\end{figure}
}

\newcommand{\coherentnplot}[1]{
\begin{figure}[!htb]
  \centering 
  \includegraphics{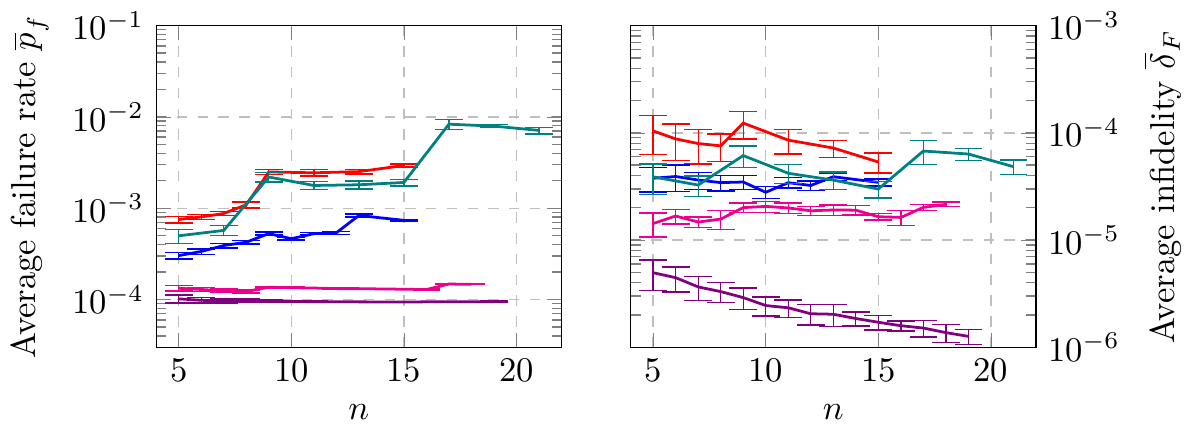}
  \tref{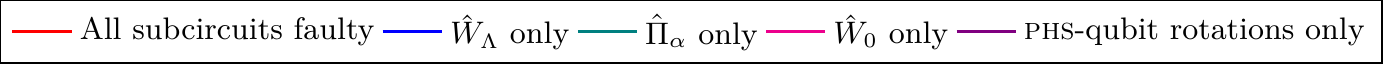}
  \fcaption{Simulation capacity plots showing the size-dependence of $\tau=20$ QSP circuits, with $\epsilon^2=\e6$ systematic amplitude errors either acting throughout the circuit, or restricted to just the $\uprep$, $\uselv$, or $\ucz$ subcircuits or \PHS/-qubit rotations. The largest system size $n$ that we can simulate in each case depends on the memory required for that faulty subcircuit}
  \label{plot:#1}
\end{figure}
}

\newcommand{\depolnplot}[1]{
\begin{figure}[!htb]
  \centering 
  \includegraphics{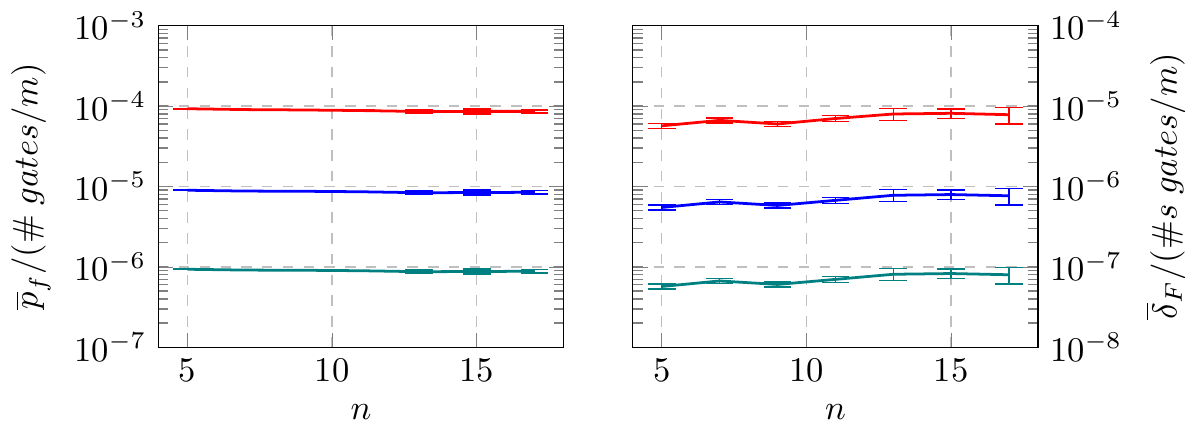}
  \tref{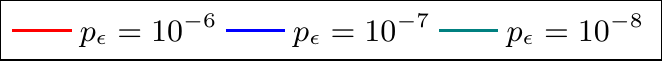}
  \fcaption{Size-dependence of failure rate and infidelity of QSP circuits subject to $\eprob\in\{\e6,\e7,\e8\}$ depolarizing noise configured well into the noise-dominated region ($\nsteps=64$, $\tau=8$).  Results are    scaled by the per-iterate gate count of the simulated circuit.  The purpose of this plot is to confirm that the parameters $\linfprob_n$ and $\linfidel_n$ in \eq{analysis:depol:hyp} quantifying the stochastic noise contribution to infidelity and failure rate is proportional to the total number of gates in the circuit}
  \label{plot:#1}
\end{figure}
}

\newcommand{\depolmplot}[1]{
\begin{figure}[!htb]
  \centering 
  \includegraphics{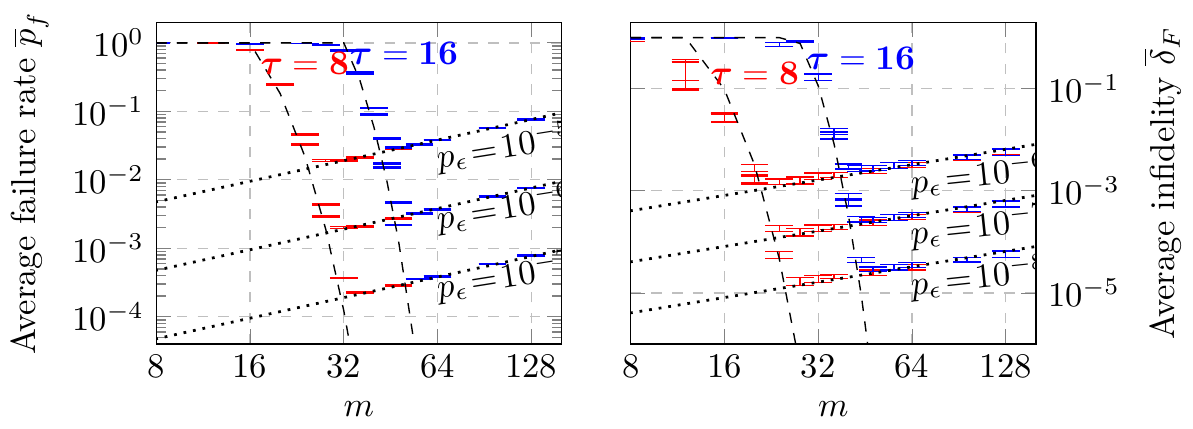}
  \fcaption{Configuration plots for $n=11$ QSP circuits subject to depolarizing noise.  Circuits are generated for $\tau=8$ and $\tau=16$, and simulated with error strengths $\eprob\in\{\e6,\e7,\e8\}$.
    Error bars indicate the deviation between $\{10^5,10^4,10^3\}$ (depending on $\eprob$) Monte Carlo trials (or the equivalent thereof, after importance sampling) of each of 32 different circuits constructed with randomly generated Hamiltonians and initial states. 
    Dashed lines show the empirical model developed from ideal (error-free) simulations (\cref{eq:analysis:th:fprob,eq:analysis:th:infid}).
    Dotted lines show linear fits $\infid=\linfidel_n\nsteps\eprob$ and $\fprob=\linfprob_n\nsteps\eprob$ computed from results in the fault-dominated region
    }
  \label{plot:#1}
\end{figure}
}

\newcommand{\depoltplot}[1]{
\begin{figure}[!htb]
  \centering 
  \includegraphics{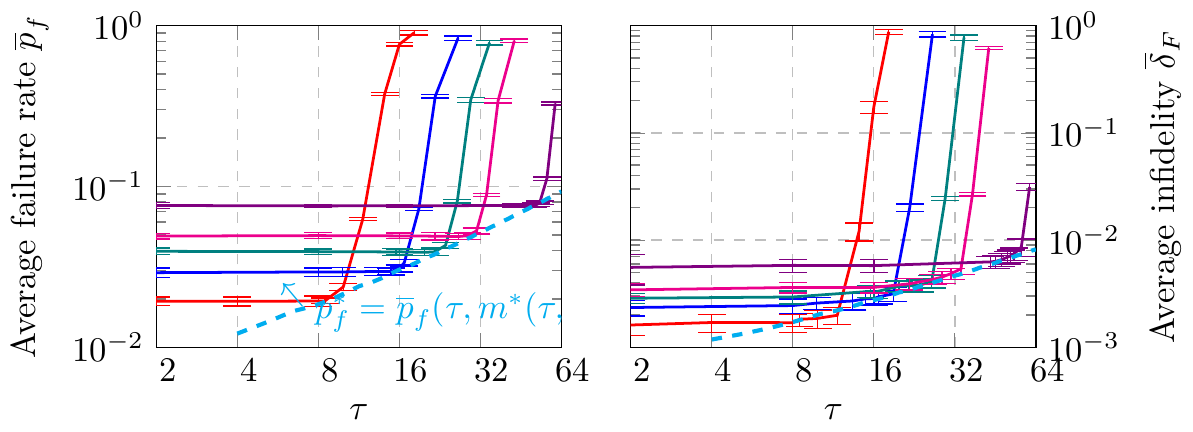}
  \tref{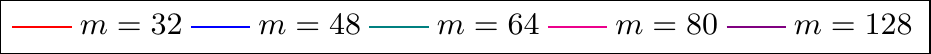}
  \fcaption{Performance of $n=11$ QSP algorithm subject to $\eprob=10^{-6}$ depolarizing noise.  Circuits are generated for $\nsteps\in\{32,48,64,128\}$ with increasing simulation times.  Dotted lines indicate the estimated simualtion capacity boundary.  Error bars indicate the deviation between $10^4$ Monte Carlo trials each of at least 64 different circuits constructed from randomly generated Hamiltonians and initial states}
  \label{plot:#1}
\end{figure}
}

\newcommand{\coherenttplot}[1]{
\begin{figure}[!htb]
  \centering 
  \includegraphics{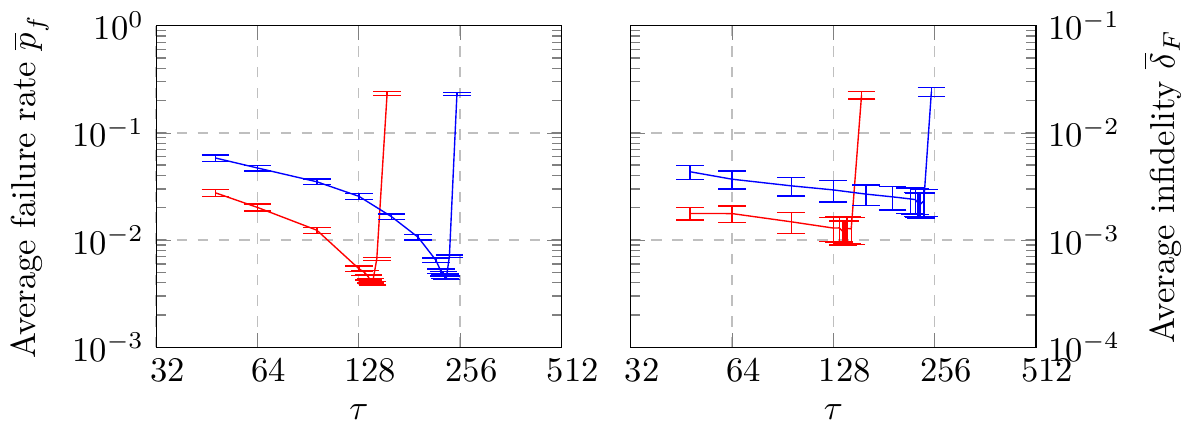}
  \tref{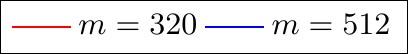}
  \fcaption{Performance of $n=11$ QSP circuits subject to systematic $\epsilon^2=\e6$ amplitude errors with query depths $\nsteps=320$ and $\nsteps=512$, in order to demonstrate the $\tau$-dependence of the error-dominated region and the sharp inflection points when $(\tau,\nsteps)$ is an optimal configuration. Error bars indicate the deviation among 8-24 randomly generated Hamiltonians and input states}
  \label{plot:#1}
\end{figure}
}

\newcommand{\coherentplot}[1]{
\begin{figure}[!htb]
  \centering 
  \includegraphics{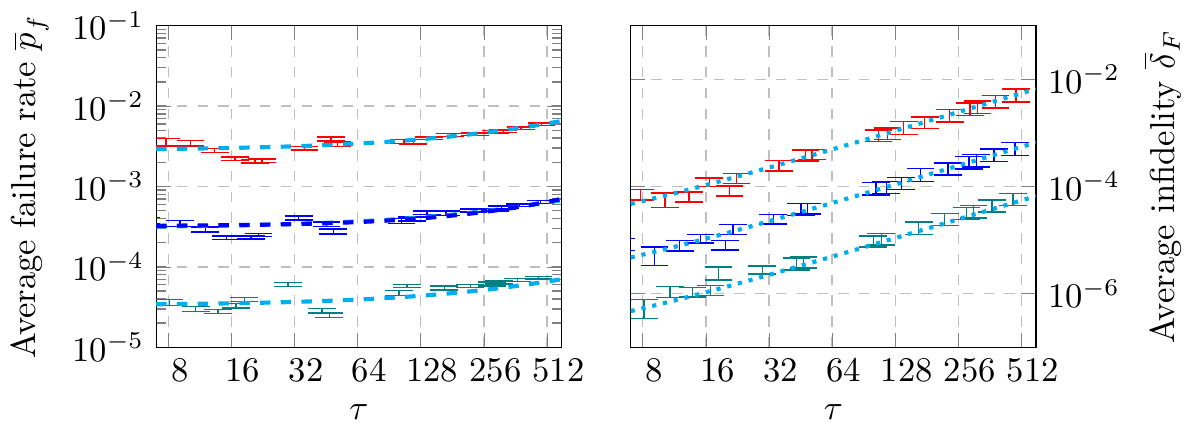}
  \tref{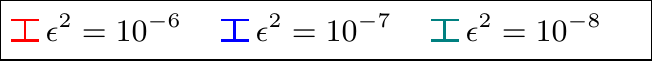}
  \fcaption{`Bootstrapped' simulation capacity plots for $n=11$ QSP circuits subject to systematic amplitude errors with strengths $\eprob\in\{\e6,\e7,\e8\}$.  Inflection points were found by sampling various values of $\tau$ for a given $\nsteps$ near $\numeps(\tau,\nsteps)\approx10\eprob$.  Error bars indicate the deviation among 24-32 randomly generated Hamiltonians and input states.  Dashed lines indicate estimates generated by fitting the error contributions of subcircuits individually
  }
  \label{plot:#1}
\end{figure}
}

\newcommand{\prepryplot}[1]{%
\begin{figure}[!htb]
  \centering 
  \includegraphics{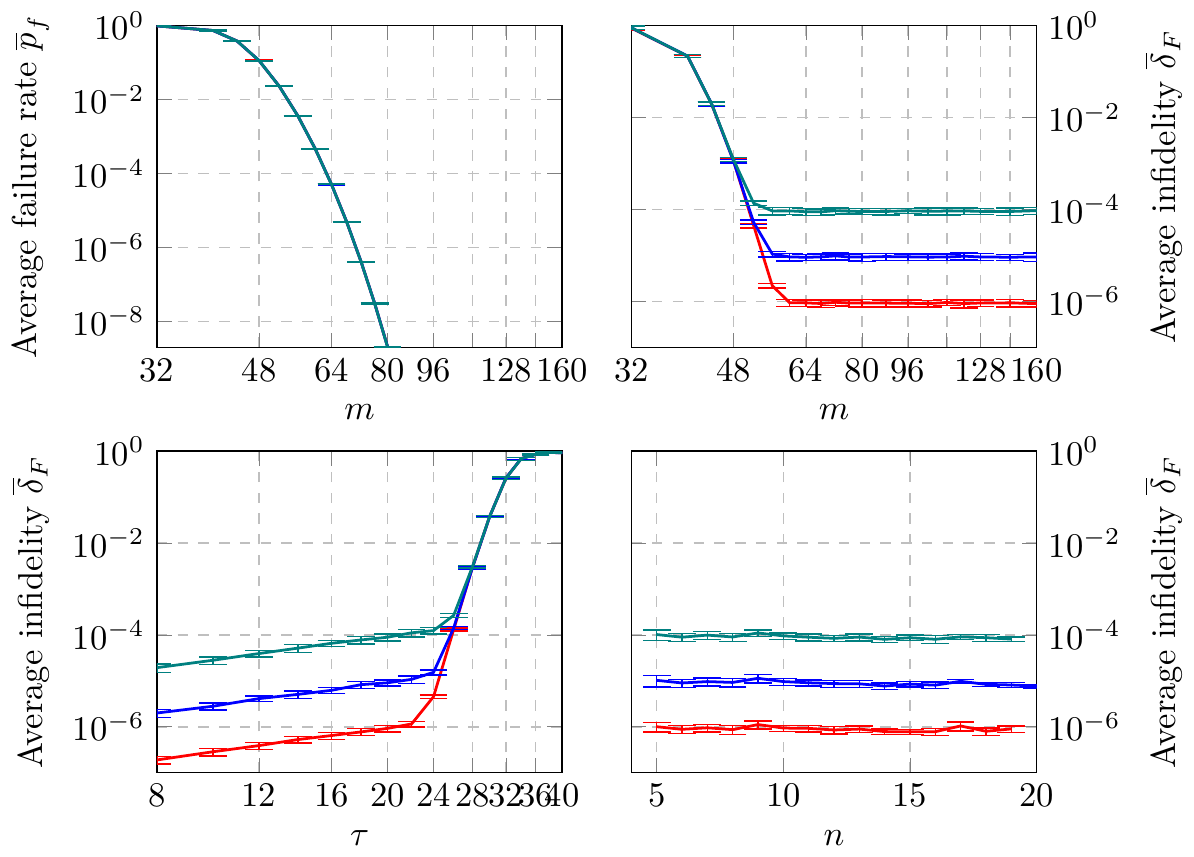}
  \tref{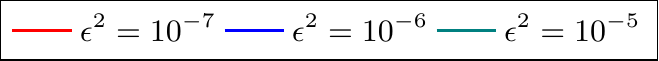}
  \fcaption{Configuration plots generated for $n=11$ QSP circuits with systematic amplitude error restricted to just the $\Ry(\cdot)$ gates in the $\uprep$ subcircuit, at constant $\tau=20$ (top row) and $\nsteps=64$ (bottom row). These errors being equivalent to perturbing the simulated Hamiltonian, in each case failure probability is unchanged from the error-free case (in the top left plot traces for $\epsilon^2\in\{\e5,\e6,\e7\}$ are overlayed). Error bars indicated the standard deviation among circuits constructed from 64 randomly generated Hamiltonians} 
  \label{plot:#1}
\end{figure}
}

\newcommand{\prepcxtplot}[1]{%
\begin{figure}[!htb]
  \centering 
  \includegraphics{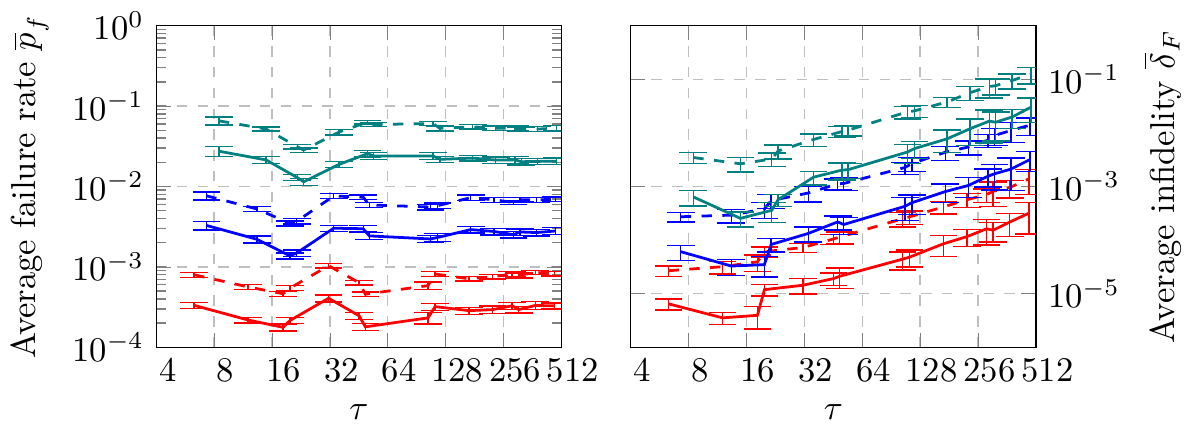}
  \tref{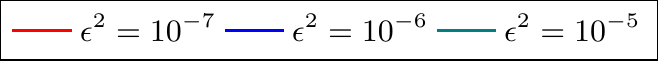}
  \fcaption{Capacity plots generated for $n=11$ QSP circuits with coherent amplitude error restricted to just $\CNOT/$ gates in the $\uprep$ subcircuit.  Solid lines represent circuits after the addition of $\corrprep=\PZ/^{\otimes\nbits}$ echo operators, while dashed lines represent the unmodified circuit (see text).  Error bars indicated the standard deviation among circuits constructed from 64 randomly generated Hamiltonians.} 
  \label{plot:#1}
\end{figure}
}

\title{Empirical determination of the simulation capacity of a near-term quantum computer}
\author{%
Rich Rines\thanks{Massachusetts Institute of Technology Lincoln Laboratory, Lexington, Massachusetts 02420, USA} \footnotemark[2],
Kevin Obenland\footnotemark[1],
and
Isaac Chuang\thanks{Massachusetts Institute of Technology, Cambridge, Massachusetts 02139, USA}
}

\begin{document}

\maketitle

\begin{abstract}
Experimentally-realizable quantum computers are rapidly approaching the threshold of quantum supremacy.  Quantum Hamiltonian simulation promises to be one of the first practical applications for which such a device could demonstrate an advantage over all classical systems.  However, these early devices will inevitably remain both noisy and small, precluding the use of quantum error correction.
We use high-performance classical tools to construct, optimize, and simulate quantum circuits subject to realistic error models in order to empirically determine
the ``simulation capacity'' of near-term simulation experiments implemented via quantum signal processing (QSP), describing the relationship between simulation time, system size, and resolution of QSP circuits which are optimally configured to balance algorithmic precision and external noise.  From simulation capacity models, we estimate
maximum tolerable error rate for meaningful Hamiltonian simulation experiments on a near-term quantum computer.

By exploiting symmetry inherent to the QSP circuit, we further demonstrate that its capacity for quantum simulation can be increased by at least two orders of magnitude if errors are systematic and unitary.
We find that a device with $\epsilon^2=\e5$ systematic amplitude errors could meaningfully simulate systems up to $n\approx16$ with an expected failure rate below $10\%$, whereas the largest system a device with a stochastic error rate of $\eprob=\e5$ could meaningfully simulate with the same rate of failure is between $n=3$ and $n=5$ (depending on the stochastic channel).
Extrapolating from empirical results, we estimate that one would typically need a stochastic error rate below $\eprob\approx{\e8}$ in order to perform a meaningful $n=50$ simulation experiment with a failure rate below $10\%$, while the same experiment could tolerate systematic unitary amplitude errors with strength $\epsilon^2=\eprob\approx\e6$ (corresponding a gate amplitude accuracy of $\epsilon\approx0.1\%$).
\end{abstract}

\begin{textblock*}{7in}(0.75in,10.3in)
\centering
\small
Distribution Statement: A. Approved for public release - distribution is unlimited
\vspace{4pt}
\\
\begin{minipage}[c]{\textwidth}
\footnotesize
This material is based upon work supported by the Assistant Secretary of Defense for Research and Engineering under Air Force Contract No. FA8721-05-C-0002 and/or FA8702-15-D-0001. Any opinions, findings, conclusions or recommendations expressed in this material are those of the author(s) and do not necessarily reflect the views of the Assistant Secretary of Defense for Research and Engineering.
\end{minipage}
\end{textblock*}


\section{Introduction}

Quantum computers were originally conceived as tools for simulating quantum systems, or studying the complex Hamiltonian dynamics of many-body systems which are inaccessible to a classical Turing machine~\cite{Manin1980,Feynman1982}. 
The first efficient quantum protocol for universal Hamiltonian simulation was shown for time-independent local Hamiltonians in 1996~\cite{Lloyd1996}, which was subsequently expanded to nonlocal sparse Hamiltonians~\cite{Aharonov2003}, and then for a variety of other special cases and applications~\cite{Szegedy2004,Berry2007,Childs2010,Low2016,Low2017}.

Today, Hamiltonian simulation remains a uniquely appealing application of quantum computing due to its possible near-term viability.
Quantum systems with as few as 50 qubits exhibit dynamics with complexity outside the capabilities of the best classical computers, while the modern framework of \emph{quantum signal processing} (QSP) establishes an elegant protocol for simulating $n$-qubit quantum systems with only $n+\order*{\log n}$ qubits and optimal resource scaling in terms of evolution time and algorithmic precision~\cite{Low2016,Low2017}.
Recent experimental progress suggests that quantum computers may soon be realized with enough physical qubits to implement an $n=50$ QSP simulation circuit~\cite{Debnath2016,Song2017,Bernien2017,Zhang2017,Wang2018,IBM2019}.

The ability of a near-term device to perform useful quantum computations is fundamentally limited by error.
Physical qubits are inescapably subject to stochastic noise due to unavoidable interactions with their environment, while quantum gate implementations are inevitably plagued with uncharacterised systematic errors due to the finite precision and bandwidth of analog control hardware.
The extensive resources necessary to implement logical qubits and error correcting codes are still
well outside the capability of foreseeable devices.
However, because Hamiltonian simulation is itself an analog task, we expect many errors to induce a steady loss of performance rather than a catastrophic failure.
QSP being a historical descendent of the development of error correcting pulse sequences, we further suggest that symmetries within the QSP circuit can be exploited to facilitate the coherent cancellation of systematic gate errors.
Important previous work~\cite{Childs2017} presents a compelling performance and scalability analysis of explicit, software-generated QSP circuits implemented with perfect gates and qubits.
The guiding question of this work is then, \emph{given a small quantum computer with imperfect gates and no error correction, what exactly will I be able to simulate?}

We approach this question empirically, using classical simulations of faulty $5\le n\le23$ QSP circuits 
in order to develop generalizable models of the expected accuracy and failure rate of optimally-configured QSP circuits subject to realistic error models.
Extrapolating from these models, we aim to resolve two subquestions:
\begin{enumerate}
\item How big of a quantum system could I meaningfully simulate given a device with a gate error rate of $\eprob=\e5$?
\item What errors could I tolerate on a hypothetical near-term device and still meaningfully simulate a $n=50$ Hamiltonian (i.e. just beyond the threshold of quantum supremacy)?
\end{enumerate}
In both cases, we take ``meaningful simulation'' to mean $\order*{n}$ in all parameters, and so explicitly require $\tau\defeq\norm{\Lam}t=n^2$ where $t$ is the evolution time modeled by the simulation and $\Lam$ is the system's Hamiltonian.

Central to this approach is our development of a series of high-performance software tools to design, optimize, and (classically) simulate explicit QSP circuits with gates afflicted by either random noise or systematic error.
Especially in the context of notoriously difficult-to-simulate coherent error models, our ability to generate models which are sufficiently predictive of the best-possible $50$-qubit simulation leans heavily on a number of low-level circuit and toolchain optimizations, both to improve the resource costs and reliability of the QSP circuit implementation and to maximize the range of QSP circuits which we can simulate and precision of the results.

Motivated by~\cite{Childs2017}, we focus on Heisenberg spin-chain Hamiltonians with periodic boundary conditions and randomized coefficients,
\begin{equation}
  \label{eq:intro:haml}
  \Lam
  = \sum_{k=0}^{n-1} \qty{ a_k\PX/^{(k)}\PX/^{(k+1)} +  b_k\PY/^{(k)}\PY/^{(k+1)} +  c_k\PZ/^{(k)}\PZ/^{(k+1)} }
  + \sum_{k=0}^{n-1} h_k \PZ/^{(k)},
\end{equation}
where $\{a_k,b_k,c_k,h_k\}$ are real-valued coefficients, $\hat{\sigma}_\eta^{(k)}$ indicates a Pauli $\hat{\sigma}_\eta\in\{\PX/,\PY/,\PZ/\}$ gate acting on the $k$th bit of the register, and for notational simplicity we take $\hat{\sigma}_\eta^{(n)} \equiv \hat{\sigma}_\eta^{(0)}$ to indicate a periodic boundary condition.
Though nothing in our procedure is unique to this particular model, it both serves to ground our analysis in a physically interesting application and enables comparisons to prior art.
For each circuit configuration, we consider the impact of both systematic over-rotation errors and various common stochastic noise channels, averaged over a set of randomly-generated spin-chain Hamiltonians and initial states. 

While with perfect gates we could construct an arbitrarily precise QSP simulation by configuring the circuit with a sufficiently large query depth $\nsteps$, in the case of faulty gates we find a tradeoff between design-induced inaccuracy at small $\nsteps$ and the additional accumulation of errors with larger $\nsteps$.  
A typical configuration-space diagram for a QSP circuit is shown on the left of \plot{ex}, in which we vary query depth while holding all experimental parameters constant and plot the average resulting resolution (measured as either infidelity or failure rate).
For every error channel $\errchan$, we observe that there is a finite, experiment-dependent and error-dependent optimal query depth $\optnsteps(n,\tau,\errchan)$ at which the error in the simulation output is minimized.  In order to determine the best-possible capacity of QSP circuits for Hamiltonian simulation under each error model, 
we therefore first generate empirical configuration plots in the form of \plot{ex} (left), in order to model the optimal query depth as a function of $n$, $\tau$, and $\errchan$.  

A typical diagram of the \emph{simulation capacity} of a quantum computing platform is shown in \plot{ex} (right).
Optimally-configured circuits with $\nsteps=\optnsteps(n,\tau,\errchan)$ are `at capacity,' tracing out a capacity boundary (highlighted in blue) as a function of either $\tau$ or $n$.  Circuits falling above this boundary are `under capacity' --- expected to underperform the best possible simulation resolution for that device and error channel due to the imperfect calibration of $\nsteps$.  The region below the capacity boundary is a no-go zone: no circuit configuration exists which can be expected to reach this resolution for the given experiment.
Extrapolating from these empirical simulation capacity models while fixing $\tau=n^2$, we can finally predict the best-possible performance of a meaningful Hamiltonian simulation experiment on a hypothetical near-term device.

\explot{ex}

A principle finding of this work is that symmetries within the QSP algorithm inhibit the coherent accumulation of the most significant contributions from systematic unitary errors. 
We find that a device with $\epsilon^2=\e5$ systematic amplitude errors could meaningfully simulate systems up to $n\approx16$ with an expected failure rate below $10\%$, whereas the largest system a device with a stochastic error rate of $\eprob=\e5$ could meaningfully simulate with the same rate of failure is between $n=5$ (under a phase-damping channel) and $n=3$ (under the bit-flip channel).
Extrapolating from empirical results, we estimate that one would typically need a stochastic error rate below $\eprob\approx{\e8}$ in order to perform a meaningful $n=50$ simulation experiment with a failure rate below $10\%$, while the same experiment could tolerate systematic unitary amplitude errors with strength $\epsilon^2=\eprob\approx\e6$ (corresponding a gate amplitude accuracy of $\epsilon\approx0.1\%$).

\ifthesis{\subsection{Outline of this chapter}}{\subsection{Outline}}

We proceed by first (\sect{alg}) providing a broad theoretical outline of the QSP algorithm and the key structures and parameters required for its circuit implementation, followed by the unique software tools, strategies, and optimizations we employ for their analysis in \sect{mthd}.  
We present an empirical analysis of explicit (classical) simulation results in \sect{analysis}.   Finally, in \sect{rslt} we summarize our findings and generalize these results to predict the simulation capacity hypothetical near-term devices.
We include implementation-specific details of the QSP algorithm, theoretical precision, circuit implementation, and circuit optimization in \cref{apx:block,apx:bounds,apx:circ,apx:opt}.  Details of the classical simulation tools used in this work have been relegated to \apx{sim}.


\section{Quantum signal processing}
\label{sec:alg}

Quantum signal processing (QSP)~\cite{Low2016,Low2017} is a generic and broadly applicable protocol for evolving eigenstates of an $n$-qubit \emph{signal} oracle $\Lam\in\cmplx^{2^n\times2^n}$ according to a Hermitian \emph{response} function $\fth:\reals\mapsto\cmplx$.
Specifically, given $\nsteps$ queries of a normal operator $\Lam$ with spectral decomposition $\Lam=\sum_\lambda\dyad\ulam$,
the QSP algorithm implements an eigenstate transformation,
\begin{equation}
  \label{eq:alg:uqsp}
  \uqsp \defeq{} \sum_\lambda \appf(\thlam) \dyad{\ulam},
  \qquad
  \thlam\defeq\asin(\Re[\lambda]/\norm{\Lam});
\end{equation}
where $\appf(\theta)$ is a
degree-$\nsteps$ Fourier series,
\begin{equation}
  \label{eq:alg:appf}
  \appf(\theta) \defeq \sum_{\mathclap{k=-\nsteps/2}}^{\nsteps/2} f_k\,e^{ik\theta},
\end{equation}
satisfying $\norm[\big]{\appf(\theta)}\le1$ with $\{f_{-\nsteps/2},...,f_{\nsteps/2}\}\in\reals^{\nsteps+1}$.

Arbitrary Hermitian response functions $\fth(\theta)$ can be approximated with QSP given a sufficiently good Fourier series approximation $\appf(\theta)\approx\fth(\theta)$.
Remarkably, the asymptotic query depth $\nsteps$ necessary for $\polyeps$-close Hamiltonian simulation in the QSP framework turns out to scale optimally with $\norm[\big]{\Lam}$ and $\abs{t}$ with only additive contributions from the precision $\polyeps$.

\subsection{Unitary QSP}
\label{sec:alg:unitary}

\newcommand{\uqry}{\hat{V}}
\WithSuffix\newcommand\uqry*{\uqry^\dagger}


The simplest (albeit somewhat contrived) instance of quantum signal processing arises when the signal oracle $\Lam=\sum_\lambda e^{i\thlam}\dyad{\ulam}$ is an $n$-bit unitary operator.
In this case, we can query $\Lam$ directly while preserving probability amplitude by applying an ($n+1$)-qubit controlled-$\Lam$ operation (\fig{unitary-query}),
\begin{equation}
  \uqry \defeq
  \dyad+\otimes\I/^{\otimes n} + \dyad-\otimes\Lam,
\end{equation}
where by convention we condition
$\Lam$ on the state $\ket-\defeq\ket0-\ket1$.
QSP then requires just a single ancilla qubit (which we label \PHS/) to control the applications of $\Lam$, in addition to the $n$-qubit input register (labeled \TGT/) containing the quantum state to be transformed (excluding any intermediary ancilla necessary for the implementation of $\Lam$).

\begin{figure}[ht]
   \centering
   \begin{minipage}[c]{0.32\textwidth}
   \centering
   \incsubcirc[3]{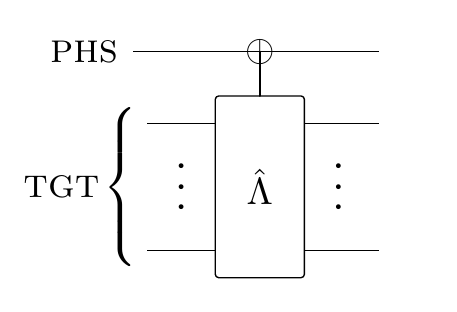}
   \vspace*{13pt}
   \incsubcirc[3]{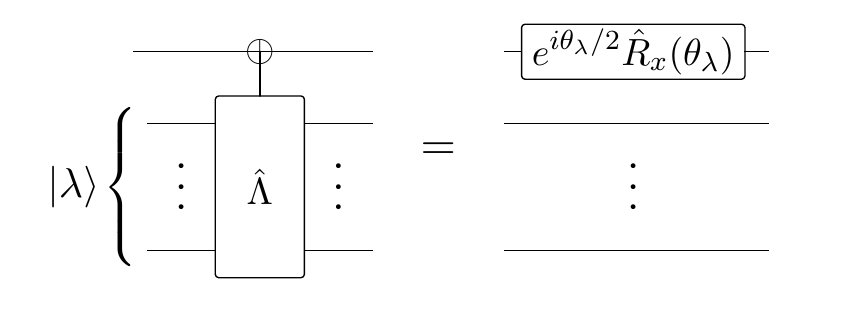}
   \end{minipage}
   \begin{minipage}[c]{0.67\textwidth}
   \centering
   \incsubcirc[3]{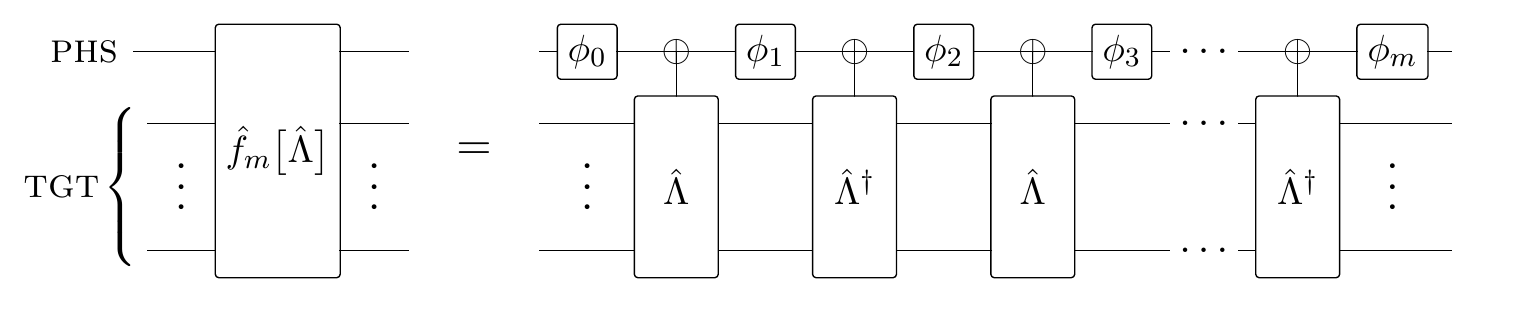}
   \vspace*{13pt}
   \incsubcirc[3]{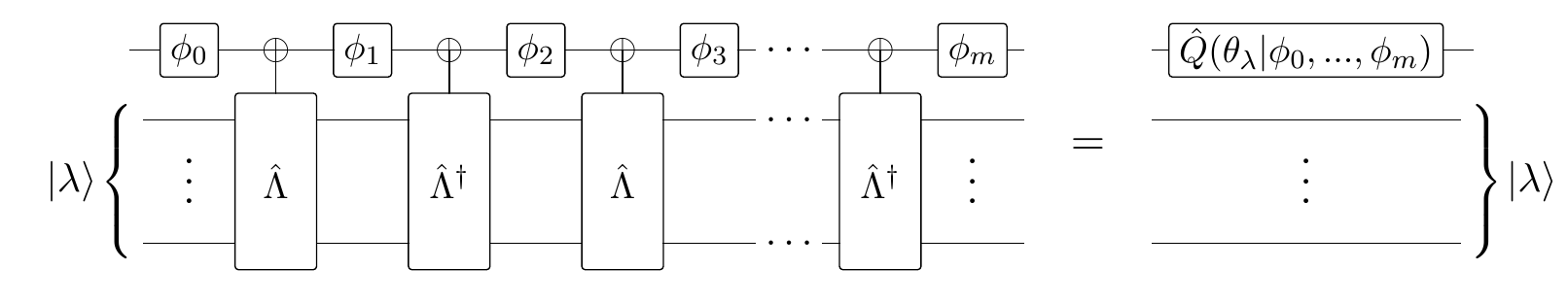}
   \end{minipage}
   \fcaption{
     QSP circuits for a unitary signal operator $\Lam=\sum_\lambda e^{i\thlam}\dyad{\ulam}$:
     \sfig{unitary-query}
     a single query of a unitary Hamiltonian $\Lam$, conditioned on the $\ket-$ state of the \PHS/ qubit;
     \sfig{unitary-query-eig}
     a single query of $\Lam$ applied to an eigenstate $\ulam$ of $\Lam$, equivalent to a rotation of just the \PHS/ qubit;
     \sfig{unitary-full}
     a complete depth-$\nsteps$ QSP circuit, comprising $\nsteps$ queries and $\nsteps+1$ \PHS/-qubit rotations; and,
     \sfig{unitary-full-eig}
      $\uqsp$ applied to eigenstate $\ket\ulam$, equivalent to the response operator $\uq(\thlam\vert\phis)$ applied to just the \PHS/ qubit}
   \label{fig:alg:unitary}
\end{figure}

Applied to an eigenstate $\ket\ulam$ of $\Lam$ in the \TGT/ register (\fig{unitary-query-eig}), $\uqry$ transparently kicks back the corresponding eigenphase $\lambda=e^{i\thlam}$ to the \PHS/ qubit's $\ket-$ state.
Each query is therefore equivalent to an eigenstate-dependent rotation gate $e^{-i\thlam/2}\Rx(\thlam)$ applied to just the \PHS/ qubit:
\begin{equation}
  \label{eq:alg:uqry:eig}
    \uqry 
  = \sum_\lambda \qty( \dyad+ + e^{i\thlam}\dyad- ) \otimes\dyad\ulam
  = \sum_\lambda e^{i\thlam/2}\Rx(\thlam) \otimes \dyad\ulam.
\end{equation}

The complete unitary-$\Lam$ QSP algorithm (\fig{unitary-full}) comprises $\nsteps$ alternating queries of $\Lam$ and $\Lam*$, interleaving $\nsteps+1$ single-qubit rotations $\{\Rz(\phi)\mid\phi=\phis\}$ acting on the \PHS/ qubit.
By alternating between queries of $\Lam$ and $\Lam*$ we eliminate the eigenstate-dependent global phases $e^{\pm i\thlam/2}$ from \eq{alg:uqry:eig}.
The full sequence is then equivalent to applying a single
eigenstate-dependent
\SU2 \emph{response operator} $\uq(\thlam\vert\phis)$ to just the \PHS/ qubit for each superimposed $\Lam$-eigenstate $\ket\lambda$ in the \TGT/ register
(\fig{unitary-full-eig}):
\begin{equation}
  \label{eq:alg:uqlam}
  \qty(\Rz(\phi_\nsteps) \otimes \I/ ) 
  \qty\Big(\uqry*)
  \qty(\Rz(\phi_{\nsteps-1}) \otimes \I/ ) 
  \qty\Big(\uqry)
  \dotsm
  \qty\Big(\uqry)
  \qty(\Rz(\phi_{0}) \otimes \I/ ) 
  =
  \sum_\lambda
  \uq\qty(\thlam\vert\phis) \otimes \dyad\ulam
\end{equation}
where,
\begin{equation}
  \label{eq:alg:uq}
  \uq\qty(\thlam\vert\phis)
  \defeq
  \Rz(\phi_\nsteps) \Rx*(\thlam) \Rz(\phi_{\nsteps-1})\Rx(\thlam) 
  \dotsc
  \Rx(\thlam)\Rz(\phi_0).
\end{equation}

With the `signal' aspect of QSP capture in the queries of $\Lam$, we are free to select phases $\phis$
in order to form the kicked-back eigenphases $e^{i\thlam}$ into the desired Fourier response function $\appf(\thlam)$.
\Eq{alg:uq} can be repartitioned into a sequence of equiangular rotation gates,
\begin{equation}
  \label{eq:alg:uq:eqa}
  \uq\qty(\thlam\vert\phis)
  =
  \Rz*\qty(\varsigma_\nsteps)
  \prod_{k=0}^{\nsteps-1} \R{\varsigma_k}\qty(\thlam);
  \qquad
  \varsigma_k \defeq \sum_{j\le k} \qty(\phi_k+\pi),
\end{equation}
where,
\begin{equation}
  \R{\phi}\qty(\theta)
  \defeq{}
  \Rz(\phi)\Rx(\theta)\Rz*(\phi)
  =
  e^{-i\theta\qty(\cos\phi\PX/+\sin\phi\PY/)/2}.
\end{equation}
Sequences in the form of \eq{alg:uq:eqa}
are thoroughly characterized in~\cite{Low2018,Haah2018}.
After $\nsteps$ queries of $\Lam$, achievable response operators can be expressed,
\begin{equation}
  \uq(\thlam) = \uf_\nsteps(\thlam) + i\PZ/\cdot\ug_\nsteps(\thlam),
\end{equation}
where $\uappf(\theta)$ and $\uappg(\theta)$ have the identical forms,
\begin{equation}
  \uf_\nsteps(\theta) \defeq \sum_{\mathclap{k=-\nsteps/2}}^{\nsteps/2} \;\; f_k \, e^{ik\theta\PX/},
  \qquad
  \ug_\nsteps(\theta) \defeq \sum_{\mathclap{k=-\nsteps/2}}^{\nsteps/2} \;\; g_k \, e^{ik\theta\PX/},
\end{equation}
with expansion coefficients $\{f_{-\nsteps/2},\dotsc,f_{\nsteps/2},g_{-\nsteps/2},\dotsc,g_{\nsteps/2}\}\in\reals^{2\nsteps+2}$ satisfying the unitarity condition,
\begin{equation}
  \label{eq:alg:ug}
  \norm[\big]{\uappg}= 1-\norm[\big]{\uappf}.
\end{equation}

Given a Fourier response function (\eq{alg:appf}),
prescriptions in~\cite{Low2018,Haah2018} demonstrate how the secondary coefficients phases $\phis$ can be efficiently chosen to generate a response operator $\uq(\thlam\vert\phis)$ which encodes the desired response \emph{function} $\appf(\thlam)$ within its $\dyad+$ matrix element:\footnote{Reference~\cite{Low2018} (used by~\cite{Childs2017}) omits the final phase rotation $\Rz(\varsigma_\nsteps)$, introducing the additional criterion $\uq(0)=\I/$ (or equivalently $\phi_0=-\phi_\nsteps$). This restriction is lifted in~\cite{Haah2018}, which appears necessary in general to avoid introducing an $\order*{\polyeps^{1/2}}$ term to the approximation error~\cite{Haah2018}}
\begin{equation}
  \label{eq:alg:response}
    \ev**{\makebig{\uq(\thlam\vert\phis)}}{+}
  = \ev**{\makebig{\uappf(\thlam)}}{+}
  = \appf(\thlam)
\end{equation}
By initializing the \PHS/ qubit in the state $\ket+$ and post-selecting the same state at the end of the circuit,
every eigenstate $\ket\ulam$ in the \TGT/ register will finally absorb the corresponding response $\appf(\thlam)$:
\begin{equation}
  \sum_\lambda \ev**{\makebig{\uq(\thlam\vert\phis)}}{+} \otimes \dyad\ulam
  = \sum_\lambda \appf(\thlam) \dyad\ulam
  = \uqsp,
\end{equation}
recovering the QSP operation defined at the outset (\eq{alg:uqsp}).

\subsection{Normal QSP}
\label{sec:alg:normal}

The unitary-$\Lam$ QSP construction described in \sect{alg:unitary} can be generalized to the case that the signal oracle $\Lam\in\cmplx^{2^n\times2^n}$ is any $n$-qubit normal operator\footnote{In fact, the recently-introduced framework of quantum singular value transformation~\cite{Gilyen2018} has generalized the results and strategies of QSP to \emph{any} complex-valued matrix $\Lam$}.
If $\Lam$ is nonunitary, direct queries of $\Lam$ are no longer trace-preserving.
To recover the deterministic behavior of the unitary algorithm we require an additional $\nbits$-qubit ancillary register (labeled \CTL/) in order to ``qubitize'' the evolution within a larger $(n+\nbits)$-qubit Hilbert space~\cite{Low2017}.

The goal of qubitization is to embed $\Lam$ within a subspace of some $(n+\nbits)$-qubit unitary propagator $\uemb$.
To construct $\uemb$, we require a reflection operator $\uselv\in\SU{2^{n+\nbits}}$ and projectors $\uprep\defeq\dprep$, $\uprep*\defeq\dprep*$ for some $\nbits$-qubit state $\ket\alpha$,
such that,
\begin{equation}
  \label{eq:alg:normal:block}
  \uprep*\uselv\uprep =
    \begin{pmatrix}
      \mqty{\Lam/\norm[\big]{\Lam}&\cdot\,\cdot \\ \rotatebox[origin=c]{90}{$\cdot\,\cdot$}&\rotatebox[origin=c]{315}{$\cdot\,\cdot$}}
    \end{pmatrix}
    = \dyad0\otimes\Lam/\norm[\big]{\Lam} + \dotsb,
\end{equation}
is an $(n+\nbits)$-qubit \emph{block encoding} of $\Lam$.  The specific implementations of $\uselv$ and $\uprep$ used in this work can be found in \apx{block}, while block encodings for a number of other circumstances can be found in \cite{Gilyen2018}.

Mirroring Grover's insight for quantum search~\cite{Grover1996}, we can then construct $\uemb$ by coupling $\uselv$ with a second reflection operator $\urefl\defeq{2\dyad\alpha-\I/^{\otimes\nbits}}$, where $\urefl=\uprep\ucz\uprep*$ is constructed from the projectors $\uprep,\,\uprep*$ and Grover's diffusion operator $\ucz\defeq2\dyad0-\I/^{\otimes\nbits}$ and acts on just the \CTL/ register.
Applied to an eigenstate $\ket\ulam$ of $\Lam$, the combined operator $\uemb=\uselv\urefl=\uselv\uprep\ucz\uprep*$ is then equivalent to the rotation $\Ry(2\thlam) = e^{-i\thlam\PY/}$ in the invariant eigenstate-dependent \SU2 subspace spanned by the ($n+\nbits$)-qubit states,
\begin{equation}
  \ket\alam\defeq\ket\alpha\ket\ulam,
  \qquad
   \ket\alamd\defeq\frac{\uselv-\lambda}{\sqrt{1-\lambda^2}}\ket\alam.
\end{equation}

As described in \apx{block}, by initializing the \CTL/ register to $\ket*\alpha$, we can ensure that each superimposed eigenstate $\ket*\ulam$ of the initial state will remain in the corresponding invariant subspace, so that queries of $\uemb$ will again kick back the eigenstate-dependent phase $e^{i\thlam}$ to the \PHS/ qubit.  We can therefore select phases $\phis$ to dial in a desired response function exactly as in the unitary case.

\begin{figure}[ht]
   \centerline{\incsubcirc[7]{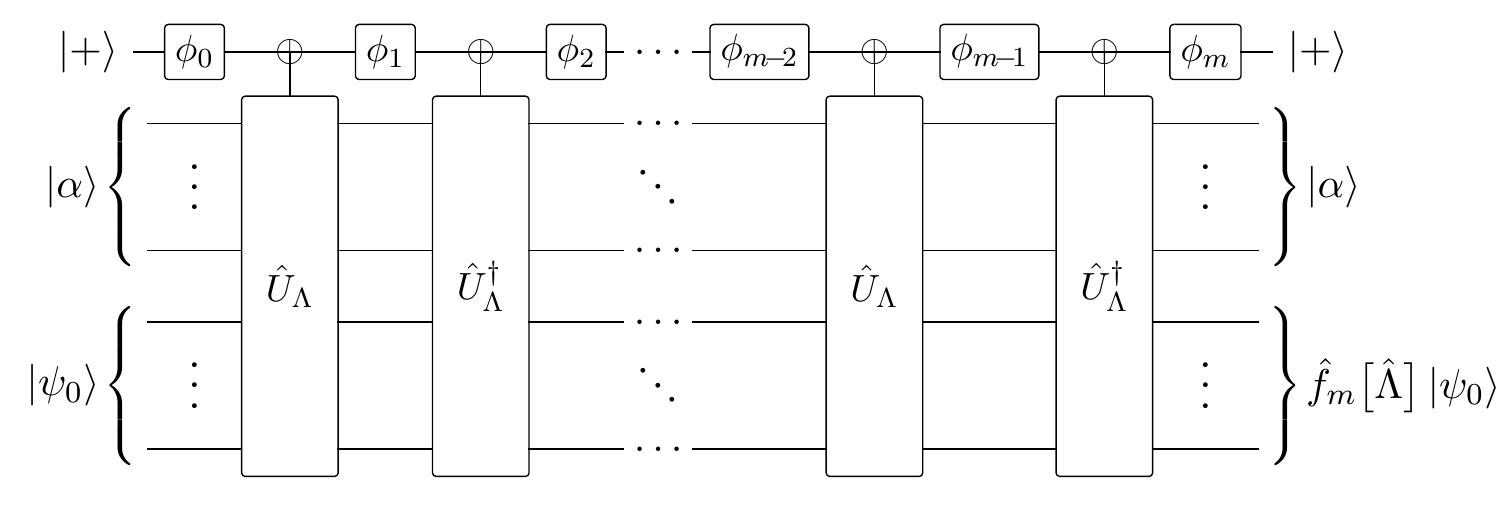}}
   \centerline{\incsubcirc[7]{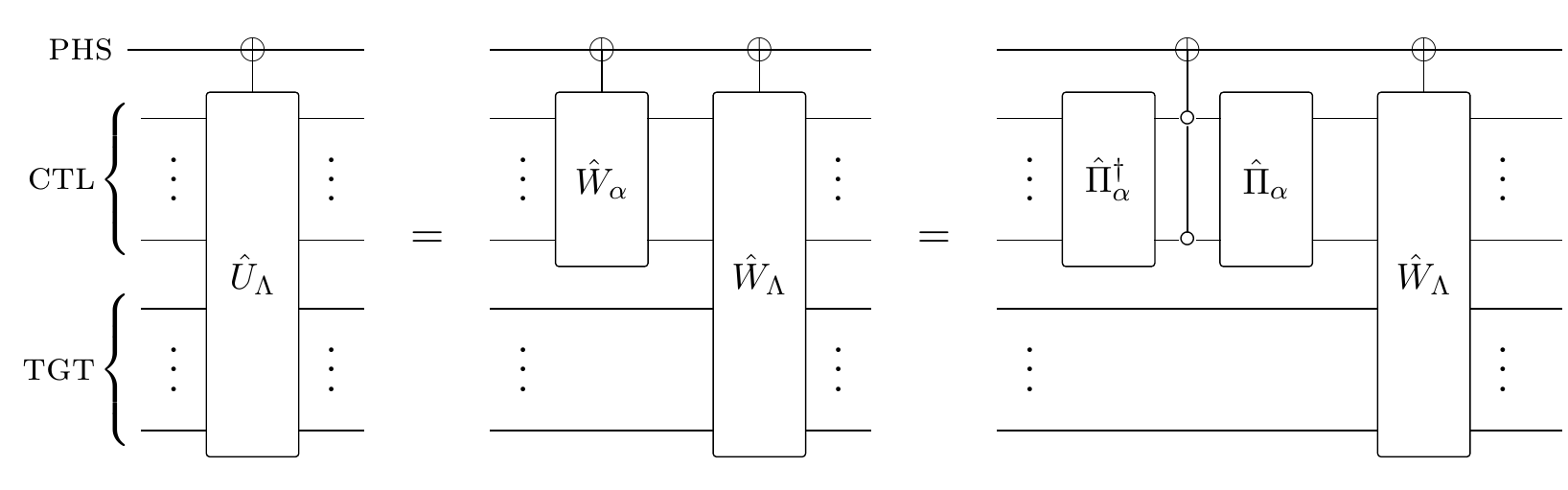}}
   \fcaption{
     \sfig{normal-full} The complete depth-$\nsteps$ QSP circuit for nonunitary $\Lam$, requiring the $\nbits$-qubit \CTL/ register initialized to the state $\ket\alpha$ in addition to the qubits required in the unitary case.
     \sfig{normal-query} A single query $\dyad+\otimes\I/+\dyad-\otimes\uemb$, of the ``qubitized'' signal operator for a nonunitary (normal) signal operator $\Lam\in\cmplx^{2^n\times2^n}$, constructed from the paired reflection operators $\uselv$ and $\urefl=2\dyad\alpha-\I/=\uprep\ucz\uprep*$.  Reflection operators $\uselv$ and $\ucz$ are conditioned on the \PHS/ qubit's $\ket-$ state. 
     The conditional version of the operator $\ucz=2\dyad0-\I/$ is equivalent to a \Toffoli/ gate controlled by all $\nbits$ qubits in the \CTL/ register and acting on the \PHS/ qubit, as drawn
   }
\end{figure}

\subsection{Hamiltonian simulation on a quantum signal processor} 
\label{sec:alg:haml}

The goal of Hamiltonian simulation is to model the Schr\"odinger evolution of an $n$-qubit input state $\ket{\psi_0}$ induced by a known Hamiltonian $\Haml(t)$.  For time-independent $\Haml(t)=\Haml$, Hamiltonian simulation amounts to efficiently approximating the unitary propagator,
\begin{equation}
	\label{eq:alg:haml:schrodinger}
	e^{-i\Haml t} 
	= \sum_{\lambda} e^{-i\lambda t} \dyad\ulam, 
\end{equation}
where $\Haml = \sum_\lambda\lambda\dyad{\ulam}$ with $\lambda\in\reals$ is the spectral decomposition of the (Hermitian) Hamiltonian operator $\Haml$.

To implement quantum simulation on a quantum signal processor,
we must reformulate \eq{alg:haml:schrodinger} in the form of \eq{alg:uqsp}.
We therefore require a degree-$\nsteps$ Fourier series approximation of the angular response function,
\begin{equation}
  \label{eq:alg:haml:angular}
  \fth(\thlam) 
  = e^{-it\lambda} 
  = e^{-i\tau\sin\thlam}
\end{equation}
where $\tau\defeq\norm{\Lam}t$ is a normalized evolution time parameter and $\thlam\defeq\asin\big(\lambda/\norm\Lam\big)$.
Expanding \eq{alg:haml:angular} using an inverse Fourier transform, we happen across Bessel's first integral,
\begin{equation}
  \label{eq:alg:haml:ifft}
  f_k' = \int_0^{2\pi}\frac{d\theta}{2\pi} e^{-ik\theta} e^{i\tau\sin\theta} = J_k(\tau),
\end{equation}
where $J_k(\tau)$ is the $k$th Bessel function of the first kind. 
The corresponding degree-$\nsteps$ Fourier series,
\begin{equation}
  \label{eq:alg:haml:appf'}
  \appf'\qty(\thlam) = \sum_{\mathclap{k=-\nsteps/2}}^{\nsteps/2} J_k(\tau) e^{ik\thlam},
\end{equation}
is then equivalent to a truncated version of the Jacobi-Anger expansion, which is known to converge to \eq{alg:haml:angular} super-exponentially in $\nsteps$~\cite{Abramowitz1965}.
We can compute the error resulting from the finite-order expansion from the excised tails of the expansion:
\begin{equation}
  \label{eq:alg:haml:maxeps}
  \maxeps(\thlam,\tau,\nsteps) \defeq
    \norm[\big]{ \appf'\qty(\thlam) - e^{-i\tau\sin\thlam} }
  = \norm[\Bigg]{ \;\;\; \sum_{\mathclap{\abs{k}>\nsteps/2}} J_{k}(\tau) e^{ik\thlam}\; }.
\end{equation}

Though the ideal response function (\eq{alg:haml:angular}) falls on the unit circle for all $\theta\in\reals$,
the truncated series in \eq{alg:haml:appf'} is only bound to $1-\maxeps(\theta)\le\norm[\big]{\appf(\theta)}\le1+\maxeps(\theta)$.  In order to coerce the approximation into the QSP framework, we therefore must rescale the expansion coefficients by $1/(1+\polyeps)$ for some $\polyeps\ge\max_{\thlam}\maxeps(\thlam)$.  The final Fourier approximation is then,
\begin{equation}
  \label{eq:alg:haml:appf}
  \appf(\thlam) =
  \qty(\frac{1}{1+\polyeps})\;\; \sum_{\mathclap{k=-\nsteps/2}}^{\nsteps/2} J_k(\tau) e^{ik\thlam},
\end{equation}
such that,
\begin{equation}
  \label{eq:alg:haml:appf:bounds}
  (1-2\polyeps)\le\norm[\big]{\appf(\thlam)}\le1.
\end{equation}
Finally, we can compute $\phivec$ from \eq{alg:haml:appf} using the techniques in~\cite{Low2018,Haah2018}.

\subsubsection{Error bounds}
\label{sec:alg:err}

The algorithmic precision of the QSP circuit is design-limited by both the finite order of the Fourier series approximation and the subsequent rescaling of the truncated series in order to fit it into the structure of a valid \SU2 response operator $\uq(\thlam)$.
From \eq{alg:haml:appf:bounds}, the trace distance between the ideal and computed states after rescaling can be bound by,
\begin{equation}
  \label{eq:alg:haml:tdist}
  \tdist 
    = \max_{\ket\psi} \frac{1}{2} \tnorm*{\big( \uqsp - e^{-i\Lam{t}} \big)\ket\psi}
    \le \max_{\thlam} \norm*{ \appf(\thlam) - e^{-i\tau\sin\thlam} }
    \le 2\polyeps.
\end{equation}

This error can be manifested in two ways.  First, because in general $\norm[\big]{\appf}<1$ for finite $\nsteps$, there is nonzero chance that we will observe $\ket{-}$ when measuring the final state $\uq(\thlam)\ket+$ of the \PHS/ qubit, such that the QSP algorithm fails in post-selection.  This algorithmic failure probability can be bound,
\begin{equation}
  \fprob \defeq
        1 - \max_{\thlam} \norm[\Big]{\ev{\uq(\thlam)}{+}}^2
      = 1 - \max_{\thlam} \norm[\big]{ \appf(\thlam) }^2
      =     \max_{\thlam} \norm[\big]{ \appg(\thlam) }^2
      \le 4\polyeps.
\end{equation}

If the algorithm does succeed in post selection, the approximation error will also limit the accuracy of the observed output state.  
Due to the maximization over all input states, empirical measurement of the maximum trace distance of a channel in the form of \eq{alg:haml:tdist} can be computationally impractical.
We will therefore characterize the accuracy of the final simulation state with the
state infidelity $\infid$, defined as one minus the state fidelity, or,
\begin{equation}
  \label{eq:mthd:fidelity}
  \infid \defeq
  = 1 - \qty(\Tr\sqrt{\sqrt\rho\rho^*\sqrt\rho})^2,
\end{equation}
where $\rho$ and $\rho^*$ are the final density matrices of the noisy and noiseless evolutions, respectively.
If we reject simulations with erroneous measurements in post-selection, the ideal simulation result will always be a pure state $\rho^*=\dyad{\psi^*}$ where $\ket{\psi^*}=e^{-i\Lam{t}}\ket{\psi_0}$.  In this case, \eq{mthd:fidelity} can be simplified to
\begin{equation}
  \infid(\rho,\psi^*)
  = 1 - \ev\rho{\psi^*}.
\end{equation}
The computational advantage of computing the state (in)fidelity is that it can be used to estimate the average channel (in)fidelity estimated via Monte Carlo sampling, rather than requiring the evaluation of all $4^n$ basis states of $\SU{2}^{\otimes n}$.

In general, state fidelity will only loosely bound trace distance.  For pure ideal state $\ket{\idealstate}$, we can compare
\begin{equation}
  \label{eq:mthd:tdist-bound}
  \infid \qty(\rho,\psi^*)
  \le \tdist\qty(\rho,{\psi^*})
  \le \sqrt{\infid\qty(\rho,\psi^*)},
\end{equation}
where the upper bound is saturated iff $\rho$ is also a pure state.
However, if we consider only the final state of the \TGT/ register in the case that the algorithm succeeds, this upper bound condition is met, so that \eq{alg:haml:tdist} guarantees $\infid\le(2\polyeps)^2$.

\subsubsection{Bounds on $\polyeps$}
\label{sec:alg:bounds}

An important subtlety in the implementation of the QSP circuit 
is that
the error bound $\polyeps\ge\max_{\thlam}\maxeps(\thlam,\tau,\nsteps)$ 
is an \emph{input} in the definition of the response function $\appf(\theta)$ (\eq{alg:haml:appf}) used to compute $\phivec$.  Whatever value we use for $\polyeps$ (provided it is sufficiently large to bound $\maxeps$) therefore gets ``baked in'' to the phases $\phis$, so that the resolution of the resulting simulation is limited by $\polyeps$ even if $\polyeps\gg\max_{\thlam}\maxeps(\thlam)$ only loosely bounds the error from truncating the Fourier expansion (\eq{alg:haml:maxeps}).

We therefore derive two separate error bounds: a closed-form, analytical \emph{asymptotic} upper bound ($\asyeps$) to verify the asymptotic behavior performance of the circuit, and a tighter but less illustrative \emph{numerical} bound ($\numeps$) used for phase calculation to minimize the error which gets baked in to the circuit construction.
In addition to optimizing the resolution of the resulting circuit, tightly configuring $\polyeps$ is also essential to our analysis in that it reduces variation in $\fprob$ and $\infid$ between circuits configured with the same $\polyeps$, making it possible to develop reliable protocols for selecting optimal configuration parameters when generating QSP circuits.

For $\nsteps\gg\tau$, we can use Bessel function properties to bound the error sum in \eq{alg:haml:maxeps} with a closed-form expression.  As derived in in \apx{bounds}, this asymptotic error bound is,
\begin{equation}
  \label{eq:alg:haml:asyeps}
  \asyeps(\tau,\nsteps)
  \defeq
  \frac{8\abs{\tau/2}^{\nsteps/2+1}}{3(\nsteps/2+1)!}
        \qty{ 1 + \abs*{\frac{\tau}{m}}^2 }^{1/2} 
  \lesssim \frac{1.68}{\sqrt{\nsteps}} \qty( \frac{e\abs{\tau}}{\nsteps+2} )^{\nsteps/2+1}, 
\end{equation}
where the final term results from the Sterling approximation.  As a closed-form expression, $\asyeps$ is useful in characterizing the asymptotic complexity and performance of the QSP protocol.  In particular, solving the r.h.s of \eq{alg:haml:asyeps} for $\nsteps$, one can bind the asymptotic query depth necessary for an error-$\polyeps$ Hamiltonian simulation~\cite{Gilyen2018},
\begin{equation}
  \nsteps\approx e\abs{\tau} + \order{\tfrac{\log(1/\polyeps)}{\log(e+\log(1/\polyeps)/\tau)}}.
\end{equation}
This (mostly) additive dependence on $\polyeps$ is a defining feature of the QSP implementation of Hamiltonian simulation.

While useful for characterizing asymptotic performance, \eq{alg:haml:asyeps} is not a tight bound on $\maxeps$ in the intermediate region considered in this work.  For the calculation of phases $\phis$ we therefore construct a tighter \emph{numerical bound} resulting from the numerical computation of \eq{alg:haml:maxeps}.  As derived in \apx{bounds}, for $\tau\le\nsteps/2$ the numerical bound can be expressed with the finite sum,
\begin{equation}
  \label{eq:alg:haml:numeps}
  \numeps(\tau,\nsteps) \defeq{} \qty{
    \abs[\Bigg]{1 - J_0(\tau) - 2\sum_{\mathclap{k=1}}^{\nsteps/4} J_{2k}(\tau)}^2
  + \abs[\Bigg]{\besselconst(\tau) - 2\sum_{\mathclap{k=2}}^{\mathclap{\nsteps/4}} J_{2k-1}(\tau)}^2
  }^{1/2},
\end{equation}
where,
\begin{equation}
  \besselconst(\tau) \defeq 
  \tau J_0(\tau) +  \frac{\pi\tau}{2}
  \Big\{ J_1(\tau)H_0(\tau) - J_0(\tau)H_1(\tau) \Big\}
\end{equation}
and $H_k(\tau)$ are the Struve functions~\cite{Struve1882}.
By construction it is always the case that $\maxeps\le\numeps\le\asyeps$.

As a demonstration of the behavior of the two limits, both the asymptotic and numerical bounds are plotted for $\nsteps=128$ in \plot{bounds} alongside numerical estimates of $\maxeps$ computed by numerically maximizing $\norm[\big]{\appf(\theta)-e^{-i\tau\sin\theta}}$ directly (dashed line).
Although it is difficult to extract the exact asymptotic form of $\nsteps$ from $\maxeps$, it can be shown that $\nsteps\sim\abs{2\tau}$ to first order in $\tau$, indicating a constant factor resource reduction of $2/e$ enabled by using $\numeps$ in the phase calculation process.  Further, for $\nsteps\ge\tau/2$ \eq{alg:haml:numeps} is monotonic in both $\tau$ and $\nsteps$, facilitating the numerical calculation of the necessary query depth $\nsteps(\tau,\numeps)$ for a given simulation time $\tau$ and precision $\numeps$.  Computed query depths for $\asyeps(\tau,\nsteps)=\numeps(\tau,\nsteps)=10^{-3}$ are compared in the righthand plot in \plot{bounds}.

\boundsplot{bounds}


\section{Implementation}
\label{sec:mthd}

Given an ideal device with infinitely precise quantum gates, the theoretical framework presented in \sect{alg} promises arbitrarily precise time-$\tau$ Hamiltonian simulation with only an additive contribution from the desired precision $\polyeps$ to the required query depth $\nsteps$.
Unfortunately, quantum computers are inherently non-ideal machines. 
To understand the capacity of real quantum devices for Hamiltonian simulation, we must transform the theoretical protocol of \sect{alg} into explicit quantum circuit elements and characterize them subject to realistic constraints.  To this end we employ a set of custom software tools for generating, optimizing, and simulating QSP circuits with an imperfect set of gates.  
Combined,
the toolchain takes as input a QSP experiment uniquely described by the configuration parameters $(\Lam,\tau,\nsteps)$, an initial state $\ket\initstate$, and an error model $\errchan$, and returns estimates of both the failure probability $\estfprob$ and infidelity $\estinfid$ of the resulting experiment.
The analysis toolchain can be described in four primary stages, summarized here and detailed in \cref{sec:mthd:response,sec:mthd:circ,sec:mthd:error,sec:mthd:sim}.

\begin{enumerate}
\item \toolstep{Phase calculation}{$(\tau,\,\nsteps)$}{$\phivec=\philist$}  \label{step:tools:ph} \\
We compute the phase sequence $\philist$ generating the degree-$\nsteps$ Fourier response function $\appf(\theta)$.  For Hamiltonian simulation the ideal response function $\fth(\theta)=e^{-i\tau\sin\theta}$ depends just on $\tau$, so $\phivec$ is determined entirely from the configuration parameters $(\tau,\nsteps)$
\item \toolstep{Circuit construction}{$\Lam,\,\phivec$}{{circuit} $\cqsp$ implementing $\uqsp$ (QASM)} \label{step:tools:circ} \\
The system Hamiltonian $\Lam$ is used to implement the unitary signal operator $\uemb$, which is then repeated between the \PHS/-qubit rotations computed in stage \step{tools:ph} in order to generate the complete QSP circuit $\cqsp$ implementing $\uqsp$ (output in QASM)
\item \toolstep{Error modeling}{$\errchan,\,\cqsp$}{faulty circuit $\ecqsp$} \label{step:tools:err} \\
Given the ideal circuit constructed in step \ref{step:tools:circ} and an error channel description, we generate a set of discrete error operators to be placed throughout the circuit in order to model the faulty circuit $\ecqsp$
\item \toolstep{Simulation}{$\ecqsp$, initial state $\ket\initstate$}{$\fprob,\,\infid$} \label{step:tools:sim} \\
Given the QSP circuit and error placements generated in steps \labelcref{step:tools:ph,step:tools:circ,step:tools:err} and an initial state $\ket{\initstate}$ of the \TGT/ register. 
The simulator executes the faulty circuit and returns both failure probability $\fprob$ and infidelity $\infid$ of its final state 
\end{enumerate}

An essential component of this analysis is our leverage of a set of low-level software optimizations implemented within each stage of our toolchain.  
These optimizations can be broadly separated into two categories. The first (labeled \qopt{} below) primarily impacts the instantiated quantum circuit itself, in order to improve the performance, resource usage, and ultimately the simulation capacity of the instantiated experiment.
The second group (labeled \copt below) only affect the performance to the classical analysis toolchain.
Though these optimizations have no effect on the capacity of the underlying circuit,
they serve to maximize the ``meta-capacity'' of our analysis procedure, or the
range of configuration parameters that we can reliably simulate and characterize via Monte Carlo analysis.
Both categories of optimizations are essential to our ability to generate models which can be extrapolated to the best-possible simulation capacity of an $n=50$ experiment.

The most significant optimizations we implement in each toolchain step are:
\begin{description}
\item[Phase calculation:]\hfill
\begin{itemize}
\item[\qopt] \textopt{Numerical error bounds:} as described in \sect{alg:bounds}, the numerical calculation of error bounds (\eq{alg:haml:numeps}) reduces both the query depth required for a given algorithmic precision and the variation between circuits constructed with the same set of resources
\item[\copt] \textopt{Phase reuse:} because the phases $\philist$ depend only on the parameters $\nsteps$ and $\tau$, we can often reuse the same phases for a number of different experiments, so that in most cases we bypass all of the complexity of phase calculation
\item[\copt] \textopt{Memoization:} by combining phase calculation protocols from~\cite{Haah2018}, existing high-performance libraries for multiprecision arithmetic and root-finding, and extensive use of memoization for high-precision subcalculations, we are able to generate circuits up to $\nsteps\le1024$
\end{itemize}
\item[Circuit construction:]\hfill
\begin{itemize}
\item[\qopt] \textopt{Subcircuit annihilation:} algorithmic symmetries in the QSP circuit allow various subcircuits to be merged or annihilated between adjacent phased iterates, reducing the overall gate count of a typical QSP circuit by 18-22\%
\item[\qopt] \textopt{Peephole optimization:} by algorithmically commuting, merging, and annihilating nearby gates we are typically able to further reduce the gate count by about 13\%
\end{itemize}
\item[Error modeling:]\hfill
\begin{itemize}
\item[\copt] \textopt{Importance sampling:} in the case of stochastic noise, we leverage importance sampling techniques in order to minimize the number of Monte Carlo trials required for randomizing error placements
\item[\copt] \textopt{Deterministic post-selection:} because in the QSP algorithm any nonzero measurement result flags an algorithmic failure, we vastly reduce the requisite number of Monte Carlo trials (to just one if error placement is also deterministic) by replacing every measurement operator with a deterministic $\dyad0$ projector, and using the amplitude of the final state to compute the total failure probability $\fprob$ in a single shot
\end{itemize}
\item[Simulation:]\hfill
\begin{itemize}
\item[\copt] \textopt{Vector-tree simulation:} we implement the parallel vector-tree data structure introduced in~\cite{Obenland1998} to reduce both memory usage (by taking advantage of sparsity of the simulated state's representation in Hilbert space) and computational complexity of gate execution
\item[\copt] \textopt{Stabilizer-basis representation:} to further increase sparsity and reduce the computational overhead of executing Clifford gates we represent states in terms of an evolving stabilizer basis (cf. \cite{Aaronson2004})
\end{itemize}
\end{description}

Between the subcircuit annihilation, peephole optimization, and query depth reduction resulting from the numerical error bounds, the circuit optimization steps reduce the overall gate count of a typical QSP experiment by roughly 47-50\%.  In addition to improving the circuit's faulty-gate performance, this reduction in complexity is also beneficial in reducing the overall simulation runtime.

Combined,
the toolchain optimizations enable us to model circuits up to $\nsteps\le1024$ and $n\le23$ when errors are nonexistent, stochastic, or constrained to certain subsets of the circuit, and $n\le16$ under a global coherent error model.
Of the toolchain-specific optimizations, the simulation strategies are most significant: had we just adopted a na\"ive state-array simulation strategy we would have been fundamentally limited to simulating QSP circuits with $n\le11$, while runtime would have been prohibitive for sufficient Monte Carlo sampling unless $n\le8$ (see \sect{mthd:sim}).
Combined with the deterministic post-selection and (in the case of stochastic errors) importance sampling optimizations, the simulation tools we employ enable us to collect reliable statistics for circuits in the range $n\le16$.

\subsection{Phase calculation}
\label{sec:mthd:response}

The first task in instantiating a QSP circuit is to calculate the phases $\phivec=\philist$ from the Fourier response series $\appf(\theta)$.  
Though in principal classically tractable~\cite{Low2016,Low2017,Haah2018}, in practice this procedure is the bottleneck of QSP circuit implementation for large $\nsteps$.
The difficulty arises from determining the Fourier expansion coefficients $\{g_{-\nsteps/2},\dotsc,g_{\nsteps/2}\}$ satisfying \eq{alg:ug}, which are required by the procedures in~\cite{Low2018,Haah2018} before solving for $\phivec$.
Further, in order to iteratively solve for phases $\philist$ with precision $\polyeps$,
these coefficients must be solved with precision at least $\order*{\polyeps/\nsteps}$. 
Known protocols for solving \eq{alg:ug} are computationally involved and numerically unstable,
requiring extensive calculation with even higher precision arithmetic.

The original efficiency claims of the QSP algorithm~\cite{Low2016,Low2017} assumed finite-time arbitrary-precision arithmetic for phase calculation.  
In~\cite{Childs2017}, the computational overhead of phase calculation lead to the introduction of a ``segmented'' algorithm, in which QSP circuits are constructed with fixed $\nsteps\le28$ and repeated in order to achieve longer evolution times.  Unfortunately, as error grows linearly with the number of individual segments, this segmented algorithm ultimately undoes the additive scaling of $\nsteps$ with $\tau$ and $\polyeps$.  
Subsequent work with more rigorous stability analysis demonstrates a prescription for solving \eq{alg:ug} in time $\order*{\nsteps^3\log\nsteps/\polyeps}$~\cite{Haah2018}, limited by the best known procedure for finding the roots of a $\order*{\nsteps}$ degree characteristic polynomial with precision $2^{-\nsteps}\polyeps/\nsteps$.

Parity observations (noted in~\cite{Haah2018}) allow us immediately to reduce the order of the characteristic polynomial by half.
By employing the well-known \texttt{mpsolve} library~\cite{Bini2014} for high-performance multiprecision polynomial root finding, we can easily complete the root-finding stage of the phase calculation procedure with sufficiently high precision for $\nsteps\lesssim1024$. 
The remainder of our phase calculation tool is written in \python{},
making use of the \texttt{mpmath} and \texttt{gmpy} libraries for high-precision arithmetic~\cite{mpmath}.
Any language-induced computational overhead resulting from \python{} is easily overcome by making extensive rational arithmetic and memoization for high-precision subcalculations, many of which are quite repetitive for high-degree polynomial manipulation.
Many of these subcalculations also turn out to be independent of either $\tau$ or $\nsteps$, enabling further speedup if we save memoized subcalculation results to disk or preemptively compute phases for batches of configurations $(\tau,\nsteps)$.
The use of a rational number representation serves to maximize the occurrence of many of these repeated calculations, and also simplifies the handling of differing precision requirements for batches of configurations.
We significantly reduce the computational overhead at this stage by extensively caching subcalculations (for example products of large binomial coefficients, which are required extensively for high-precision polynomial manipulation), which are often independent of $\tau$ and can therefore be reused throughout the calculation and between batches of calculations.

With these tools we are typically able to compute phases for query depths up to $\nsteps\le128$ in $\order*{\text{seconds}}$, $\nsteps\le256$ in $\order*{\text{minutes}}$ and $\nsteps\le1024$ in $\order*{\text{hours}}$, which is sufficient for our analysis.
Finally, because the phases depend only on the tuple $(\tau,\nsteps)$, we store every unique solution so that it can be reused for whenever a new circuit is generated with the same parameters; in practice, for any query depth $\nsteps$ we only need to generate phases for $\order*{10}$ values of $\tau$ to sufficiently cover the region $10^{-12}\le\polyeps(\tau,\nsteps)\le1$.

As described in \sect{alg:bounds}, the parameter chosen for $\polyeps\ge\maxeps$ ultimately determines the resolution of QSP circuits constructed from the resulting phases. 
Our software uses the numerical bound $\numeps$ defined in \eq{alg:haml:numeps}.  Accordingly, the algorithmic resolution of circuits generated with our software will always be predicted and bound in terms of $\polyeps=\numeps(\tau,\nsteps)$.  Because it comprises finite sums of depth $\order*{\nsteps}$, calculating $\numeps$ is never a computational bottleneck in computing $\phivec$.

\subsection{Circuit construction}
\label{sec:mthd:circ}

After phase calculation, the toolchain uses the calculated phases $\phivec$ and the input system Hamiltonian $\Lam$ in order to construct an explicit quantum circuit $\cqsp$ implementing the desired QSP experiment $\uqsp$.  
This circuit construction occurs in two stages.  First, the signal operator $\Lam$ is used to generate the circuit components necessary to implement the ``qubitized'' signal unitary $\uemb$.
Using the qubitization strategy outlined in \sect{alg:normal} (\fig{normal-query}), the construction of $\uemb$ in turn requires constructions of the individual reflection and projection subcircuits $\uselv$, $\ucz$, and $\uprep$.
Because $\uemb$ is independent of the response parameters $(\tau,\nsteps)$ used to compute $\phivec$, this stage can occur independently from the phase calculation step.

In the second stage of the circuit construction procedure, the software compiles the newly-generated subcircuit implementations and the previously-computed phases $\phivec$ into a complete depth-$\nsteps$ circuit implementation of $\uqsp$ expressed in quantum assembly (QASM).
At this stage we also employ a couple of circuit optimizations to reduce its overall resource usage: first by annihilating subcircuits between adjacent queries of $\uemb$ and $\uemb*$, and then more granularly by merging and annihilating gates using peephole optimization software.

\subsubsection{Signal unitary construction}

Our software tool for generating $\uemb$ from $\Lam$ descends from the tools used in~\cite{Childs2017,Childs2017b}, written in the Quipper programming language~\cite{Green2013}.
It takes as input a Pauli-decomposed system Hamiltonian,
\begin{equation}
  \Lam
  = \sum_{k=0}^{\nops-1}\alpha_k\Lam_k, 
\end{equation}
expressed as a series of $n$-qubit Pauli operations $\Lam_k \in \paulis^{n}$ and corresponding real-valued coefficients $\alpha_k=\Tr[\makebig{\Lam_k\Lam}]/2^n\in\reals$ and passed to the software in a standalone text file.
The tool additionally allows the user to specify specific gatesets for circuit decomposition, for example by enabling or disabling three-qubit \Toffoli/ gates or fully decomposing each rotation gate into a Clifford+\T/ sequence.
Though we focus on spin-chain Hamiltonians, the software itself is agnostic to the form of $\Lam$ provided that it has an efficient (as in $\nops\ll2^n$) Pauli decomposition.

\begin{table}[hbt]%
\begin{richtabular}{llllll}
        &           &         & \multicolumn{3}{l}{Gate counts:} \\
Circuit & Operation & Ancilla & Rotation & \Toffoli/ & \CNOT/ \\
\midrule
$\uselv$
    & $\sum_{k<\nops}\dyad{k}\otimes\Lam_k$
    & $\nbits$   & 0 & $3\cdot2^{\nbits-1}-4$  & $3\cdot2^{\nbits-1}+\omega_\Lambda$ \\
$\ucz$
    & $2\dyad0 - \I/^{\otimes\nbits}$
    & $\nbits-2$
    & 0 & $2d-2$ & 0 \\
$\uprep,\,\uprep*$
    & $\dyad{\alpha}{0},\,\dyad{0}{\alpha}$
    & 0
    & $2^\nbits-1$ & 0 & $2^\nbits-2$  \\
\midrule
$\uemb$
    & $\uselv\uprep\ucz\uprep*$
    & $\nbits$ 
    & $2^{d+1}$ & $3\cdot2^{\nbits-1}+2\nbits-6$  & $7\cdot2^{\nbits-1}-2$ \\
\multicolumn{2}{l}{(\emph{adjacent queries} $\uemb,\,\uemb*$)}
    & $\nbits$ 
    & $2^{d+1}$ & $3\cdot2^{\nbits}+2\nbits-8$  & $5\cdot2^{\nbits}-2$ \\
\end{richtabular}%
\tcaption{\label{tab:mthd:circ}Circuit elements and corresponding resource costs required to construct each element of a QSP circuit for a Hamiltonian expressed as a sum of $\nops$ Pauli operators, where $\nbits=\clog2\nops$ is the number of qubits required for the \CTL/ register and $\omega_\Lambda$ is the total number of single-qubit Pauli gates in the decomposition of $\Lam$.  For an $n$-qubit spin-chain Hamiltonian (\eq{intro:haml}), $\nops=4n$, $\nbits=2+\clog2n$, and $\omega_\Lambda=7n$. The final line indicates the total resource costs for a pair of adjacent $\uemb,\,\uemb*$ queries, after the subcircuit elimination optimizations (\sect{mthd:circ:opt})}
\end{table}%

Using the sum-of-unitaries construction described in \apx{block}, the reflection and projection operators employed to implement a unitary block-encoding of $\Lam$ are,
\begin{equation}
  \uselv \defeq \sum_{k=0}^{\nops-1}\dyad{k}\otimes\Lam_k,\qquad
  \uprep \defeq \dprep = \sum_{k=0}^{\nops-1}\sqrt{\alpha_k}\dyad{k}{0},
\end{equation}
requiring $\nbits=\clog2{\nops}$ qubits for the \CTL/ register.
The explicit constructions used by the circuit generation software to implement the qubitization subcircuits $\uselv$, $\ucz$, and $\uprep$ are described in \apx{circ}, with resources costs summarized in \tab{mthd:circ}.

\subsubsection{Subcircuit annihilation}
\label{sec:mthd:circ:opt}

While a standalone application of the signal unitary $\uemb=\uselv\uprep\ucz\uprep*$ requires implementations of all four subcircuits, in the context of the QSP circuit the operator can be simplified.   As shown in \fig{mthd-signal-pair}, between every pair of adjacent queries of $\uemb$ and $\uemb*$ a pair of projectors $\uprep*\uprep=\I/$ can be annihilated.  Further, using the tree implementation of $\ucz$ described in \apx{circ:refl}, we can annihilate about half of each pair of adjacent $\ucz$ circuits by neglecting to uncompute ancilla bits between their application (each $\ucz$ is drawn as a single $\nbits$-qubit \Toffoli/ gates in \fig{mthd-signal-pair}, so this optimization is not pictured).
The final line in \tab{mthd:circ} indicates the combined resource costs of a pair of adjacent queries, taking these optimizations into account.

Additionally, the QSP algorithm requires that the \CTL/ register be prepared and measured in the state $\ket\alpha$.
Noting that $\uemb\ket\alpha = \uselv\uprep\ucz\uprep*\ket\alpha = \uselv\uprep\ket0$ (and identically that $\bra\alpha\uemb* = \bra0\uprep*\uselv*$),
if we start the QSP sequence by querying $\uemb$ and conclude with the conjugated circuit $\uemb*$ we can absorb the preparation and uncomputation of $\ket\alpha$ into the first and final iterates.  After subcircuit annihilation, the complete QSP circuit therefore requires $\nsteps$ applications each of the $\uselv$ and $\uprep$ subcircuits but only $\nsteps-2$ applications of $\ucz$.

Given the summaries in \tab{mthd:circ}, the execution bottleneck of QSP is dependent on the particular platform
and gate set: the $\uselv$ circuit spans all $n+\nbits+1$ qubits and is dominated by 3-qubit Toffoli
gates, whereas each $\uprep$ projector involves only single- and two-qubit gates but contains $2^\nbits-1\le\nops$ arbitrary
rotations.  Implemented with error-correcting logical qubits, these rotations will likely need to be further decomposed
into $\order*{\log n}$ Hadamard and $T$ gates~\cite{Ross2014}, making $\uprep$ the asymptotic bottleneck of the fault-tolerant QSP circuit.

\begin{figure}[ht]
   \includecircuit[6]{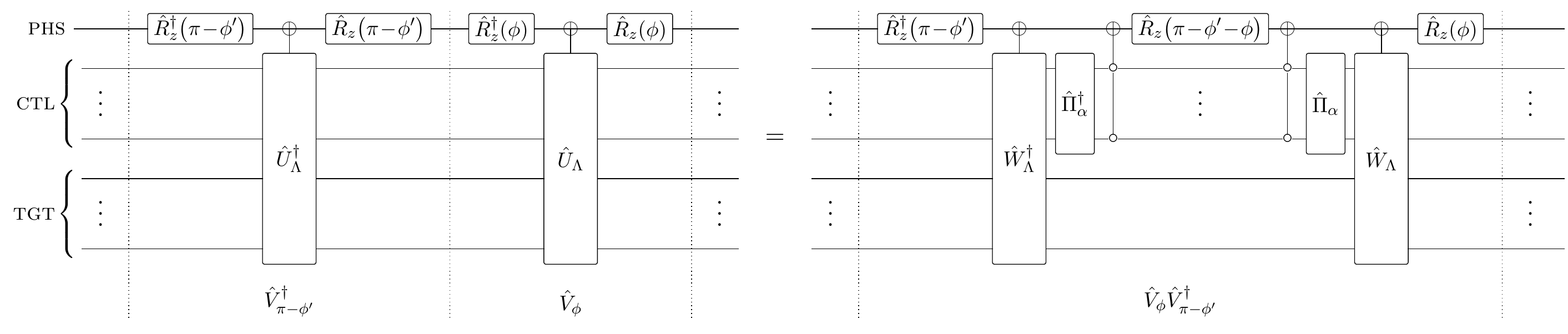}
   \fcaption{\label{fig:mthd-signal-pair}Cancellation between adjacent queries of $\uemb$ and $\uemb*$ (compare to the single-iterate decomposition shown in \fig{normal-query}).  In addition to the pictured annihilation of projectors $\uprep*\uprep=\I/^{\otimes\nbits}$, the complexity of the pair of conditional $\ucz=2\dyad0-\I/$ operations (drawn as $\nbits$-control \Toffoli/ gates) is also reduced by half by preserving garbage bits between their execution}
\end{figure}

\subsubsection{Compilation and peephole optimization}
\label{sec:mthd:circ:peephole}

To complete the QSP circuit $\cqsp$, we repeat the generated $\uemb$ subcircuit $\nsteps$ times (with every second application reversed), interleaving the \PHS/-qubit rotations $\philist$.  The combined circuit is then exported from Quipper to a \python{} script, which translates Quipper's unique output format to QASM and injects various pragmas and directives necessary to control the software workflow in the coming simulation stage.

As part of the translation from Quipper to QASM, we also run multiple peephole optimization passes in order to reduce the total number of gates in the circuit.
In each pass, the optimizer (written in \python{} as part of the translation tool) iterates through every gate in the circuit, and uses a simple set of commutation rules to attempt to shift the gate backward in time until it either (1) annihilates a previous gate, (2) can be merged into an existing gate (for example $\hat{S}^2\mapsto\hat{Z}$), or (3) none of the given rules allow it to be commuted any further.  In the third case, the gate is returned to its original time slot.  We repeat passes through the entire circuit until two consecutive runs return the same circuit; for a QSP circuit this typically amounts to three passes and results in about a 13\% reduction in the overall gate count.

\subsection{Error modeling}
\label{sec:mthd:error}

The simulation tool enables us to model a variety of randomized and systematic error channels.  Each
channel is specified by its Kraus decomposition.  To model a given channel, prior to simulation we generate a set of discrete error operators, which are placed after faulty gates throughout the circuit.

If a channel is stochastic, after every faulty gate we randomly select discrete single-qubit error operators for each qubit involved in that gate, according to the channel's Kraus decomposition.  For every channel we use $\eprob$ to describe the total probability of \emph{not} selecting the channel's first Kraus operator; that is, if a given channel is described by Kraus operators $\{\E/_0,\E/_1,...\}$ the probability of selecting $\E/_0$ is $(1-\eprob)$.

The decomposition of a systematic error channel consists of just a single unitary error $\E0\big[\hat{G}\big]$ that is deterministically placed after each applied gate $\hat{G}$ (and which typically depends on the underlying gate being afflicted).  Because the divergence between systematic errors and random noise results from the former's potential for coherent accumulation, we focus on error modes for which $\big[\E0,\hat{G}\big]=\emptyset$, maximizing the likelihood of constructive combination.  In particular, we consider a multiplicative ``amplitude error'' channel, defined by the gate-dependent unitary,
\begin{equation}
  \E0\big[\hat{G}\big] \defeq
  \exp\big\{\epsilon\log\hat{G}\big\}.
\end{equation}
For consistent comparison to the stochastic channels, we can also characterize the strength of coherent errors in terms of $\eprob=\sin^2{\epsilon}$.

A summary of some of the error channels supported by the simulator is shown in \tab{mthd:models}.

\begin{table}[hbt]%
\begin{richtabular}{llw{1.6in}w{2.2in}}
& \textbf{Channel} & \textbf{Kraus operators} & \textbf{Description} \\
\midrule
\textbf{Stochastic} &
     \emph{Amplitude-damping} & ${\E0=\dyad0+\sqrt{1-\eprob}\dyad1}$, ${\E1=\sqrt{\eprob}\dyad{0}{1}}$ 
                              & Models spontaneous decay, giving rise to the device's characteristic $T_1$ time
\\ \addlinespace[4pt]
   & \emph{Phase-damping}     & ${\E0=\sqrt{1-\eprob}\,\I/}$, ${\E1=\sqrt{\eprob}\dyad0}$, ${\E2=\sqrt{\eprob}\dyad1}$
                              & Decoherence into classical states, giving rise to the device's characteristic $T_2$ time
\\ \addlinespace[4pt]
   & \emph{Bit-flip}          & ${\E0=\sqrt{1-\eprob}\,\I/}$, ${\E1=\sqrt{\eprob}\,\PX/}$
                              & Spontaneous bit flip occurring with probability $\eprob$
\\ \addlinespace[4pt]
   & \emph{Phase-flip}        & ${\E0=\sqrt{1-\eprob}\,\I/}$, ${\E1=\sqrt{\eprob}\,\PZ/}$
                              & Spontaneous phase flip occurring with probability $\eprob$
\\ \addlinespace[4pt]
   & \emph{Depolarization}    & ${\E0=\sqrt{1-\eprob}\,\I/}$, ${\E1=\sqrt{\eprob/3}\,\PX/}$,
                                ${\E2=\sqrt{\eprob/3}\,\PY/}$, ${\E3=\sqrt{\eprob/3}\,\PZ/}$
                              & Spontaneous faults occurring with probability $\eprob$
\\ \addlinespace[8pt]
\textbf{Systematic} &
     \emph{Amplitude}         & $\E0\big[\hat{G}\big] = \exp\big\{\epsilon\log\hat{G}\big\}$
                              & Coherent on-axis over-rotation error
\end{richtabular}
\tcaption{\label{tab:mthd:models}Description and Kraus decomposition of various error models}
\end{table}

\subsubsection{Importance sampling}

Because we place every error operator prior to simulation, we can easily leverage importance sampling techniques for stochastic channels.  For a given circuit, if $\nslots$ is the total number of slots into which an error operator can be placed, the probability that exactly $\nerr$ slots contain an error other than $\E/_0$ is determined by the binomial distribution:
\begin{equation}
  \label{eq:mthd:error:binom}
  p(\nerr\vert \nslots) = \mqty(\nslots\\\nerr) \eprob^{\nerr} (1-\eprob)^{1-\nerr}.
\end{equation}
In most cases we consider, $\eprob\ll1/\nslots$, so that $p(\nerr\vert\nslots)$ quickly falls off and becomes negligible.  It is often more efficient to do separate Monte Carlo analysis for each $\nerr\in(0,1,2,\dotsc)$, so that we can later combine the resulting failure rates and infidelities according to \eq{mthd:error:binom} to estimate their expected values for any sufficiently small error strength $\eprob$.

\subsection{Simulation}
\label{sec:analysis:procedure}
\label{sec:mthd:sim}

Finally, the generated circuit and error placements are used by the simulator in order to return both a failure probability and final state infidelity.
Details of the \cpp{} quantum state simulation software we employ can be found in \cref{apx:sim}.  The heart of the simulator is the space-efficient parallel vector-tree structure introduced in~\cite{Obenland1998}.  The software performs a full-state simulation, returning complex amplitudes for every occupied basis state.
To improve the efficiency of Clifford operations, we also implement well-known techniques for stabilizer-basis simulation~\cite{Aaronson2004} as a front end to the full-state simulator.  In this hybrid strategy Clifford gates are executed in linear-time as updates to a set of stabilizer and destabilizer generators, while non-Clifford gates are mapped to equivalent gates in the destabilizer basis and executed in the vector-tree.

Runtimes for simulations of $\nsteps=64$ QSP circuits are compared in \plot{timing}.  On our single-node system\footnote{\label{foot:comp}Dell Precision Tower 5810 with Intel Xeon E5-1660 v3 at 3.00GHz and 64GB RAM at 2133 MHz}, both the hybrid and parallel tree software can handle error-free QSP circuits up to $n=23$ (38 total qubits), and full coherent error simulations up to $n\le16$ (28 qubits).  
The advantage of the hybrid simulator is about a factor of two for error-free simulations and simulations of stochastic error channels (in which case the \TGT/ register of the QSP circuit undergoes only Clifford operations\footnote{This is not strictly true of amplitude-damping and phase-damping noise, but we can nonetheless implement their error operators in the stabilizer framework by tracking a global probability amplitude}).  For coherent error channels the difference between the simulators becomes negligible (in this case all gates are effectively non-Clifford).
For comparison we also plot the performance of a naive quantum state simulator, implemented as a single array of $2^N$ complex values (where $N$ is the total number of qubits in the simulated circuit) and otherwise using the same \cpp{} routines as the other tools.  This naive model becomes prohibitively slow on our system at $n\ge11$, demonstrating the importance of these high-performance tools.
Runtime is always directly proportional to $\nsteps$, so these runtimes are predictive of those at any query depth.

\timingplot{timing}

\subsubsection{Deterministic post-selection and failure rate computation}

The QSP algorithm is considered to have failed when any measurement result is nonzero.  We therefore only need to consider the simulator's post-measurement state when the measured qubit is $\ket0$.  We can therefore replace measurement operators with deterministic $\dyad{0}$ projectors, so that the norm of the simulator's final state $\ket\finalstate$ can be used to determine the total probability $\fprob=1-\norm[\big]{\braket*\finalstate}$ of a post-selection failure occurring anywhere in the circuit.
After this optimization, the simulation becomes completely deterministic for a given QSP circuit and error placement,
bypassing any Monte Carlo sampling that would otherwise be necessary for randomized quantum measurements.
For non-stochastic error channels, the error placement itself is also deterministic, so we can completely characterize an experiment with a single simulator trial.
In the stochastic case, we still must loop over randomized error placements, but the total number of trials required to resolve $\eprob$ is much smaller.
Most significantly, because the simulation result is deterministic whenever the total number of placed errors is zero, in conjunction with importance sampling this optimization means that we can characterize the $\nerr=0$ case with a single trial.  
Typically this zero-error case accounts for the majority of the trials in a naive Monte Carlo implementation---for example, 90\% of the trials in the case of an $n=11$, $\nsteps=128$ QSP circuit with $\eprob=\e6$ stochastic noise will have zero placed errors.

\subsubsection{Final state infidelity}

In order to measure the infidelity of the final state, alongside each simulation we also compute the ideal final state $\ket\idealstate\defeq e^{-i\tau\Lam}\ket\initstate{}$ from initial state $\ket\initstate{}$ by explicit matrix exponentiation (done relatively quickly in \python{} using \texttt{SciPy} sparse matrix library).  The final state $\ket{\finalstate}$ of each simulation trial is then used to to measure the trial overlap $\olap=\norm[\big]{\bra{\finalstate}\ket{\idealstate}}$.
The overall final state infidelity of a deterministic simulation is then simply $\infid=1-\olap^2$.
For stochastic error channels, the true final state $\rho$ is mixed, so infidelity is estimated via Monte Carlo sampling over randomized error placements:
\begin{equation}
  \infid
  = 1 - \ev{\rho}{\idealstate}
  = 1 - \frac{1}{N}\sum \norm[\big]{\braket{\idealstate}{\makebig{{\psi}_k}}}^2
  = 1 - \sum_k {p}_k \olap_k^2 \,/\, \sum_k p_k, 
\end{equation}
where we sum over individual Monte Carlo trials with final state $\ket*{{{\psi}_k}}$, norm $p_k=\norm[\big]{\braket*{{\psi_k}}}^2$, and ideal-state overlap $\olap_k=\norm[\big]{\braket*{{\psi_k}}{\idealstate}}$.


\section{Empirical analysis}
\label{sec:analysis}

In this section we present and characterize results generated by our software tools.  Details about the procedures we employ for circuit construction and coherent error optimization can be found in \cref{apx:circ,apx:opt}.

\subsection{Methodology}

Motivated by~\cite{Childs2017}, we focus on simulating the periodic-boundary Heisenberg spin-chain Hamiltonian introduced in \eq{intro:haml}.
We randomize the system by sampling each coupling coefficient $\{a_k,b_k,c_k\}$ uniformly from the interval $(0,2)$, and each external field coefficient $h_k$ from the uniform interval $(-1,1)$. Though nothing in our procedure is unique to this model, it provides a useful basis for our resource and error analyses.

For every error configuration $\errchan\ne\mathcal{I}$, there turns out to be an optimal configuration which minimizes either $\avgfprob$ or $\avginfid$.  We define the optimal query depth $\optnsteps$ as that minimizing failure rate\footnote{in principle we could just as well define $\optnsteps$ as minimizing infidelity, which in general is near but not exactly equal to the minimum in failure rate},
\begin{equation}
  \optnsteps(n,\tau,\errchan)
    \defeq \argmin_\nsteps \avgfprob\qty(n,\tau,\nsteps,\errchan),
\end{equation}
with corresponding at-capacity resolution,
\begin{align}
  \optfprob(n,\tau,\errchan)
    & \defeq         \avgfprob\qty(n,\tau,\optnsteps\qty(\tau,\errchan),\errchan)
    = \min_{\nsteps} \avgfprob\qty(n,\tau,\nsteps,\errchan), \\
  \optinfid(n,\tau,\errchan)
    & \defeq         \avginfid\qty(n,\tau,\optnsteps\qty(\tau,\errchan),\errchan).
\end{align}

Our first task for each error channel is therefore to use the software toolchain to generate configuration plots in the form of \plot{ex}(left) in order to develop a model of $\optnsteps(n,\tau,\errchan)$.  From the generated model, we zero in on the simulation capacity boundary, which we can further characterize with additional simulations in order to generate capacity plots from which to extrapolate to a hypothetical $n=50$ experiment.

\subsection{Resource requirements}
\label{sec:analysis:gates}

Using just the circuit generation tools, we first characterize the gate and qubit requirements of explicit QSP circuits for simulating spin-chain Hamiltonians.  
\Tab{analysis:gates} summarizes empirical per-query resource counts for QSP circuits with system sizes $3\le n\le50$.  Gate counts in \tab{analysis:gates} are are averages of 64 circuits constructed from randomly generated spin-chain Hamiltonians; typical variation between circuits is on the order of $\pm0.5\%$ after peephole optimization.

\Plot{gates} presents a more detailed breakdown of the quantum gates required per $\uemb$ query for $5\le n\le50$.
Total gate counts are shown both before and after peephole optimization (\sect{mthd:circ:peephole}); on average we find that the peephole optimizer reduces the number of gates by 13-14\%, with almost all of the reduction coming from single qubit Pauli and Clifford gates.  The steps visible in \plot{gates} after $n=8$ and $n=16$ are due to the size of the \CTL/ register, which for the spin-chain Hamiltonian is $\nbits=\clog2{4n}$.

\begin{table}[hbt]%
\begin{richtabular}{ccc|ccc}
    $n$   & Qubits
    & $(\#\;gates)/\nsteps$
    & ${R_{xyz}}$
    & ${\Toffoli/}$
    & ${\CNOT/}$
\\ \midrule
  3  & 12  & 120  & 16  & 24  & 51   \\ 
  5  & 16  & 213  & 32  & 41  & 93   \\ 
  7  & 18  & 263  & 32  & 50  & 119  \\ 
  9  & 22  & 392  & 64  & 70  & 177  \\ 
  11 & 24  & 440  & 64  & 78  & 203  \\ 
  13 & 26  & 489  & 64  & 90  & 229  \\ 
  15 & 28  & 532  & 64  & 98  & 255  \\ 
  50 & 67  & 1819 & 256 & 318 & 902  \\ 
\end{richtabular}%
\tcaption{Average per-query resource estimates for QSP circuits at various system sizes ($n$), built using the spin-chain Hamiltonian in \eq{intro:haml}}
\label{tab:analysis:gates}
\end{table}

From \tab{mthd:circ} we expect the gate count of each subcircuit to be a linear combination of $n$, $\nbits=\clog2n$, and $2^{\nbits}$. Using a least-squares regression to fit the total post-optimization per-query gate counts in \plot{gates} to the three parameters, we find,
\begin{equation}
  \label{eq:analysis:gates}
  (\#\;gates)/\nsteps \approx 24.0n + 2.3\nbits + 2.9\cdot2^\nbits
                      \approx 35.6n + 2.3\log_2n + 4.6,
\end{equation}
where second approximation takes $d\approx 2+\log2n$.

\gatesplot{gates}{Gate count comparison of QSP circuits for simulating $4\le n\le50$ spin-chain Hamiltonians (\eq{intro:haml}), using the circuit constructions described in \apx{circ}.  Counts are averaged over multiple randomized Hamiltonians and shown \emph{per query}; the expected number of gates required for the complete QSP implementation is found by multiplying values on the $y$-axis by $\nsteps$}

\subsection{Ideal gates}
\label{sec:analysis:th}

In order to verify the our procedure and toolchain and establish a baseline for subsequent analysis, we begin by observing the performance of error-free QSP circuits in comparison to the theoretical bounds established in \sect{alg:bounds}.
On a perfect system, we can continuously increase query depth so as to decrease $\fprob$ and $\infid$ indefinitely, making the simulation capacity of the ideal circuit infinite.
Empirical configuration plots for error-free $n=11$ QSP circuits with fixed $\tau\in\{16,32,64,128,256\}$ are shown in \plot{th:m}, which demonstrate this unbounded super-exponential drop in failure rate and infidelity with increasing query depth. 

\theorymplot{th:m}

The theoretical limits $\fprob\le4\polyeps$ and $\infid\le(2\polyeps)^2$ are drawn with dashed lines in \plot{th:m}, where (here and throughout the remainder of this section) $\polyeps=\numeps(\tau,\nsteps)$ is the numerical bound established in \eq{alg:haml:numeps} which has been ``baked in'' to the circuit through the calculation of $\phivec$.
These bounds turn out to be fairly tight: empirically, for $\polyeps\le0.1$ we observe,
\begin{align}
  \label{eq:analysis:th:fprob}
  \avgfprob\qty(\tau,\nsteps,\noerr) 
  &\approx (2.93\pm0.23)\,\numeps(\tau,\nsteps), \\
  \label{eq:analysis:th:infid}
  \avginfid(\tau,\nsteps,\noerr)
  &\approx (0.833\pm0.074)\,\numeps^2(\tau,\nsteps),
\end{align}
with both slightly smaller at higher $\polyeps$.
The stability of $\avgfprob(\tau,\nsteps)$ and $\avginfid(\tau,\nsteps)$ and the consistency with which they can be predicted from $\numeps(\tau,\nsteps)$ is a result of using the numerical bound for phase calculation, which minimizes the functional dependencies which get baked in to the response algorithm and thereby narrows the distribution of possible design-induced error effects.

A second and related benefit of the numerical error bound is a constant-factor improvement in query depth as a function of $\tau$ and $\polyeps$.  As described in \sect{alg:bounds}, when the $\numeps(\tau,\nsteps)$ is numerically solved for $\nsteps(\tau,\numeps)$ we find $\nsteps\sim2\tau$ asymptotically, which is observable in the spacing of traces in \plot{th:m}.  Had we instead used the 
asymptotic error bound (\eq{alg:haml:asyeps}) when computing $\phivec$, we would instead expect $\nsteps(\tau,\asyeps)\sim e\tau$ to leading order, corresponding to a factor of $e/2$ increase in the query depth and gate count for same infidelity and failure rate.  These improvements will be essential to our
ability to reliably model and optimally configure faulty QSP circuits in order to determine their best-possible simulation capacity.

\subsection{Stochastic noise}
\label{sec:analysis:depol}

We now consider the performance of QSP circuits in the presence of various stochastic noise models.  The descriptive analysis in this subsection focuses on just the depolarizing channel; we subsequently repeat the procedure while substituting other stochastic models to generate the remainder of the results presented in \sect{rslt}.

As described in \sect{mthd:error}, if a given circuit has $\nslots$ positions in which a random error can occur, the probability that at least one fault occurs anywhere in the circuit is $1-p(0\vert\nslots)\approx\eprob\nslots$ for error strengths $\eprob\ll\nslots$.  As the number of possible error positions is directly proportional to the query depth $\nsteps$, 
 we hypothesize a simple linear model for fixed-$n$ systems undergoing stochastic noise:
\begin{equation}
\begin{split}
  \label{eq:analysis:depol:hyp}
  \avgfprob(\tau,\nsteps,\eprob) &= \avgfprob(\tau,\nsteps,\noerr) + \linfprob_n\nsteps\eprob,
  \\
  \avginfid(\tau,\nsteps,\eprob) &= \avginfid(\tau,\nsteps,\noerr) + \linfidel_n\nsteps\eprob,
\end{split}
\end{equation}
where the first term in each expression captures the inherent resolution of the error-free circuit (\cref{eq:analysis:th:fprob,eq:analysis:th:infid}, respectively), and $\linfprob_n$ and $\linfidel_n$ are unknown $n$-dependent parameters quantifying the respective contributions of stochastic noise to failure rate and infidelity.
By design we are guaranteed $\linfprob_n+\linfidel_n\le\nslots/\nsteps$: if any single fault causes the circuit to fail in post selection, we would expect $\linfprob_n=\nslots/\nsteps$, whereas if 
individual faults are never detectable in post selection but always make $\infid\to1$ for that trial we would expect $\linfprob_n=0$ and $\linfidel_n=\nslots/\nsteps$.

\Plot{depol:m} shows empirical configuration plots for $n=11$, $\tau\in\{8,16\}$ QSP circuits subject to depolarizing noise with strengths $\eprob\in\{\e8,\e7,\e6\}$. Each average was taken over 32 randomly generated circuits, with the equivalent of 100000, 10000, or 1000 (corresponding to the three values of $\eprob$) Monte Carlo trials apiece for error placement randomization.

\depolmplot{depol:m}

As suggested in the introduction (e.g. \plot{ex}), each configuration $(\tau,\eprob)$ in \plot{depol:m} has a distinct optimal query depth $\optnsteps(\tau,\eprob)$ minimizing either $\avgfprob$ or $\avginfid$.  For circuits configured with too few queries ($\nsteps<\optnsteps$), resolution is design-dominated and closely matches that observed in the fault-free case (\plot{th:m}).
With too many queries ($\nsteps>\optnsteps$), the circuit becomes dominated by noise, such that every additional query adversely impacts simulation performance.  In this region both failure rate and infidelity exhibit $\tau$-independent linear behavior consistent with that hypothesized in \eq{analysis:depol:hyp}.  The spacing between traces generated with different error rates is also consistent with the linear dependence on $\eprob$ in \eq{analysis:depol:hyp}.

\depoltplot{depol:tau}

In \plot{depol:tau}, failure rate and infidelity are plotted against $\tau$ at constant query depths $\nsteps\in\{32,48,64,80,128\}$ (shown with $\eprob=10^{-6}$ depolarizing noise).  Here, the inflection points of the constant-$\nsteps$ traces begin to map the platform's \emph{simulation capacity}, indicating the largest $\tau$ which can be simulated on the platform with a given resolution. 
In this case the $\tau$-independence of the noise-dominated region is manifest in the horizontal traces when $\tau$ is below capacity.

We can measure the constants $\linfprob_{n}$ and $\linfidel_{n}$ from the slope of linear fits to $\avgfprob/\eprob$ and $\avginfid/\eprob$ in the noise-dominated region (as shown with dotted lines in \plot{depol:m}).  Empirical values of $\linfprob_{11}$ and $\linfidel_{11}$ are presented for each stochastic error channel in \tab{analysis:depol}.
From these values (and exploiting the monotonicity of $\numeps(\tau,\nsteps)$), we can estimate the optimal query depth $\optnsteps(\tau,\eprob)$ and the corresponding failure rate $\optfprob(\tau,\eprob)$ and infidelity $\optinfid(\tau,\eprob)$ by direct numerical minimization of \eq{analysis:depol:hyp}.  Both $\optfprob(\tau,\eprob=\e6)$ and $\optinfid(\tau,\eprob=\e6)$ are plotted against $\tau$ with dashed cyan lines in \plot{depol:tau}. 
For a given platform configuration $(n,\eprob)$, these boundaries indicate the platform's \emph{simulation capacity}, or the minimum achievable average failure rate as a function of $\tau$.

Though our error model (\eq{analysis:depol:hyp}) depends linearly on the gate error rate $\eprob$, the optimal configuration performance is slightly sublinear in $\eprob$.  This is because $\optnsteps(\tau,\eprob)$ is slightly smaller for larger error rates, as can be observed in location of the minima of \plot{depol:m}.  Intuitively, if the stochastic error rate is increased, we can tolerate a comparable increase in design-induced error before it can contribute anything over the stochastic noise floor.  In turn, this decreased query depth results in decreased error accumulation, and accordingly a smaller contribution to $\avgfprob$ and $\avginfid$.  This effect is mostly insignificant, however:
because the design-induced falls super-exponentially with $\nsteps$ while noise is accumulated linearly, the optimal query depth turns generally remains within a narrow range. 
As can be seen in the horizontal spread of traces in \plot{th:m}, for a given $\tau$ only a handful of possible query depths exist in the vicinity of possible noise contributions---for example, at $\tau=400$ the entire range of query depths satisfying $\e7\le\numeps(\tau,\nsteps)\le\e2$ falls within the interval $832\le\nsteps\le884$, corresponding to a $\pm3\%$ fluctuation in the noise contribution to $\avgfprob$ in our model.
A rough power-law fit finds $\avgfprob\sim\eprob^{0.96}$ with the parameters used in \plot{depol:m}.

\subsubsection{System size}
\label{sec:analysis:depol:n}

Finally, in order to predict the performance of an $n=50$ platform we need to understand how circuit performance scales with system size.  While the resolution of the error-free circuit depends only on the response function parameters $\nsteps$ and $\tau$ and is therefore independent of $n$, in the stochastic case the expected number of faults will grow with $n$ as the number of faulty gates in the circuit increases. We confirm that the noise contribution to $\avgfprob$ and $\avginfid$ is proportional to the number of gates 
by simulating circuits of sizes $5\le n\le17$ configured with $\tau=8$ and $\nsteps=64$ (i.e. well into the noise-dominated region). 
As shown in \plot{depol:n}, both $\avgfprob/(\#\;gates)$ and $\avginfid/(\#\;gates)$ (where $(\#\;gates)$ is the total number of gates counted in the simulated circuit) are found to be roughly constant in $n$ in this domain.

\depolnplot{depol:n}

So far we have absorbed this scaling into the $n$-dependence of the constants $\linfprob_n$ and $\linfidel_n$. 
In principle, we can now use the empirical gate count model in \eq{analysis:gates} to estimate $\linfprob_n,\,\linfidel_n$ for any $n$ from the measured values of $\linfprob_{11}$ and $\linfidel_{11}$.  However, we have already explicitly measured average gate counts for both $n=11$ and $n=50$ circuits in \sect{analysis:gates}.  To model the hypothetical $n=50$ experiment, we read their respective values directly from \tab{analysis:gates} to find
$\linfprob_{50} \approx (1819/440)\linfprob_{11}$ and $\linfidel_{50} = (1819/440)\linfidel_{11}$.

\subsection{Systematic error}
\label{sec:analysis:coherent}

We now repeat the analysis of \sect{analysis:depol}, but using the systematic amplitude error model.  As described in \sect{mthd:error}, coherent amplitude errors are modeled with gate-dependent error operators $\E/[\hat{G}]=\exp\{\epsilon\log\hat{G}\}$, characterized by the multiplicative strength $\epsilon=\asin\sqrt\eprob$.  Throughout this section we use the optimized circuit elements described in \apx{opt}, which significantly reduce the failure rate and infidelity of QSP circuits subject to coherent errors.

In the coherent error case we find that $\avgfprob$ depends strongly on $\tau$ in both the design-dominated and fault-dominated regions.
\Plot{coherent:tau} shows the $\tau$-dependence of $n=11$ QSP circuits with fixed query depths $\nsteps=320$ and $\nsteps=512$ and subject to systematic $\epsilon=\e3$ (i.e. $\eprob=\epsilon^2=\e6$) amplitude error.  Unlike the near right angle at the transition between noise-dominance and design-dominance under the depolarizing channel (\plot{depol:tau}), in the coherent case there is a sharp inflection point such that for a given query depth there is a narrow band of simulation times outside of which failure rate increases rapidly.
We also observe a modest dependence of the final state infidelity on $\tau$ in the fault-dominated region, albeit significantly less pronounced than for failure probability.
This $\tau$-dependent error response is perhaps somewhat surprising: as discussed in \sect{alg}, the only way in which $\tau$ impacts the overall QSP circuit implementation is in the set of phases $\phivec$, which in \plot{opt:all} we find do not contribute significantly to the overall faulty circuit resolution.

\coherenttplot{coherent:tau}

The best-possible simulation performance occurs when circuits are configured at the narrow minima of the inflection points.  
Without a rigorous prior as to the location of these minima, we instead construct a bootstrapped simulation capacity model with an iterative estimation procedure:
for randomly selected query depths $32\le\nsteps\le1024$, we simulate circuits constructed with various values of $\tau$ in order to manually minimize $\avgfprob(\tau,\nsteps)$, and then use the coordinates $(\tau,\nsteps)$ of the resulting minima in order to better predict the minimizing $\tau$ for subsequent query depths.
As a rough guiding heuristic we find that the optimal configurations consistently occur in the narrow band satisfying $\eprob\le\numeps(\tau,\nsteps)\le100\eprob$.
Capacity plots constructed from the observed minima are shown for error strengths $\epsilon^2\in\{\e6,\e7,\e8\}$ in \plot{coherent:cap}.

\coherentplot{coherent:cap}

As observed in \apx{opt}, when coherent amplitude errors are restricted to the $\uprep$ subcircuit the expected at-capacity failure rate is (on average) constant in $\tau$, whereas (after optimizing for coherent errors) the contributions from the $\uselv$ and $\ucz$ subcircuits and \PHS/-qubit rotation gates have similar (positive) $\tau$-dependence.
We therefore use the coordinates $(\tau,\optnsteps)$ of each minima determined by the bootstrapping procedure (i.e. the points in \plot{coherent:cap}) to construct two new partial capacity plots, with errors restricted to (1) just the gates of the $\uprep$ subcircuit, and (2) with errors placed everywhere \emph{expect} in the $\uprep$ circuit.  The empirical subcircuit-restricted capacity plots are shown for $\epsilon^2=\e6$ in \plot{coherent:sub}.  The mean contribution from the $\uprep$ subcircuits (dotted blue line in \plot{coherent:sub}) is measured to be,
\begin{equation}
  \label{eq:analysis:prep}
  \avgfprob\Big\rvert_{\uprep} \approx \qty( 2470.\pm5. ) \cdot \e6.
\end{equation}

\subcircplot{coherent:sub}

We are again without a rigorous prior as to the form of the non-$\uprep$ contribution to failure rate.  
Empirically, after the symmetrization optimization described in \apx{opt} it appears to contain contributions from both a constant term and a $\tau$-dependent term, with the latter appearing somewhat sublinear throughout the observable range $16\le\nsteps\le1024$.
We therefore consider both a generic power-law model $\avgfprob\sim\alpha+\beta\tau^\gamma$ and a more conservative linear model $\avgfprob'\sim\alpha'+\beta'\tau$.  Using a least-square regression to fit the model parameters, we find,
\begin{equation}
  \label{eq:analysis:noprep:pow}
  \avgfprob/\eprob\Big\rvert_{\PHS/,\uselv,\ucz} \;\approx 19.4\tau^{0.82}+330.0,
\end{equation}
using the power-law model, and,
\begin{equation}
  \label{eq:analysis:noprep:lin}
  \avgfprob'/\eprob\Big\rvert_{\PHS/,\uselv,\ucz} \;\approx 6.3\tau+449.4,
\end{equation}
with the conservative linear model.  Both fits are plotted for $\epsilon^2=\e6$ in \plot{coherent:sub}, in which visually the power-law (dark blue, dashed) is a much better fit to the data than the linear model (light blue, solid).

\subsubsection{System size}

In order to
extrapolate to larger systems, we also need to model the coherent-error simulation capacity as a function of $n$.  Again we observe differing dependencies among the different subcircuits.
Because error effects depend on $\tau$ at every configuration, to characterize the at-capacity $n$-dependence of $\optfprob$ and $\optinfid$ of each subcircuit, we must generate circuits exactly at the simulation capacity boundary (unlike the stochastic case, in which case it was sufficient to simulate circuits will into the noise-dominated region).  In practice, we find that $\optnsteps(n,\tau,\epsilon)$ varies very little with $n$, making this boundary relatively simple to discover for each size.

Unfortunately, whenever the $\uselv$ subcircuit is subject to coherent errors, the
simulator ends up needing to allocate memory for every qubit in the system (including all $\nbits$ ancilla bits in the $\uselv$ implementation).  Our system is therefore memory-limited to $n\le16$ (29 total qubits, compared to 32 qubits for the $n=17$ circuit).  This restriction disappears when errors are instead restricted to the $\uprep$ or $\ucz$ subcircuits.  The $\uprep$ circuit involves no ancilla qubits, enabling simulations up to $n\le23$ (31 non-ancilla qubits) before filling 64GB RAM.  The $\ucz$ circuit involves two fewer ancilla qubits than $\uselv$, allowing for simulations up to $n\le18$.

Empirical capacity plots of $\tau=20$ circuits with amplitude errors restricted to each subcircuit are shown in \plot{coherent:n}.
In the range $5\le n\le15$ in which we can compare all subcircuits, we observe that the size-dependence of the overall circuits is overwhelmingly dominated by the faulty $\uprep$ projectors.  The distinct steps in failure probability at $n=9$ and $n=17$ indicate that the contribution of the $\uprep$ circuit is proportional to $4^{\nbits}\sim n^2$, where $\nbits$ is the size of the \CTL/ register.
This is reasonable given the circuit implementation $\uprep$, in which $2^{\nbits+1}$ gates always target the most significant bit of the \CTL/ register.
It is likely that the circuit implementation of $\uprep$ could optimized so as to improve this result significantly.
However, because the contribution of the $\uprep$ term is constant in $\tau$, this quadratic growth will turn out not to drive the overall resolution of meaningful Hamiltonian simulations.

\coherentnplot{coherent:n}

In the size range in which we can simulate the faulty $\uselv$ subcircuit, we find that the size dependence of its contribution to $\avgfprob$ scales roughly with the total number of gates in the circuit, as was the case for random noise.  Importantly, we do not find any indication of the coherent accumulation of systematic errors (which would lead to a quadratic increase failure rate with $n$), nor do we observe a step when the size of the \CTL/ register increases at $n=9$.

Combining the size dependencies with \cref{eq:analysis:prep,eq:analysis:noprep:pow} (and using that $(\#\;gates/\nsteps)=440$ and $\nbits=6$ for $n=11$), we finally arrive at a complete simulation capacity model,
\begin{equation}
  \label{eq:analysis:fprob:pow}
  \avgfprob(\tau,n,\eprob=\e6) \approx 
  \underbrace{0.603\cdot4^\nbits\cdot\e6}_{\uprep\;\text{contribution}} +
  \underbrace{(\#\;\text{gates}/\nsteps)\cdot(0.0441\tau^{0.82}+0.750)}_{\text{non-$\uprep$\;\text{contribution}}},
\end{equation}
or, using the linear $\tau$-dependence model (\eq{analysis:noprep:lin}) for the non-$\uprep$ contribution,
\begin{equation}
  \avgfprob(\tau,n,\eprob=\e6) \approx 
  \underbrace{0.603\cdot4^\nbits\cdot\e6}_{\uprep\;\text{contribution}} +
  \underbrace{(\#\;\text{gates}/\nsteps)\cdot(0.0143\tau+1.021)}_{\text{non-$\uprep$\;\text{contribution}}}.
\end{equation}

The final state infidelity contributed by any of the subcircuits does not appear to grow significantly with $n$, with the exception of the jumps observed at $n=9$ and $n=17$ when errors are restricted to the $\uprep$ circuit.  The increased failure rate resulting from these jumps is subsequently reversed as we continue to increase $n$.
Because the contributions of the different circuit components also grow similarly with $\tau$ (\plot{coherent:sub} (albeit by different magnitudes), we fit the generic power-law model to the unrestricted simulation capacity plot (\plot{coherent:cap}), finding,
\begin{equation}
  \avginfid/\eprob \approx 
  11.3 + 3.70\tau^{1.17}.
\end{equation}

\section{Results and conclusions}
\label{sec:rslt}

For each stochastic noise channel, we repeat the analysis of \sect{analysis:depol} in order to empirically determine the constants $\linfprob_{11}$ and $\linfidel_{11}$ in our hypothesized linear error model (\eq{analysis:depol:hyp}), from which we extrapolate to the hypothetical $n=50$ experiment using relative numbers of gates as in \sect{analysis:depol:n}.  Measured values of $\linfprob_{11}$ and $\linfidel_{11}$ and corresponding estimates of $\linfprob_{50}$ and $\linfidel_{50}$ are shown in \tab{analysis:depol}.
A notable distinction between the stochastic models can be seen in the relationship between $\linfprob_n$ and $\linfidel_n$.  For example, for the bit-flip channel has a smaller $\linfidel_n$ and greater $\linfprob_n$ when compared to the phase-flip channel, indicating that for the same experiment the bit-flip channel will be more likely to fail in post-selection, but if it succeeds the output state will on average be closer to the ideal simulation result.

\begin{table}[hbt]%
\begin{richtabular}{rllll}
Channel
& $\linfprob_{11}$ & $\linfidel_{11}$ 
& $\linfprob_{50}$ & $\linfidel_{50}$ 
\\ \hline
    \emph{Depolarization}     & $594.\pm4$ & $51.\pm6$ 
                              & $2456.\pm12$ & $210\pm25$
\\  \emph{Bit-flip}           & 786.7  & 38.8
                              & 3249. & 160.4
\\  \emph{Phase-flip}         & 425.0  & 48.2
                              & 1757. & 199.3
\\  \emph{Phase-damping}      & 255.5 & 27.2
                              & 1056. & 112.4
\end{richtabular}%
\tcaption{\label{tab:analysis:depol}Measured values of constants $\linfprob_n$ and $\linfidel_n$ (as defined in \eq{analysis:depol:hyp}) for various stochastic noise channels at $n=11$, with corresponding predictions for an $n=50$ system}
\end{table}%

By numerically minimizing \eq{analysis:depol:hyp} using the values in this table, we can finally generate our empirical estimates of the optimal query depth $\optnsteps(n,\tau,\eprob)$ and corresponding at-capacity circuit performance $\optfprob(n,\tau,\eprob)$ and $\optinfid(n,\tau,\eprob)$ for an $n=50$ system subject to each stochastic error channel.
In \plot{rslt}, we compare the resulting $n=50$ capacity plots (dashed lines) to the empirical capacities on the $n=11$ system (solid lines).
The results shown are rescaled by the system's gate error rate $\eprob$, so that multiplying the $y$-axis by $\eprob$ returns the expected at-capacity failure rate and infidelity of a system with that gate error rate.
For comparison, we also plot the simulation capacity of the $n=11$ and $n=50$ experiments (\eq{analysis:fprob:pow}) subject to systematic amplitude errors.

\rsltsplot{rslt}

Meaningful Hamiltonian simulation (i.e. that which is sufficiently large in all parameters to be classically intractable) requires $t\sim\norm{\Lam}\sim n$, and therefore implies that $\tau\sim n^2$.  
In \plot{rslt:n}, we use our models to estimate the expected failure rate of $\tau=n^2$ simulation experiments under each error channel as a function of $n$.
For the stochastic channels, we know that (to first order) $\linfprob_n\propto(\#\;gates)\propto n$ and $\optnsteps\propto\tau\propto n^2$, so that the overall failure rate of a meaningful simulation
is asymptotically cubic in $n$. 
In the coherent case, we have two terms to consider.
The failure rate contribution of the 
$\uprep$ circuit grows with $4^\nbits\sim n^2$ while being independent of $\tau$ so that its overall contribution to a meaningful Hamiltonian simulation is just quadratic in $n$.
Using the more aggressive power-law fit for the contributions of the remaining subcircuits (\eq{analysis:noprep:pow}), the overall $n$-dependence of a meaningful simulation is reduced to $\avgfprob\propto(\#\;gates)\tau^{0.82}\propto n^{2.64}$.
With the more conservative linear model (\eq{analysis:noprep:lin}), the dependence is again cubic in $n$.  However, because the asymptotic probability of failure is driven by that of the $\uselv$ subcircuit, after the coherent error optimizations described in \apx{opt} the failure rate is more than two orders of magnitude less than that under a stochastic channel with comparable error strength $\eprob=\epsilon^2$.

\rsltsnplot{rslt:n}

Finally, we can return to our motivating questions.
For a hypothetical $n=50$ experiment with $\tau=50^2$, we can read $\optfprob/\eprob$ and $\optinfid/\eprob$ for each error channel directly from \plot{rslt}.
In the stochastic case, the error rate necessary for an expected failure probability $\avgfprob\approx10\%$ is between $\eprob\approx2\cdot\e{8}$ (phase damping channel) and $\eprob\approx5\cdot\e{9}$ (bit flip channel).
For the coherent amplitude error channel, the same failure rate would require $\eprob=\epsilon^2\approx1\cdot\e6$, or gate amplitudes accurate to $\epsilon\approx0.1\%$.  In both cases, we would expect the final state infidelity of the resulting experiment to be $\avginfid\sim0.1$.

Conversely, if we are given a devices with known gate error rate $\eprob$, 
we can read the maximum system size $n$ which we could meaningfully simulate with a target resolution from \plot{rslt:n}.
For $\eprob=\e5$ stochastic noise, the largest possible meaningful simulation with $\avgfprob\le10\%$ is $n\approx5$ for phase-damping noise, $n\approx4$ with depolarizing noise, and just $n\approx3$ under the bit-flip channel.
With systematic amplitude error of the same strength, we expect to be able to be able to simulate systems up to $n=16$ with the same failure rate.
If we managed to reduce the stochastic error rate to $\eprob=\e6$, we would expand the range of possible experiments to $n=9$ in the depolarizing case and $n\approx13$ in the phase-damping case.

\newpage{}
\appendix


\section{QSP implementation details}

In this appendix we provide further details specific to the implementation of QSP used in this work.
\Apx{block} outlines operators and algorithmic considerations for QSP constructed for a signal operator $\Lam$ which can be efficiently decomposed in a sum-of-unitaries representation.
The corresponding quantum circuit components are explicitly described \apx{circ}.
Derivations for the error bounds defined in \sect{alg:bounds} are provided in \apx{bounds}.
\ifthesis{%
Finally, \apx{opt} presents additional simulation data and resulting circuit optimizations for mitigating systematic unitary error models.}{%
\Apx{opt} presents additional simulation data and resulting circuit optimizations for mitigating systematic unitary error models. Finally, in \apx{sim} we provide further implementation details of our simulation software.}

\secorsub{QSP algorithm for linear combination of unitaries}
\label{apx:block}

As outlined in \sect{alg:normal}, we can ``qubitize'' a normal signal operator $\Lam$ by constructing a pair of reflection operators $\uselv$ and $\urefl=\uprep\ucz\uprep*=2\dyad\alpha-\I/$, where $\ucz=2\dyad0-\I/$ is Grover's diffusion operator and $\uprep=\dprep,\,\uprep*=\dprep*$ are projectors such that $\uprep*\uselv\uprep = \dyad0\otimes\Lam/\norm[\big]{\Lam} + \dotsb$
forms a block encoding of $\Lam$. 
Efficient block encodings for a variety of other operator structures are described in detail in \cite{Gilyen2018}.

Hamiltonians describing a number of physical systems, including the Heisenberg spin-chain model used in this work (\eq{intro:haml}), are often naturally expressed as a sum of unitary elements, i.e.,
\begin{equation}
  \label{eq:apx:haml}
  \Lam = \sum_{k=0}^{\nops-1} \alpha_k\Lam_k,
\end{equation}
where each $\Lam_k\in\SU{2^{n}}$ is a unique unitary operator.
In this case, we can implement the projection and reflection operators,
\begin{equation}
  \label{eq:apx:block}
  \uprep = \dprep 
    \defeq \sum_{k=0}^{\nops-1} \sqrt{\abs{\alpha_k}} \dyad{k}{0}, \qquad
  \uselv 
    \defeq \sum_{k=0}^{\nops-1} e^{i\arg{\alpha_k}} \dyad{k} \otimes \ULam_k.
\end{equation}
where $(\alpha_k/\abs{\alpha_k}) = e^{i\arg\alpha_k}$ absorbs the sign or phase of $\alpha_k$.
For Hermitian $\Lam$, it is simple to check that $\uselv^2=\I/$ and that $\uprep*\uselv\uprep=\sum_k\alpha_k\dyad0\otimes\ULam_k$ is a block encoding of $\Lam$.

For each eigenstate $\ket\ulam$ of $\Lam$, the \SU2 subspace generated by the paired reflections $\uselv$ and $\urefl=\uprep\ucz\uprep*=2\dyad\alpha-\I/$ is spanned by the orthogonal basis states,
\begin{align}
  \ket{\alam}
  & \defeq{}
  \ket{\alpha}\ket{\ulam}
  = \sum_k \sqrt{\alpha_k}\ket{k} \otimes \ket{\ulam}, \\
  \ket{\alamd}
  & \defeq{} \frac{1}{\sqrt{1-\lambda^2}}\sum_k \sqrt{\alpha_k}\ket{k} \otimes \qty(\ULam_k-\lambda)\ket{\ulam},
\end{align}
such that,
\begin{align}
  \urefl\ket\alampm &= \pm\ket\alampm, \\
  \uselv\ket\alampm &= \lambda\ket\alampm + \sqrt{1-\lambda^2}\ket\alammp.
\end{align}
Combining the two reflections, we construct an eigenstate-specific \SU2 rotation operator,
\begin{equation}
  \uemb = \uselv\urefl = \uselv\uprep\ucz\uprep* =
  \begin{pmatrix}
    \lambda & -\sqrt{1-\lambda^2} \\
    \sqrt{1-\lambda^2} & \lambda
  \end{pmatrix}
  = e^{-i\acos\lambda\PY/}
  = \Ry\qty(\pi-2\thlam),
\end{equation}
with eigenphases $\mp ie^{\pm i\thlam}$.  The imaginary phase can be eliminated by adding a $\Rx(\pi/2)=\H/\S/\H/$ gate to the \PHS/ qubit above each query (or equivalently substituting \PHS/-qubit rotations $\Rz(\phi_k)\mapsto\Ry(\phi_k)$ for odd $k$), leaving eigenphases $e^{i\thlam}$ and $e^{\pi-i\thlam}$. 
 
Applied to the eigenstate $\ket{\blam+}\defeq\ket\alam+ i\ket\alamd$ of $\uemb$, each query of $\uemb$ would kick back the corresponding eigenphase $e^{i\thlam}$ so that we would immediately recover the unitary-$\Lam$ QSP algorithm. 
However, in general we do not have access to the eigenstate-dependent states $\ket{\blam\pm}$.
Instead, we prepare $\ket\alpha$ in the \CTL/ register in order to generate decoupled basis states $\ket\alam=\ket{\blam+}+\ket{\blam-}$ for each $\Lam$ eigenstate $\ket\ulam$ superimposed in the \TGT/ register, and rely on the Hermiticity of the response function $\appf(\thlam)$ to ensure that the \CTL/ register is returned to the known state $\ket\alpha$ to be unprepared and measured at the end of the algorithm:
\begin{equation}
  \uqsp \ket\alam
  = \appf(\thlam) \qty( \ket{\blam+} + \ket{\blam-})
  = \appf(\thlam) \ket{\blam+} - \appf(\pi-\thlam) \ket{\blam-}
  = \appf(\thlam) \ket{\alam}.
\end{equation}

\secorsub{Error bounds}
\label{apx:bounds}

\newcommand{\reeps}{\polyeps_{re}}
\newcommand{\imeps}{\polyeps_{im}}
Here we derive the error bounds introduced in \sect{alg:bounds}.
Both the asymptotic and numerical bounds approximate \eq{alg:haml:maxeps} by maximizing the real and imaginary terms separately,
\begin{align}
  \imeps
  & \defeq \max_\theta \norm*{ \appa(\theta) - \cos(\tau\sin\theta) }
      \le 2 \sum_{\mathclap{k=0}}^{\infty} \abs*{J_{2k+\nsteps/2+2}(\tau)}
  \label{eq:bounds:real}
  \\
  \reeps
  & \defeq \max_\theta \norm[\big]{ \appc(\theta) - \sin(\tau\sin\theta) }
      \le 2 \sum_{\mathclap{k=0}}^{\infty} \abs*{J_{2k+\nsteps/2+1}(\tau)},
  \label{eq:bounds:imag}
\end{align}
so that the total error is bound by $\maxeps \le \qty{\reeps^2 +  \imeps^2}^{1/2}$.

The asymptotic bound
$\asyeps$ is calculated using the Bessel function property $\abs{J_k(\tau)}\le\abs{\tau/2}^{\abs{k}}/\abs{k}!$~\cite{Abramowitz1965}. 
Sums in the form of \cref{eq:bounds:real,eq:bounds:imag} can then be bound~\cite{Berry2015,Gilyen2018},
\begin{equation}
  \label{eq:bounds:sum}
      \sum_{\mathclap{k=0}}^{\infty} \abs[\big]{J_{2k+q}(\tau)}
  \le \sum_{\mathclap{k=0}}^{\infty} \frac{\abs{\tau/2}^{2k+q}}{(2k+q)!}
  <   \frac{\abs{\tau/2}^q}{q!} \sum_{\mathclap{k=0}}^{\infty} \qty(\frac{\tau}{2q})^{2k}
  =   \frac{\abs{\tau/2}^q/q!}{1-\qty(\tau/2q)^2}.
\end{equation}
Asserting that $\tau\le\nsteps/2$, we combine \cref{eq:bounds:sum,eq:bounds:real,eq:bounds:imag} to establish a closed-form asymptotic upper bound for $\maxeps$:
\begin{equation}
\label{eq:bounds:asyeps}
  \maxeps
  <   \qty{ \polyeps_a^2 + \polyeps_c^2 }^{1/2}
  \le \frac{8\abs{\tau/2}^{\nsteps/2+1}}{3(\nsteps/2+1)!}
             \qty{ 1 + \qty(\frac{\tau}{\nsteps+4})^2 }^{1/2} 
  \le \frac{4\sqrt5}{3\sqrt{\pi\nsteps}} \qty( \frac{e\abs{\tau}}{\nsteps+2} )^{\nsteps/2+1} \defeq \asyeps,
\end{equation}
where the final inequality results from the Sterling approximation.

Solving the r.h.s of \eq{bounds:asyeps} for $\nsteps$, one can derive QSP's optimal asymptotic query depth $\nsteps\approx e\abs{\tau} + \order{\makebig{\tfrac{\log(1/\asyeps)}{\log(e+\log(1/\asyeps)/\tau)}}}$~\cite{Gilyen2018}.  This(mostly) additive dependence on $\asyeps$ serves as the proof of the optimal resource scaling of the QSP implementation of Hamiltonian simulation.

We calculate the tighter numerical bound by computing the maxima in \cref{eq:bounds:real,eq:bounds:imag} exactly.
The first root of Bessel function $J_k(\tau)$ is known to occur outside $\abs{\tau}\ge k+1.85576k^{1/3}$~\cite{Krasikov2006}.  For $\nsteps\ge2\tau$, every Bessel function evaluation in \cref{eq:bounds:real,eq:bounds:imag} will be inside that function's first root, and will share the same sign.
We can then move the absolute value operation outside the sum, and 
compute \cref{eq:bounds:real,eq:bounds:imag} exactly as finite sums. Exploiting the identities $\sum_{k\in2\ints} J_{k}(\tau)=1$ and $\sum_{k\in2\ints+1}J_{k}(\tau)=\besselconst{}$ where 
$\besselconst{} \defeq \int_0^\tau d\tau'J_0(\tau') = (\pi\tau/2)(J_1(\tau)H_0(\tau)-J_0(\tau)H_1(\tau))+\tau J_0(\tau)$ and $H_k(\tau)$ are the Struve functions~\cite{Struve1882,Spanier1987}, we have, 
\begin{align}
  \polyeps_a
  &
  = \sum_{\mathclap{k=0}}^{\infty} \abs*{J_{2k+\nsteps/2+2}(\tau)}
  = \abs[\Bigg]{\sum_{\mathclap{k=0}}^{\infty} J_{2k+\nsteps/2+2}(\tau)}
  = \abs[\Bigg]{1 - J_0(\tau) - 2\sum_{\mathclap{k=1}}^{\nsteps/4} J_{2k}(\tau)},
  \\ \polyeps_c
  &
  = \sum_{\mathclap{k=0}}^{\infty} \abs*{J_{2k+\nsteps/2+1}(\tau)}
  = \abs[\Bigg]{\sum_{\mathclap{k=0}}^{\infty} {J_{2k+\nsteps/2+1}(\tau)}}
  = \abs[\Bigg]{\besselconst{} - 2\sum_{\mathclap{k=1}}^{\mathclap{\nsteps/4}} J_{2k-1}(\tau)}.
\end{align}
We can then take the numerical bound to be $\numeps\defeq\qty{\polyeps_a^2 + \polyeps_c^2}^{1/2}$ exactly, and differs from $\maxeps$ only in that $\polyeps_a$ and $\polyeps_c$ are maximized separately.

Because the numerical bound is an exact computation of the terms bounded by $\asyeps$, it is always the case that $\maxeps\le\numeps\le\asyeps$)

\secorsub{Circuit Implementation}
\label{apx:circ}

The heart of the normal-$\Lam$ quantum signal processor is the ``qubitized'' unitary signal operator $\uemb$.
Using the sum-of-unitaries Hamiltonian encoding from \apx{block}, the circuit spans three qubit registers: the $n$-qubit \TGT/ register containing the state being evolved, the $\nbits={\clog2\nops}$-qubit \CTL/ register required for qubitization, and the single \PHS/ qubit.
As described in \sect{alg:normal}, we can construct $\uemb$ from pairs of reflectors $\uselv,\,\ucz$ and projectors $\uprep,\,\uprep*$, where both $\uselv$ and $\ucz$ are conditioned on the $\ket-$ state of the \PHS/ qubit, and $\uselv$ acts on both the \CTL/ and \TGT/ registers while $\ucz$ and $\uprep$ act on just the \CTL/ qubits.  The specific circuit constructions we use for each subcircuit (which descend from those used in~\cite{Childs2017,Childs2017b}) are described here.

\suborsubsub{Projectors $\uprep$, $\uprep*$}
\label{apx:circ:prep}

As defined in \eq{apx:block}, the \PREP/ and \UPREP/ subroutines act entirely within the \CTL/ register to encode the coefficients $\vec\alpha\defeq\{\alpha_0,...,\alpha_{\nops-1}\}$ in the unitary decomposition of $\Lam$ (\eq{apx:haml}) into the state $\ket{\alpha}$:
\begin{equation}
  \label{eq:apx:circ:prep}
  \ket{0} \lra[\PREP/] \ket{\alpha} = \sum_{k=0}^{\nops-1}\sqrt{\alpha_k}\ket{k}
  \lra[\UPREP/] \ket{0}.
\end{equation}
Provided that \eq{apx:circ:prep} is satisfied and $\uprep*\uprep=\I/^{\otimes\nbits}$, the action of \PREP/ and \UPREP/ on other states in the \CTL/ register does not affect the behavior of the $\uemb$ circuit, and so can be left unspecified.

We implement $\uprep$ with the zero-ancilla recursive procedure outlined in \cite{Shende2006} (adapted from the implementation in \cite{Childs2017,Childs2017b}).  We require that $\nops=2^\nbits$ exactly, zero-padding $\vec\alpha$ if its length is not already a power of two.
Beginning with the $\nbits$-qubit state $\ket{\alpha}_\nbits$ (where the subscript indicates the number of qubits in the register)
and defining $\alpha'_k \defeq{} \abs{\alpha_k}+\abs{\alpha_{k+L/2}}$ and $\ang_k\defeq{}2\acos{\sqrt{\alpha_k/\alpha'_k}}$, we can decouple the most significant bit of $\ket{\alpha}_\nbits$:
\begin{equation}
  \sum_{k=0}^{\nops-1}\sqrt{\alpha_k}\ket{k}_\nbits
  = \sum_{k=0}^{\mathclap{\nops/2-1}}
        \bigg( \sqrt{\frac{\alpha_k}{\alpha'_k}}\ket0 
             + \sqrt{\frac{\alpha_{k+\nops/2}}{\alpha'_k}}\ket1 \bigg)
    \otimes \sqrt{\alpha'_k}\ket{k}_{\nbits-1}
  = \sum_{k=0}^{\mathclap{\nops/2-1}}
    \Ry\qty(\ang_k) \ket0
    \otimes
    \sqrt{\alpha'_k}\ket{k}_{\nbits-1}.
\end{equation}
We therefore define the operator,
\begin{equation}
  \label{eq:circ:prep:upr}
  \upr_{\nbits} 
  \defeq{} \sum_{k=0}^{\mathclap{\nops/2-1}} \Ry(\ang_k) \otimes \dyad{k}_{\nbits-1},
\end{equation}
which constructs $\ket{\alpha}_\nbits=\upr_{\nbits}\ket0\ket{\alpha'}_{\nbits-1}$ from the $(\nbits-1)$-qubit state $\ket{\alpha'}_{\nbits-1}\defeq\sum_{k<\nops/2}\alpha'_k\ket{k}$.
The state $\ket{\alpha'}_{\nbits-1}$ then has identical structure to $\ket{\alpha}_\nbits$, and so can similarly be constructed using an $(\nbits-1)$-qubit operator $\upr_{\nbits-1}$.  Beginning with 
the all-zeros state $\ket{0}_\nbits$, the full state $\ket{\alpha}_\nbits$ can be constructed with the recursive sequence of operators $\{\upr_1,...,\upr_{\nbits}\}$ shown in \fig{circ-prep}.

\begin{figure}[htb!]
   \centerline{\incsubcirc[5]{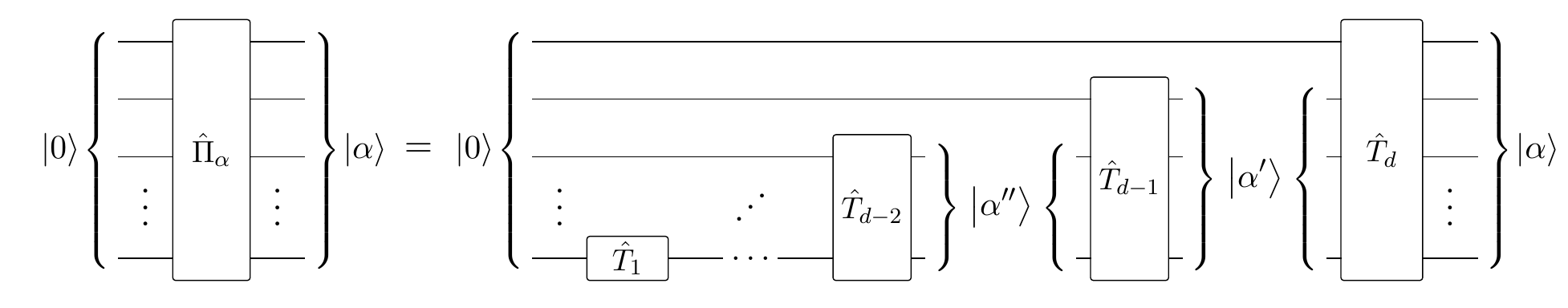}}
   \hfill \\
   \centerline{\incsubcirc[5]{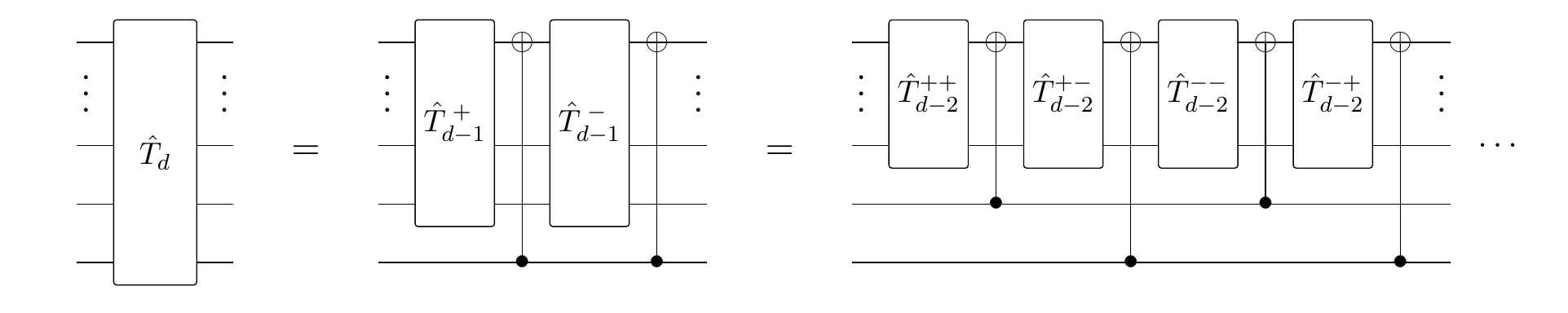}}
   \fcaption{\sfig{circ-prep} Recursive implementation of the \PREP/ subcircuit, using the $\upr_d$ operator defined in \eq{circ:prep:upr}, \sfig{circ-prep-td} recursive implementation of $\upr_d$}
\end{figure}

\newcommand{\uprcnot}{\Big(\PX/^{(\nbits\sm-1)}\otimes\dyad1 + \I/\otimes\dyad0\Big)}
The single-qubit instance $\upr_1$ requires just a single rotation gate $\upr_1=\Ry(\ang)$.  Larger instances $\upr_\nbits$ for $\nbits>1$ can themselves be derived recursively.
Defining the $(\nbits-1)$-qubit instances,
\begin{equation}
  \upr_{\nbits-1}^{\,(\pm)}
  \defeq{}
  \sum \Ry(\ang_k\pm\ang_{k+\nops/4}) \otimes\dyad{k}_{\nbits-1},
\end{equation}
and noting that conjugation of the target bit by \PX/ negates each rotation angle $\ang_k$ in the operator, we can use a pair of \CNOT/ gates to construct the $\nbits$-qubit operator,
\begin{align}
  \label{eq:apx:upr:decomp}
  \upr_d 
  & =
  \Big(\upr_{d-1}^{\,+}\otimes\I/\Big)
  \uprcnot
  \Big(\upr_{\nbits-1}^{\,-}\otimes\I/\Big)
  \uprcnot
  \nonumber \\ & =
  \uprcnot
  \Big(\upr_{d-1}^{\,+}\otimes\I/\Big)
  \uprcnot
  \Big(\upr_{\nbits-1}^{\,-}\otimes\I/\Big).
\end{align}
As shown in \fig{circ-prep-td}, for all but the first step of this recursive decomposition we can alternate between the two equivalent constructions in \eq{apx:upr:decomp}  operators can be arranged such that one of the two \CNOT/ gates is annihilated.

The full operator $\upr_\nbits$ therefore requires $2^{\nbits-1}$ single-qubit rotations and $2^{\nbits-1}$ \CNOT/ gates.  The total resource costs of the ancilla-free $\uprep$ circuit are then,
\begin{itemize}
  \item $2^\nbits-1$ single-qubit rotation gates,
  \item $2^\nbits-2$ \CNOT/ gates.
\end{itemize}

\suborsubsub{\SELV/ reflection}
\label{apx:circ:selv}

The second subcircuit we require is the reflection operator $\uselv$ introduced in \sect{alg:normal}.
As defined in \eq{apx:block}, for $\Lam$ represented as a weighted sum of unitaries the $\uselv$ operator selectively applies the unitary component $\ULam_k$ for each binary index state $\ket{k}$ in the \CTL/ register.  We additionally require that the $\uselv$ circuit be controlled by the $\ket-=\H/\ket1$ state of the \PHS/ qubit which is equivalent to expanding the index to the $(\nbits+1)$-qubit state $\ket-\ket{k}=\H/\ket1\otimes\ket{k}$.

Naively (\fig{circ-selv}), conditioning on a $(\nbits+1)$-qubit binary state would require a pair of $(\nbits+1)$-control \Toffoli/ gates for each of the $2\nops$ unitary elements $\ULam_k$ in $\Lam$.
However, as insightfully noticed in~\cite{Childs2017}, it turns out that we can better utilize ancilla bits to remove most of the complexity between consecutive indices.

This optimization arises when each $(\nbits+1)$-qubit \Toffoli/ is first constructed suboptimally, using $\nbits$ sequential 3-qubit \Toffoli/s and $\nbits$ ancilla qubits to construct a ``staircase'' AND operator.  For each index state $\ket-\ket{k}$ we implement both an `ascent' staircase (beginning with the MSB) prior to the application of $\ULam_k$, followed by a reversed, ancilla-clearing `descent' staircase.  This construction initially costs a total of $2\nops\nbits$ unparallelizable  3-qubit gates for the complete $\uselv$ circuit.  However, all but the final descent is immediately followed by the next \Toffoli/ gate's ascent, which differs only in whether individual controls are activated on the $\ket0$ or $\ket1$ state (as determined by the binary decomposition of the index value $k$).  For each MSB shared by the binary indices $\ket-\ket{k}$ and $\ket-\ket{k+1}$, we can therefore annihilate a pair of \Toffoli/ gates.

Every other step between indices can then be implemented with just a single \CNOT/ gate, while every fourth step (after some circuit optimizations described in~\cite{Childs2017}) requires a three-qubit \Toffoli/ and two \CNOT/ gates, every eighth step requires three \Toffoli/s and two \CNOT/s, etc.---in general the  number of steps requiring a descent of depth $w$ scales with $2^{-w}\nops$ so that the overall number of gates required for the \SELV/ circuit is linear in ${\nops}$.

The total resource costs of the \SELV/ circuit are then,
\begin{itemize}
\item $3\cdot2^{\nbits-1}-4$ Toffoli gates,
\item $3\cdot2^{\nbits-1}-2+\npauli$ \CNOT/ gates,
\item $\nbits$ ancilla qubits.
\end{itemize}

\cc{
We first replace each logarithmic-depth tree with a suboptimal linear-depth ``staircase'' AND operation, requiring $\nbits-1$ ancilla qubits and temporarily increasing the total \SELV/ depth to $\order*{2^\nbits\nbits}$.  Each Toffoli gate involves both a $\nbits$-gate ascent and an ancilla-clearing $\nbits$-gate descent. Because all but the final descent is immediately followed by the next Toffoli gate's ascent, which differs only in whether individual controls are activated on the $\ket0$ or $\ket1$ state (determining which binary index value $\dyad{k}$ that Toffoli is selecting for.  Stepping between indices $\dyad{k}$ and $\dyad{k+1}$, we only need to descend until we reach all the differing bits in the binary representation of $k$ and $(k+1)$.  

Assuming we compute the initial staircase in order of decreasing significance, every other step can then be implemented with just a CNOT gate, while every fourth step requires a three-qubit Toffoli and two CNOT gates, etc. -- in general the  number of steps requiring a descent of depth $w$ scales with $2^{-w}\nops$, so that the overall number of gates required for the \SELV/ circuit is linear in ${\nops}$.

Finally, we must implement the controlled $\Lam_k$ operators themselves.  If each $\Lam_k$ can be expressed as an $n$-qubit Pauli operator this requires at most $\npauli$ two-qubit Clifford gates, where $\npauli$ is the total number of single-qubit Pauli gates in the decomposition of $\Lam$.
}
\begin{figure}[htb!]
   \includecircuit[8]{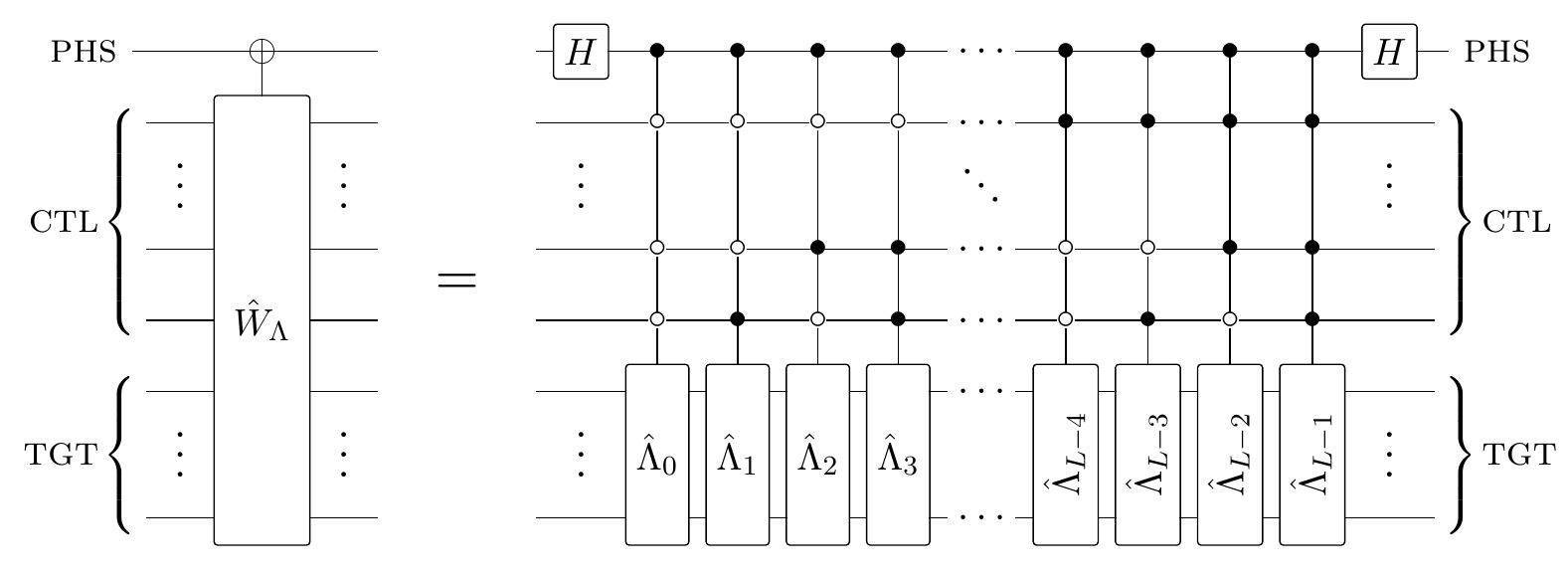}
   \fcaption{Na\"ive implementation of the \SELV/ circuit, implementing the unitary operation $\uselv=\sum_{k<\nops}\dyad{k}\otimes\Lam_k$ with $\nops$ sequential $\nbits$-Toffoli operators.}
   \label{fig:circ-selv}
\end{figure}

\suborsubsub{$\ucz$ reflection}
\label{apx:circ:refl}

Finally, we require an implementation of the diffusion operator $\ucz=2\dyad0-\I/$, again conditioned on the \PHS/-qubit's $\ket-$ state.  This controlled-$\ucz$ operation is equivalent to a single $\nbits$-control \Toffoli/ gate, conditioned on the $\nbits$-qubit zero state in the \CTL/ register and targeting the \PHS/ qubit.  In this case we implement the multi-control \Toffoli/ with a logarithmic-depth tree of 3-qubit \Toffoli/s using $\nbits-2$ ancilla qubits, as shown for $\nbits=4$ in \fig{circ-w0}.  As a standalone operation, we would then have to reverse all but the final \Toffoli/ gate in order to uncompute the intermediary ancilla bits.  However, as described in \sect{mthd:circ:opt}, in the context of the QSP circuit each pair of adjacent queries $\uemb,\,\uemb*$ (after the annihilation of projectors $\uprep\uprep*$) contains a pair of $\ucz$ circuits separated by just a \PHS/qubit rotation.  We can therefore elide the uncomputation and recomputation of the ancilla bits between the two applications (\fig{circ-w0-2}).  Each $\ucz$ component of each query then requires at most,
 $\nbits-1$ 3-qubit
 \begin{itemize}
  \item $\nbits-1$ 3-qubit \Toffoli/ gates,
  \item $\nbits-2$ ancilla qubits,
\end{itemize}
and is parallelizable to depth $\clog2\nbits$.

\begin{figure}[ht]
   \incsubcirc[5]{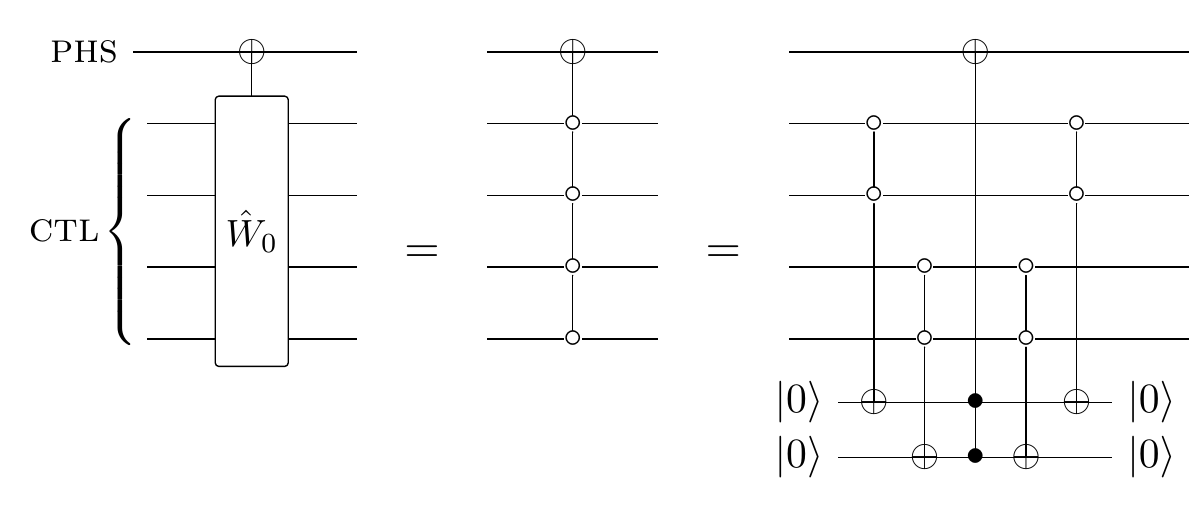}
   \incsubcirc[5]{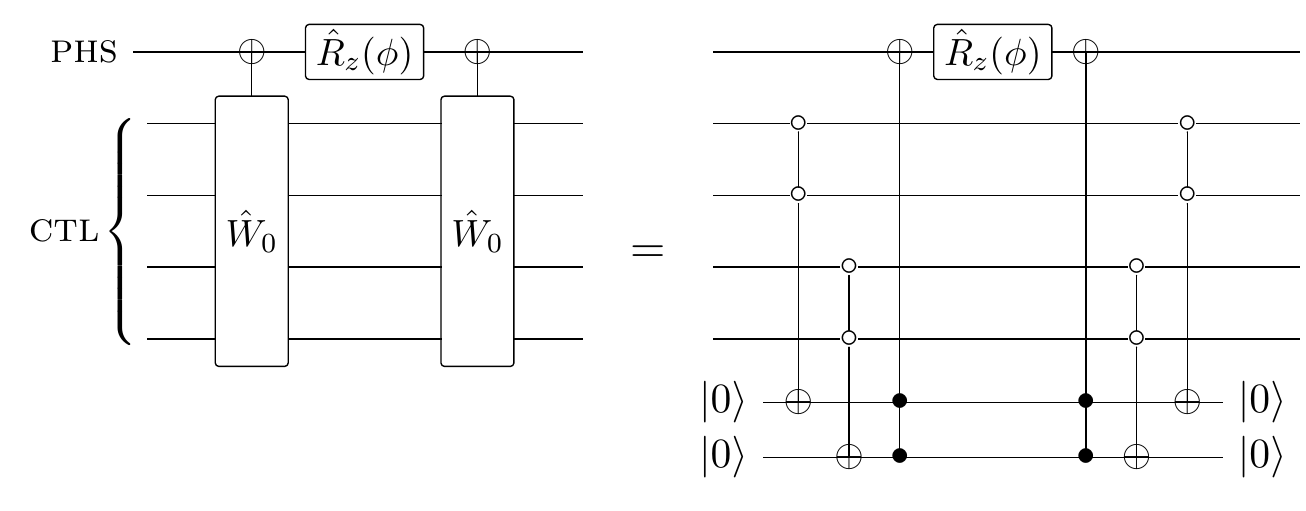}
   \fcaption{Circuit implementation of the diffusion operator $\ucz=2\dyad0-\I/$. Conditioned on the $\ket-$ state of the \PHS/ qubit \sfig{circ-w0}, $\ucz$ amounts to a $\nbits$-control \Toffoli/ gate with ``active low'' controls on the bits of the \CTL/ register, and can be implemented in logarithmic depth with $\nbits-2$ ancilla qubits. Between adjacent queries of $\uemb$ and $\uemb*$, $\ucz$ circuits are separated by just a \PHS/-qubit rotation \sfig{circ-w0-2}; in this case we can forgo the clearing of ancilla bits between the pair so that the combined complexity is the same as a single $\ucz$ (plus one \Toffoli/ gate)}
\end{figure}

\suborsubsub{Total resources}

The full QSP circuit comprises $\nsteps$ queries of $\uemb$, where each queries contains a \SELV/ reflection and a $\uprep$ projection, and all but the first and last queries contain half of a $\ucz$ subcircuit (as described in \sect{mthd:circ:opt}, we can bypass the reflection entirely in the first and final queries).
  In terms of primitive gates, the complete QSP circuit therefore comprises,
\begin{itemize}
\item $2^\nbits\nsteps+1$ single-qubit Rotation gates
\item $3\cdot2^{\nbits-1}\nsteps+\nbits\nsteps-5\nsteps$ Toffoli gates,
\item $13\nsteps2^{\nbits-2}-3\nsteps\nbits-\nsteps$ \CNOT/ gates,
\item $\nbits$ ancilla qubits.
\end{itemize}

We can further decompose the $\ucz$ and $\uselv$ circuits into Clifford+\T/ operations.  Because we only care about the target qubit state after any Toffoli gate in the tree, and further noting any relative phases will be undone when we reverse the tree to uncompute the ancilla bits, we can map each three- and four-qubit \Toffoli/ gate to the circuits shown in \fig{alg:circ:toffoli}.   Both constructions require no ancilla and are equivalent to the standard gates except for phase.  The three-qubit circuit (originally due to~\cite{Barenco1995}) requires four \T/ gates, two Hadamard gates, and three \CNOT/s, while the four-qubit circuit (borrowed from~\cite{Childs2017b}) requires eight \T/ gates, four Hadamards, and six \CNOT/s.
The total per-query resource costs of the QSP circuit implemented with Clifford+\T/ gates are then,
\begin{itemize}
  \item $\nsteps2^\nbits$ single-qubit Rotations
  \item $1.5\cdot2^\nbits-2$ \CNOT/ gates,
  \item $2$ \NOT/ gates,
  \item $\nbits$ ancilla qubits.
\end{itemize}

\begin{figure}[ht]
   \includecircuit[7]{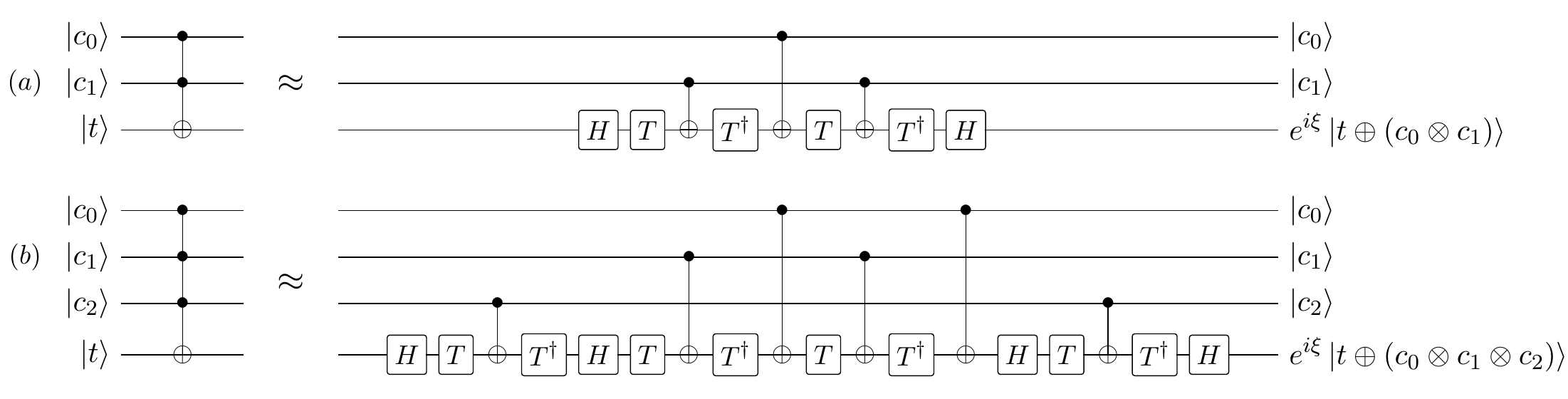}
   \fcaption{Pseudo-decompositions of (a) three-qubit~\cite{Barenco1995} and (b) four-qubit Toffoli gates, correct up to some input-dependent phase $e^{-\xi}$.  The four-qubit construction is borrowed from~\cite{Childs2017b}}
   \label{fig:alg:circ:toffoli}
\end{figure}


\secorsub{Coherent error optimization}
\label{apx:opt}

In this section, we consider the effects of systematic unitary errors on the QSP circuit or subcircuits in detail by independently enabling errors on specific subcircuits and gate subsets.  These steps elucidate a few simple circuit optimizations, which together significantly reduce the leading-order impact of coherent errors.
These optimized circuits are used for the general coherent error characterization in \sect{analysis:coherent}.
In practice, combined with
the simulation tools outlined in \sect{mthd:sim} (further described in \apx{sim}),
the separate consideration of errors on particular gate or circuit subsets also greatly
reduces simulation runtime and enables a more thorough characterization of the deleterious effects of coherent errors.

As a baseline, we first generate rough capacity plots of unoptimized QSP circuits with systematic amplitude errors restricted to each subcircuit.  As seen in \plot{opt:all}, for circuits configured near the expected optimal query depth, the dominant sources of error are the $\uprep$ and $\uselv$ subcircuits, which will therefore be the focus of our optimization.  

\optplot{opt:all}

\suborsubsub{Projectors $\uprep,\,\uprep*$ }
\label{apx:opt:prep}

Using the circuit implementation described in \apx{circ:prep}, the $\uprep=\dprep$ subcircuit consists of just $\CNOT/$ gates and single-qubit $\Ry(\cdot)$ rotations.  We therefore consider the impact of errors on each gate type separately.

If just the single-qubit $\Ry(\cdot)$ rotation gates are subject to multiplicative amplitude errors, the faulty circuits $\eprep,\,\eprep*$ will still satisfy $\eprep*\eprep=\I/$.
Accordingly, we expect the failure rate of the faulty circuit to be unchanged from that of the error-free circuit.
Because the angles of the single-qubit rotations are chosen to encode the coefficients $\{\alpha_0,...\alpha_{\nops-1}\}$ in the unitary decomposition $\Lam=\sum_{k<\nops}\alpha_k\Lam_k$ of the system Hamiltonian $\Lam$, the circuit constructed from the faulty gates is exactly equivalent to the ideal circuit constructed from a perturbed Hamiltonian $\eLam=\sum_\lambda\tilde{\alpha}_k\ULam_k$ for $\tilde{alpha}\approx\alpha_k$.
However, the infidelity of the final state is impacted due to the divergence of the faulty and ideal propagators,
\begin{equation}
  \infid
    \approx 1 - \norm[\big]{e^{i\Lam t}e^{-i\eLam t}}^2
    \sim \order{\epsilon^2\tau^2}.
  \label{eq:opt:prep-ry}
\end{equation}

We confirm this behavior by generating configuration plots.  At constant $\tau=20$ (\plot{prep:ry}, top), we observe that failure rate decays super-exponentially with increasing query depth independent of error rate, while infidelity becomes constant in the error-dominated performance region.  If we instead fix $\nsteps=64$ (bottom left), the observed infidelity is consistent with \eq{opt:prep-ry} in the fault-dominated region.  Finally (bottom right), we note that infidelity is remarkably independent of system size $n$ when errors are restricted to just these gates.

\prepryplot{prep:ry}

If we instead limit coherent errors to the $\uprep$ circuit's \CNOT/ gates, we break the symmetry between the faulty $\uprep$ and $\uprep*$ circuits so that in practice $\eprep*\eprep\ne\I/$. In this case both failure rate and infidelity to grow with $\nsteps$ as the evolving eigenstates leak out of their invariant subspace.

Some of this leakage can be constained using a standard echo technique.  We define $\corrprep\in\paulis^{\otimes\nbits}$ to be a $\nbits$-qubit Pauli correction operator, and $\uprep'\defeq\corrprep\uprep\corrprep$ to be the corresponding projector obtained by conjugating $\uprep$ by $\corrprep$.  As each bit in $\uprep$ is targeted by an even number of \CNOT/ gates, this conjugation amounts to reversing the signs of every $\Ry(\cdot)$ rotation acting on bits for which $\corrprep$ contains a $\PX/$ or $\PZ/$ gate (crucially, this means that we preserve the symmetry which benefited the $\Ry(\cdot)$-restricted error case).
By substituting $\uprep\mapsto\corrprep\uprep'\corrprep$ for every $\uprep$ (but not $\uprep*$) circuit in the QSP circuit, we can contain the component of the error which anti-commutes with $\corrprep$ so that it remains constant with increasing query depth.  Any additional error introduced by a faulty $\nbits$-gate $\corrprep$ operator is negligible in comparison to the $\order*{2^\nbits}$ faulty \CNOT/ gates making up the $\uprep$ circuit.

\Plot{prep:cx} compares capacity plots of the unmodified circuit (dashed lines) and that with $\corrprep=\PZ/^{\otimes\nbits}$ (solid lines).  We observe about a factor of three reduction in both error probability and average infidelity with this modification for optimally-configured QSP circuits when coherent errors are limited to the $\uprep$ circuit's \CNOT/ gates.

\prepcxtplot{prep:cx}

\suborsubsub{Reflection $\uselv$}
\label{apx:opt:selv}

The $\uselv$ operator is predominantly composed of logical operators in order to select the unitary component $\Lam_k$ given the index state $\ket{k}$ in the \CTL/ register while (ideally) leaving the state $\ket{k}$ unchanged.  Using the implementation in \apx{circ:selv}, the $\uselv$ subcircuit additionally requires $\nbits$ ancilla qubits, which are returned to $\ket0$ at the end of the operation.

After subcircuit elimination (\sect{mthd:circ:opt}), every pair of queries $\uemb,\,\uemb*$ contains a pair of $\uselv$ subcircuits separated by just a \PHS/-qubit rotation.
As we did for the $\uprep$ subcircuit, we can use a simple echo technique to promote the coherent cancellation of a component of the error term generated by each query.  Letting $\corrselv\in\paulis^{\otimes\nbits}$ be a second $\nbits$-qubit Pauli operator acting in the \CTL/ register, we define the conjugated reflection circuit, $\uselv'\defeq\corrselv\uselv\corrselv$.  We can construct $\uselv'$ by noting that $\PZ/$ gates acting on \CTL/ qubits trivially commute with $\uselv$, while a $\PX/$ or $\PY/$ gate acting on the $j$th qubit has the effect of swapping the indices $\ket*{k}\leftrightarrow\ket*{k\oplus2^{j}}$ and therefore requires swapping the unitary components $\Lam_k\leftrightarrow\Lam_{k\oplus2^j}$ between the $\uselv$ and $\uselv'$ implementations.

At $\tau=32$, by conjugating the first of each pair of $\uselv$ circuits by $\corrselv=\PZ/^{\otimes\nbits}$ we reduce at-capacity failure rate by about an order of magnitude and corresponding infidelity by about two orders of magnitude.  However, as seen in the capacity diagrams shown in \plot{selv:tau}, while the fidelity improvement is consistent for all $\tau\le512$, the asymptotic failure rate grows much more quickly, surpassing that of the unmodified circuit for $\tau>256$.

We can mitigate this accumulation by exploiting symmetry in the QSP sequence.
The Bessel function property $J_k(\tau)=(-1)^kJ_{-k}(\tau)$ results in an analogous symmetry $\phi_k=(-1)^{k}\phi_{\nsteps-k}$ (to first order in $\polyeps$).  To promote cancellation between corresponding terms in the resulting Fourier expansion, we modify the first $\nsteps/2$ $\uemb$ queries using the $\corrprep,\,\corrselv$ echo operators as described, but for the remaining $\nsteps/2$ queries we instead conjugate the \emph{second} of each pair of adjacent $\uselv$ circuits by $\corrselv$ and similarly substitute $\uprep*$ with $(\uprep*)'=\corrprep\uprep*\corrprep$ while leaving $\uprep$ unmodified. 
As shown in \plot{selv:tau}, the infidelity of the ``symmetric'' circuit is the same as that before this final modification, but growth of failure rate on $\tau$ is significantly diminished (at the cost of slightly worse performance for small $\tau$).

\selvplot{selv:tau}

\suborsubsub{Reflection $\ucz$}

Like the $\uselv$ circuit, the $\ucz$ operators appear in pairs in the QSP circuit, where in this case nearly half of each circuit can be annihilated (\sect{mthd:circ:opt}).
The small gate count of this subcircuit (relative to $\uselv$ and $\uprep$) means that it does not contribute significantly to the resolution of the overall circuit.  However, for $\corrprep=\PZ/^{\otimes\nbits}$ we can mitigate a piece of its small contribution by commuting the $\corrprep$ operator from a neighboring $\uprep'$ or $(\uprep*)'$ circuit to the middle of the pair of half-operators (as constructed in \apx{circ:refl}, the qubits in the \CTL/ register are only involved in the $\ucz$ circuit as controls to \Toffoli/ gates, and so each \PZ/ gate in $\corrprep$ trivially commutes).


\secorsub{Classical simulation}
\label{apx:sim}

As detailed in \apx{circ}, the QSP circuit as implemented requires $n+2\nbits$ qubits (and even the most qubit-efficient construction would require a minimum of $n+\nbits+1$ qubits).  Though for applicable Hamiltonians this converges to $n$ in leading order, it presents relatively large second-order terms for small $n$.  Our test Hamiltonian (\eq{intro:haml}) contains a modest $4n$ terms in its Pauli decomposition, which translates to $n+2\clog2n+4$ required qubits, quickly overwhelming the capability of a naive state-array style classical simulator.
\cc{A naive array representation of the full qubit state then becomes cumbersome at relatively small $n$.  By $n=15$ (requiring 27 qubits), an array of double-precision complex values would require 2GB of memory, all of which must be acted upon with every quantum gate.}

Instead, we adopt the vector-tree structure introduced in~\cite{Obenland1998}, in which quantum states are distributed into a tree structure with a subset of qubits represented in smaller quantum state arrays at the leaves and the remaining qubits held in single-binary-index interior link nodes.  This structure has a number of advantages over the simplistic state array:
\begin{itemize}
  \item Because branches are created and destroyed adaptively, the system size (i.e. the memory required to represent the state and the number of primitive operations to required to execute a quantum gate) can better conform to the size of the \emph{occupied} Hilbert space, rather than the total Hilbert space of an arbitrary $N$-qubit system
  \item Diagonal gates and some logical gates that act exclusively on qubits represented by internal links can be executed simply by updating the phase values at link level, rather than multiplicatively across an entire state array,
  \item The remaining arithmetic operations are broadcast to smaller state array nodes, where their action on different subspaces of the overall Hilbert space can be evaluated in a parallel, distributed, or executed in a more cache-friendly way.
\end{itemize}

A remarkable feature of the QSP circuit is that qubit requirement is dominated by the \TGT/ register, which for a Pauli-decomposed Hamiltonian is subject only to two-qubit Clifford operations.  
We are therefore motivated to further adapt our system to take advantage of well-known techniques for stabilizer-state simulation.
By the Gottesman-Knill theorem, quantum circuits comprising exclusively Clifford operations can be simulated classically in polynomial time by tracking updates to a set of $n$ stabilizer generators which together stabilize a unique stabilizer state $\ket{\psi_s}$.  Many techniques have extended this model to arbitrary circuits, with runtimes growing exponentially only in the number of non-Clifford gates~\cite{}.  These systems tend to be special purpose, performing poorly for circuits with a high number of non-Clifford gates or which do not easily fit in the stabilizer methodology.  It turns out that we can overcome many of these limitations by tying the stabilizer methods to the general-purpose vector-tree structure.

We adapt the stabilizer/destabilizer construction introduced in \cite{Aaronson2004}, in which a set of $n$ orthogonal ``destabilizers''  $\{\hat{D}_0,...,\hat{D}_{n-1};\;\hat{D}_k\in\mathcal{P}^n\}$ are updated along with the $n$ stabilizers generators.  We use the destabilizers to define a new logical basis: given the stabilizer state $\ket{\psi_s}$, we define the destabilizer state,
\begin{equation}
 \ket{\psi_d}=\ket{d_{n-1}...d_{0}}
 \defeq \prod_k \hat{D}_k^{d_k} \ket{\psi_s}.
\end{equation}
By design, the $n$ destabilizers and $\ket{\psi_s}$ form a complete basis of $\mathcal{H}^{2^n\times2^n}$, so that any quantum state can be written as a linear combination of destabilizer states $\ket{\psi_d}$.
Using this basis for our simulation, we can continue to use the vector-tree structure to apply non-Clifford gates and hold non-stabilizer states.  Stabilizer simulation then proceeds as in~\cite{Aaronson2004}: we define a binary tableau representing the stabilizer and destabilizer generators, which is updated in polynomial time via bitwise operations with each Clifford gate application.  The set of stabilizers and destabilizers then defines both the state $\ket{\psi_s}$ and the basis for the vector tree, so that non-Clifford gates can be mapped to corresponding gates in the destabilizer basis by checking which of the $n$ destabilizer generators the gate does not commute with.
Overall, the stabilizer front-end of our simulator serves to absorb any Clifford operations in the circuit (crucially, including discrete Pauli errors), while maintaining a destabilizer basis with which to map every non-Clifford operation to an equivalent gate in the tree structure and therefore preserving the advantages of of that construction.  The adaptive basis can further the extend these advantages to a more general class of circuits, for example eliminating unnecessary branch creation and annihilation by absorbing Hadamard gates.

As described in the main text (\sect{mthd:sim}), for circuits with only Clifford errors, we generally find that simulation times can be reduced by about factor by using the hybrid stabilizer-tree simulator in comparison to the vector-tree simulator alone.  This advantage diminishes when the circuit comes to be dominated by non-Clifford operations, as in the case of a coherent error model, but remains beneficial if these errors are applied only to subsets of the circuit.
(Note that all of these methods were spot-checked against each other to confirm equivalent behavior.)

\

To model faulty evolution, the simulator affixes single-qubit error operators to each qubit involved in each executed gate.  Errors are determined and placed at runtime from parameterized error models specified in a JSON configuration file.  The tool allows each error model to be restricted to a subset of qubits, gate types, or subcircuit (separated via directives incorporated into the QASM code).  Stochastic errors are placed according to a characteristic distribution and specified error rate, while coherent errors will deterministically be placed after every relevant gate.

Because we are post-selecting runs without any projected error, we significantly reduce the requisite number of Monte Carlo trials to characterize success probability by replacing the randomized quantum measurement operators with deterministic $\dyad{0}$ projectors.  With each projection, the simulator computes the total norm of the remaining state, renormalizes the state vector, and updates a running probability for the full simulation run (computed as the running product of the remaining probability after each projection).

\end{document}